\documentclass[twocolumn,times,twocolappendix]{aastex631}
\newcommand{\LCO}{\affiliation{Las Cumbres Observatory, 6740 Cortona Drive, Suite 102, Goleta, CA 93117-5575, USA}}
\newcommand{\UCSB}{\affiliation{Department of Physics, University of California, Santa Barbara, CA 93106-9530, USA}}

\newcommand{\STScI}{\affiliation{Space Telescope Science Institute, 3700 San Martin Drive, Baltimore, MD 21218-2410, USA}}
\newcommand{\UT}{\affiliation{University of Texas at Austin, 1 University Station C1400, Austin, TX 78712-0259, USA}}

\newcommand{\Itagaki}{\affiliation{Itagaki Astronomical Observatory, Yamagata 990-2492, Japan}}
\newcommand{\Einstein}{\altaffiliation{NASA Einstein Fellow}}

\newcommand{\CfA}{\affiliation{Center for Astrophysics \textbar{} Harvard \& Smithsonian, 60 Garden Street, Cambridge, MA 02138-1516, USA}}
\newcommand{\UA}{\affiliation{Steward Observatory, University of Arizona, 933 North Cherry Avenue, Tucson, AZ 85721-0065, USA}}

\newcommand{\TAU}{\affiliation{School of Physics and Astronomy, Tel Aviv University, Tel Aviv 69978, Israel}}

\newcommand{\Columbia}{\affiliation{Department of Physics and Columbia Astrophysics Laboratory, Columbia University, Pupin Hall, New York, NY 10027, USA}}
\newcommand{\Flatiron}{\affiliation{Center for Computational Astrophysics, Flatiron Institute, 162 5th Avenue, New York, NY 10010-5902, USA}}
\newcommand{\CIERA}{\affiliation{Center for Interdisciplinary Exploration and Research in Astrophysics (CIERA) and Department of Physics and Astronomy, \\Northwestern University, 1800 Sherman Avenue, 8th Floor, Evanston, IL 60201, USA}}

\newcommand{\TAMU}{\affiliation{Department of Physics and Astronomy, Texas A\&M University, 4242 TAMU, College Station, TX 77843-4242, USA}}
\newcommand{\Mitchell}{\affiliation{George P.\ and Cynthia Woods Mitchell Institute for Fundamental Physics \& Astronomy, College Station, TX 77843, USA}}

\newcommand{\IAIFI}{\affiliation{The NSF AI Institute for Artificial Intelligence and Fundamental Interactions, USA}}

\newcommand{\Catalyst}{\altaffiliation{LSSTC Catalyst Fellow}}
\newcommand{\TAPIR}{\affiliation{TAPIR, Mailcode 350-17, California Institute of Technology, Pasadena, CA 91125-0001, USA}}
\newcommand{\Tokyo}{\affiliation{Research Center for the Early Universe, School of Science, The University of Tokyo, 7-3-1 Hongo, Bunkyo-ku, Tokyo 113-0033, Japan}}
\newcommand{\MIT}{\affiliation{MIT-Kavli Institute for Astrophysics and Space Research, 77 Massachusetts Ave., Cambridge, MA 02139, USA}}
\newcommand{\Konkoly}{\affiliation{Konkoly Observatory, CSFK, MTA Center of Excellence, Konkoly-Thege M. \'ut 15-17, Budapest, 1121, Hungary}}
\newcommand{\ELTE}{\affiliation{ELTE E\"otv\"os Lor\'and University, Institute of Physics and Astronomy, P\'azm\'any P\'eter s\'et\'any 1/A, Budapest, 1117 Hungary}}
\newcommand{\Szeged}{\affiliation{Department of Experimental Physics, University of Szeged, D\'om t\'er 9, Szeged, 6720, Hungary}}
\newcommand{\Yonsei}{\affiliation{Department of Physics, Yonsei University, Seoul 03722, Republic of Korea}}

\usepackage{xcolor}

\usepackage{amsmath}

\def\Ang{\textup{\AA}}

\begin{document}

\shorttitle{From Discovery to the First Month of the Type II Supernova 2023ixf}
\shortauthors{Hiramatsu et al.}

\title{\bf \Large From Discovery to the First Month of the Type II Supernova\,2023ixf: High and Variable Mass Loss in the Final Year before Explosion}


\author[0000-0002-1125-9187]{Daichi Hiramatsu}
\CfA\IAIFI

\correspondingauthor{Daichi~Hiramatsu}
\email{daichi.hiramatsu@cfa.harvard.edu}

\author[0000-0002-6347-3089]{Daichi~Tsuna}
\TAPIR\Tokyo

\author[0000-0002-9392-9681]{Edo~Berger}
\CfA\IAIFI

\author{Koichi~Itagaki}
\Itagaki

\author[0000-0003-1012-3031]{Jared~A.~Goldberg}
\Flatiron

\author[0000-0001-6395-6702]{Sebastian~Gomez}
\STScI

\author[0000-0002-8989-0542]{Kishalay~De}
\Einstein\MIT

\author[0000-0002-0832-2974]{Griffin~Hosseinzadeh}
\UA

\author[0000-0002-4924-444X]{K.~Azalee~Bostroem}
\Catalyst\UA


\author[0000-0001-6272-5507]{Peter~J.~Brown}
\TAMU\Mitchell

\author[0000-0001-7090-4898]{Iair~Arcavi}
\TAU

\author[0000-0001-6637-5401]{Allyson~Bieryla}
\CfA

\author[0000-0003-0526-2248]{Peter~K.~Blanchard}
\CIERA

\author[0000-0002-9789-5474]{Gilbert~A.~Esquerdo}
\CfA

\author[0000-0003-4914-5625]{Joseph~Farah}
\LCO\UCSB

\author[0000-0003-4253-656X]{D.~Andrew~Howell}
\LCO\UCSB

\author[0000-0002-9350-6793]{Tatsuya~Matsumoto}
\Columbia

\author[0000-0001-5807-7893]{Curtis~McCully}
\LCO\UCSB

\author[0000-0001-9570-0584]{Megan~Newsome}
\LCO\UCSB

\author[0000-0003-0209-9246]{Estefania~Padilla~Gonzalez}
\LCO\UCSB

\author[0000-0002-7472-1279]{Craig~Pellegrino}
\LCO\UCSB

\author[0000-0001-9214-7437]{Jaehyon~Rhee}
\CfA\Yonsei

\author[0000-0003-0794-5982]{Giacomo~Terreran}
\LCO\UCSB

\author[0000-0001-8764-7832]{J\'ozsef~Vink\'o}
\UT\Konkoly\ELTE\Szeged

\author[0000-0003-1349-6538]{J.~Craig~Wheeler}
\UT

\begin{abstract}

We present the discovery of the Type II supernova SN~2023ixf in M101 and follow-up photometric and spectroscopic observations, respectively, in the first month and week of its evolution. 
Our discovery was made within a day of estimated first light, and the following light curve is characterized by a rapid rise ($\approx5$ days) to a luminous peak ($M_V\approx-18.2$\,mag) and plateau ($M_V\approx-17.6$\,mag) extending to $30$ days with a fast decline rate of $\approx0.03$\,mag\,day$^{-1}$. During the rising phase, $U-V$ color shows blueward evolution, followed by redward evolution in the plateau phase. 
Prominent flash features of hydrogen, helium, carbon, and nitrogen dominate the spectra up to $\approx5$ days after first light, with a transition to a higher ionization state in the first $\approx2$ days.
Both the $U-V$ color and flash ionization states suggest a rise in the temperature, indicative of a delayed shock breakout inside dense circumstellar material (CSM). From the timescales of CSM interaction, we estimate its compact radial extent of $\sim(3-7)\times10^{14}$\,cm. 
We then construct numerical light-curve models based on both continuous and eruptive mass-loss scenarios shortly before explosion. For the continuous mass-loss scenario, we infer a range of mass-loss history with $0.1-1.0\,M_\odot\,{\rm yr}^{-1}$ in the final $2-1$ yr before explosion, with a potentially decreasing mass loss of $0.01-0.1\,M_\odot\,{\rm yr}^{-1}$ in $\sim0.7-0.4$ yr toward the explosion.
For the eruptive mass-loss scenario, we favor eruptions releasing $0.3-1\,M_\odot$ of the envelope at about a year before explosion, which result in CSM with mass and extent similar to the continuous scenario.
We discuss the implications of the available multiwavelength constraints obtained thus far on the progenitor candidate and SN~2023ixf to our variable CSM models. 

\end{abstract}

\keywords{
\href{https://vocabs.ardc.edu.au/repository/api/lda/aas/the-unified-astronomy-thesaurus/current/resource.html?uri=http://astrothesaurus.org/uat/1668}{Supernovae (1668)}; 
\href{https://vocabs.ardc.edu.au/repository/api/lda/aas/the-unified-astronomy-thesaurus/current/resource.html?uri=http://astrothesaurus.org/uat/304}{Core-collapse supernovae (304)}; 
\href{https://vocabs.ardc.edu.au/repository/api/lda/aas/the-unified-astronomy-thesaurus/current/resource.html?uri=http://astrothesaurus.org/uat/1731}{Type II supernovae (1731)}; 
\href{https://vocabs.ardc.edu.au/repository/api/lda/aas/the-unified-astronomy-thesaurus/current/resource.html?uri=http://astrothesaurus.org/uat/732}{Massive stars (732)};
\href{https://vocabs.ardc.edu.au/repository/api/lda/aas/the-unified-astronomy-thesaurus/current/resource.html?uri=http://astrothesaurus.org/uat/1661}{Supergiant stars (1661)};
\href{https://vocabs.ardc.edu.au/repository/api/lda/aas/the-unified-astronomy-thesaurus/current/resource.html?uri=http://astrothesaurus.org/uat/1375}{Red supergiant stars (1375)};
\href{https://vocabs.ardc.edu.au/repository/api/lda/aas/the-unified-astronomy-thesaurus/current/resource.html?uri=http://astrothesaurus.org/uat/1613}{Stellar mass loss (1613)};
\href{https://vocabs.ardc.edu.au/repository/api/lda/aas/the-unified-astronomy-thesaurus/current/resource.html?uri=http://astrothesaurus.org/uat/241}{Circumstellar matter (241)}
}

\section{Introduction}
\label{sec:intro}

The majority of massive stars (zero-age main-sequence (ZAMS) masses of $M_{\mathrm{ZAMS}} \approx 8-25\,M_{\odot}$) end their lives when their iron cores collapse, leading to explosions as hydrogen-rich (H-rich) Type II supernovae (SNe~II; see, e.g., \citealt{Smartt2009ARA&A..47...63S, Smartt2015PASA...32...16S, Arcavi2017hsn..book..239A}, for reviews). 
While there is consensus that the progenitors of SNe~II are red supergiants (RSGs) from direct progenitor identifications (e.g., \citealt{VanDyk2017hsn..book..693V}), evidence is mounting that they experience elevated mass loss in the final months to decades before explosion, which is not predicted from standard stellar evolution theory. The presence of circumstellar material (CSM) from such mass loss can be inferred from early light-curve excess above shock-cooling emission (e.g., \citealt{Morozova2018ApJ...858...15,Moriya2018MNRAS.476.2840M,Forster2018NatAs...2..808F,Hiramatsu2021NatAs...5..903H}), narrow high-ionization emission lines, so-called flash features excited by the radiation from shock breakout or the SN--CSM shock interface (e.g., \citealt{Gal-Yam2014Natur.509..471G,Yaron2017NatPh..13..510Y,Dessart2017A&A...605A..83D,Boian2019A&A...621A.109B,Boian2020MNRAS.496.1325B,Terreran2022ApJ...926...20T,Jacobson-Galan2022ApJ...924...15}), and/or bremsstrahlung X-ray and synchrotron radio/millimeter emission from SN--CSM shock interaction (e.g., \citealt{Chevalier2017hsn..book..875C}).

The recent discovery of Type~II~SN~2023ixf in M101 (\S\ref{sec:disc}) has allowed intensive multiwavelength observations. A dust-obscured RSG progenitor candidate with a possible periodic variability ($\approx1000$-day period over the past $\approx13$ yr) has been identified in pre-explosion \textit{Hubble Space Telescope}, \textit{Spitzer Space Telescope}, and ground-based optical to infrared images \citep{Pledger2023ApJ...953L..14P,Kilpatrick2023ApJ...952L..23K,Jencson2023ApJ...952L..30J,Neustadt2023arXiv230606162N,Soraisam2023arXiv230610783S}.
Post-explosion observations have revealed early light curve excess in the optical bands \citep{Jacobson-Galan2023arXiv230604721J,Hosseinzadeh2023ApJ...953L..16H,Teja2023ApJ...954L..12T,Sgro2023RNAAS...7..141S}, flash features in low-to-high-resolution optical spectra \citep{Yamanaka2023PASJ..tmp...66Y,Jacobson-Galan2023arXiv230604721J,Smith2023arXiv230607964S,Bostroem2023arXiv230610119B,Teja2023ApJ...954L..12T}, polarization measurements in spectropolarimetry \citep{Vasylyev2023arXiv230701268V}, early X-ray detections \citep{Grefenstette2023ApJ...952L...3G}, and millimeter nondetections \citep{Berger2023ApJ...951L..31B}.

Here we report our discovery of SN~2023ixf in \S\ref{sec:disc} and follow-up photometric and spectroscopic observations, respectively, in the first month and week of its evolution in \S\ref{sec:obs}. In \S\ref{sec:ana}, we analyze the light curves and spectra. We then construct numerical light-curve models based on both continuous and eruptive mass-loss scenarios to extract the CSM formation history shortly before explosion in \S\ref{sec:mod}. Using this information, we discuss a consistent picture among the available observational constraints. Finally, we summarize our findings in \S\ref{sec:sum}.

\section{Discovery and Classification}
\label{sec:disc}

\begin{figure*}
    \centering
    \includegraphics[width=1.0\textwidth]{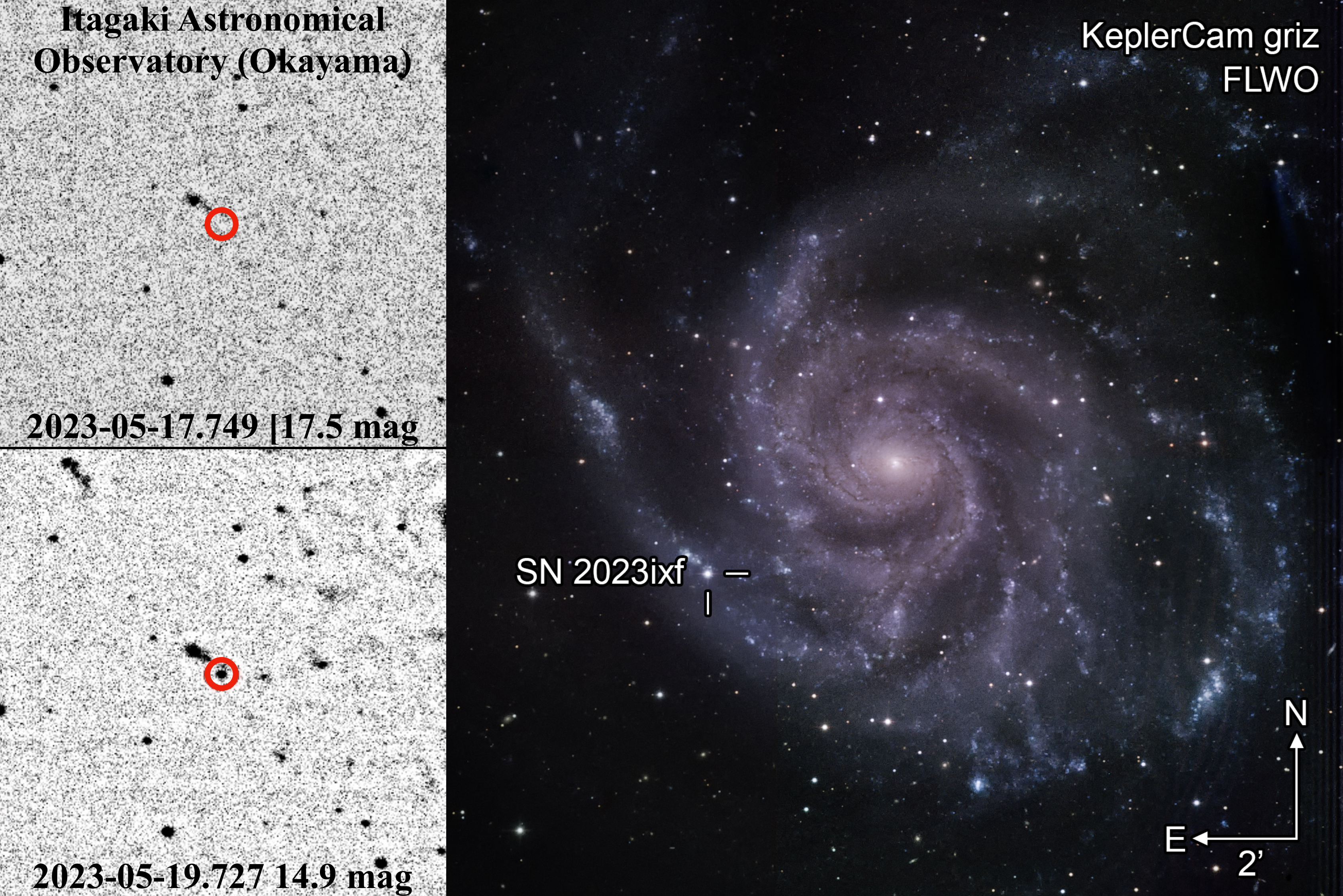}
\caption{\textit{Left}: last nondetection (\textit{top}) and discovery (\textit{bottom}) images taken by the Itagaki Astronomical Observatory. The red circle marks the position of SN~2023ixf.
\textit{Right}: color-composite ($griz$-band) image of SN~2023ixf and its host galaxy M101 taken by the Fred Lawrence Whipple Observatory. SN~2023ixf is located on a spiral arm with a nearby H~{\sc ii} region roughly at solar metallicity and young ($10^{6}-10^{7}$ yr) stellar population \citep{Hu2018ApJ...854...68H}. 
}
\label{fig:m101}
\end{figure*}


We \citep{Itagaki2023TNSTR1158....1I} discovered SN~2023ixf in M101 (Figure~\ref{fig:m101}) on 2023 May 19.727 (UT dates are used throughout; ${\rm MJD} = 60083.727$) at an unfiltered magnitude of $14.9\pm0.1$ at $\text{R.A.}=14^{\text{h}}03^{\text{m}}38^{\text{s}}.580$ and $\text{decl.}=+54^{\circ}18'42".10$ with the Itagaki Astronomical Observatory 0.35\,m telescope (Okayama, Japan) + KAF-1001E CCD. The last nondetection was obtained with the same setup at 17.5 mag on 2023 May 17.749 (${\rm MJD} = 60081.749$).
Subsequently, several other observers have reported limits and detections in the time window between our last nondetection and discovery \citep{Perley2023TNSAN.120....1P,Filippenko2023TNSAN.123....1F,Fulton2023TNSAN.124....1F,Mao2023TNSAN.130....1M,Koltenbah2023TNSAN.144....1K,Chufarin2023TNSAN.150....1C}. By taking the midpoint of the $\gtrsim 20.4$ mag limit on 2023 May 18.660 (${\rm MJD}=60082.660$) from \citet{Mao2023TNSAN.130....1M} and the detection with $18.76\pm0.25$ mag on 2023 May 18.826 (${\rm MJD}=60082.826$) from \citet{Chufarin2023TNSAN.150....1C}, we estimate the epoch of SN first light (i.e., the moment when the SN light emerges from the stellar/CSM surface)\footnote{This estimate is somewhat limited by the instrument sensitivity since the magnitude limit of $20.4$ corresponds to a luminosity limit of $\sim2\times10^{39}$\,erg\,s$^{-1}$, roughly five times the estimated progenitor luminosity \citep{Neustadt2023arXiv230606162N}. We assume that the SN light reached the luminosity limit from the progenitor luminosity within the error in the epoch of SN first light ($\approx2$ hr).} to be ${\rm MJD} = 60082.743 \pm 0.083$, corresponding to $0.98$ days before our discovery.

\citet{Perley2023TNSAN.119....1P} obtained an optical spectrum of SN~2023ixf on 2023 May 19.93 (${\rm MJD} = 60083.93$; 4.9 hr after discovery) with the SPectrograph for the Rapid Acquisition of Transients (SPRAT; \citealt{Piascik2014SPIE.9147E..8HP}) on the Liverpool Telescope (LT; \citealt{Steele2004SPIE.5489..679S}), classifying it as an SN~II with flash ionization features.
Subsequent spectra have been obtained and reported by \cite{Teja2023TNSCR1233....1T} with the Himalaya Faint Object Spectrograph (HFOSC) on the 2 m Himalayan Chandra Telescope (HCT),\footnote{\url{http://www.iiap.res.in/iao/hfosc.html}} \cite{Stritzinger2023TNSAN.145....1S} with the Alhambra Faint Object
Spectrograph and Camera (ALFOSC) on the Nordic Optical Telescope (NOT), \cite{Zhang2023TNSAN.132....1Z} with the Beijing Faint Object Spectrograph and Camera (BFOSC) on the Xinglong 2.16 m Telescope \citep{Fan2016PASP..128k5005F}, and \cite{Yaron2023TNSCR1233....1T} with the Spectral Energy Distribution Machine (SEDM; \citealt{Blagorodnova2018PASP..130c5003B}) on the Palomar 60-inch telescope, confirming the SN~II classification.

In this work, we adopt a redshift of $z=0.000804\pm0.000007$ \citep{deVaucouleurs1991rc3..book.....D}\footnote{Via the NASA/IPAC Extragalactic Database: \url{http://ned.ipac.caltech.edu/}} and luminosity distance of $d_L=6.9\pm0.1$\,Mpc (distance modulus of $\mu=29.194\pm0.039$ mag; \citealt{Riess2022ApJ...934L...7R}) to M101. We use the estimated first light (${\rm MJD} = 60082.743 \pm 0.083$) as the zeropoint reference for all phases unless otherwise specified.

\section{Observations and Data Reduction}
\label{sec:obs}

\subsection{Photometry}
\label{sec:phot}

Following the discovery, we continued monitoring SN~2023ixf with the Itagaki Astronomical Observatory 0.35, 0.5, and 0.6\,m telescopes (Kochi, Okayama, and Yamagata, Japan) + unfiltered KAF-1001E CCD. Using our custom software, the aperture photometry was extracted and calibrated to Vega magnitudes from the Fourth US Naval Observatory CCD Astrograph Catalog \citep{Zacharias2013AJ....145...44Z}.
Additionally, we include the prediscovery amateur points listed in \S\ref{sec:disc} in the subsequent analysis.

Through the Global Supernova Project \citep{Howell2017AAS...23031803H}, we obtained Las Cumbres Observatory (LCO; \citealt{Brown2013PASP..125.1031B}) \textit{UBVgriz}-band imaging with the SBIG and Sinistro cameras on the network of 0.4 and 1.0\,m telescopes at the Haleakal$\bar{\text{a}}$ Observatory (Hawaii, USA), McDonald Observatory (Texas, USA), and Teide Observatory (Canary Islands, Spain) starting on 2023 May 19.97 (${\rm MJD}=60083.97$; 5.8 hr after discovery). The initial photometry up to 2023 May 30 (${\rm MJD}=60094$) has been presented in \cite{Hosseinzadeh2023ApJ...953L..16H}, and we extend it to 2023 June 18 (${\rm MJD}=60113$; 30 days after discovery) following the same reduction procedures; \textit{UBV} and \textit{griz} point-spread function (PSF) photometry was calibrated to Vega \citep{Landolt1983AJ.....88..439L,Landolt1992AJ....104..340L} and AB \citep{Albareti2017ApJS..233...25A} magnitudes, respectively, using \texttt{lcogtsnpipe}\footnote{\url{https://github.com/LCOGT/lcogtsnpipe}} \citep{Valenti2016MNRAS.459.3939V}.

In addition, we triggered the Neil Gehrels \textit{Swift} Observatory Ultraviolet/Optical Telescope (UVOT) follow-up imaging starting on 2023 May 20.27 (${\rm MJD}=60084.27$; 13 hr after discovery).
The aperture photometry was conducted and calibrated to Vega magnitudes using the pipeline for the \textit{Swift} Optical Ultraviolet Supernova Archive (SOUSA; \citealt{Brown2014Ap&SS.354...89B}), including the zero-points from \cite{Breeveld2011AIPC.1358..373B} and the sensitivity correction from 2020 September. The initial photometry up to 2023 May 21 (${\rm MJD}=60085$) has been presented in \cite{Hosseinzadeh2023ApJ...953L..16H}, and we extend it to 2023 June 9 (${\rm MJD}=60104$). Unfortunately, the majority of the early \textit{Swift} UVOT observations were saturated, and we only include the unsaturated observations analyzed by the standard SOUSA procedures.

Through our FLEET program \citep{Gomez2020ApJ...904...74G,Gomez2023ApJ...949..114G}, we also obtained \textit{griz}-band imaging with  KeplerCam \citep{Szentgyorgyi2005AAS...20711010S} on the 1.2 m Telescope at the Fred Lawrence Whipple Observatory (FLWO; Arizona, USA) starting on 2023 May 21.32 (${\rm MJD}=60085.32$; 1.59 days after discovery). The PSF photometry was extracted and calibrated to AB magnitudes from the Pan-STARRS1 \citep{Chambers2016arXiv161205560C} Data Release 2 \citep{Flewelling2020ApJS..251....7F}.

To explore possible pre-explosion variability of SN~2023ixf, we also processed and examined the Zwicky Transient Facility (ZTF; \citealt{Bellm2019PASP..131a8002B, Graham2019PASP..131g8001G}), Asteroid Terrestrial-impact Last Alert System (ATLAS; \citealt{Tonry2018PASP..130f4505T, Smith2020PASP..132h5002S}), and Wide-field Infrared Survey Explorer (WISE; \citealt{Wright2010AJ....140.1868W,Mainzer2014ApJ...792...30M}) survey data.  
ZTF and ATLAS photometry were retrieved, respectively, from the ZTF forced-photometry service\footnote{\url{https://ztfweb.ipac.caltech.edu/cgi-bin/requestForcedPhotometry.cgi}} \citep{Masci2019PASP..131a8003M,Masci2023arXiv230516279M} in the $g$, $r$, and $i$ bands (from 2018 March 21; ${\rm MJD}=58198$) and the ATLAS forced photometry server\footnote{\url{https://fallingstar-data.com/forcedphot/}} \citep{Shingles2021TNSAN...7....1S} in the $c$ and $o$ bands (from 2015 December 30; ${\rm MJD}=57386$). Then, we stacked them with a time window of every 10 days ($\lesssim1\%$ of the observed progenitor period; \citealt{Kilpatrick2023ApJ...952L..23K,Jencson2023ApJ...952L..30J,Soraisam2023arXiv230610783S}) for each filter to obtain deeper measurements.

For the ongoing NEOWISE all-sky survey in the \textit{W1} ($3.4$\,$\mu$m) and \textit{W2} ($4.5$\,$\mu$m) bands, we retrieved time-resolved coadded images of the field created as part of the unWISE project \citep{Lang2014AJ....147..108L, Meisner2018AJ....156...69M,unWISE}. With the custom code \citep{De2020PASP..132b5001D} based on the ZOGY algorithm \citep{Zackay2016ApJ...830...27Z}, we performed image subtraction on the NEOWISE images using the full-depth coadds of the WISE and NEOWISE mission (obtained during 2010--2014) as reference images. Photometric measurements were obtained by performing forced PSF photometry at the transient position on the subtracted WISE images until the epoch of the unWISE data release (data acquired until 2022 May).

\subsection{Spectroscopy}
\label{sec:spec}

Through our FLEET program, we obtained optical spectra of SN~2023ixf with the FAST spectrograph \citep{Fabricant1998PASP..110...79F,Tokarz1997ASPC..125..140T} on the FLWO 1.5 m Telescope starting on 2023 May 21.33 (${\rm MJD}=60085.33$; 1.60 days after discovery). The combination of the 300 grating with a $3"$-wide slit was used for dispersion, resulting in a wavelength coverage of $4000\,\Ang$ with a resolution $R\approx1000$. One-dimensional spectra were extracted, reduced, and calibrated following standard procedures using PyRAF and flux-calibrated to a standard taken during the same night. Additionally, we retrieved the public LT/SPRAT, HCT/HFOSC, NOT/ALFOSC, Xinglong/BFOSC, and P60/SEDM spectra (listed in \S\ref{sec:disc}) via the Transient Name Server (TNS)\footnote{\url{https://www.wis-tns.org/}} and include them in the subsequent analysis.

We also obtained high-resolution optical spectra on 2023 May 25 and 26 (${\rm MJD}=60089$ and $60090$; 5.6 and 6.5 days after discovery, respectively) with the Tillinghast Reflector Echelle
Spectrograph (TRES; \citealt{Szentgyorgyi2007RMxAC..28..129S}) on the FLWO 1.5 m Telescope ($3850-9100\,\Ang$ coverage with $R\approx44$,$000$), and on 2023 May 27, 28, and 29 (${\rm MJD}=60091$, $60092$, and $60093$; 7.4, 8.4, and 9.4 days after discovery, respectively) with Hectoechelle \citep{Szentgyorgyi1998SPIE.3355..242S} on the 6.5 m MMT Observatory ($6460-6660\,\Ang$ coverage centered on H$\alpha$ with $R\approx34$,$000$). The reduced data products were provided by the Smithsonian Astrophysical Observatory Optical/Infrared Telescope Data Center \citep{Mink2005ASPC..347..228M,Mink2007ASPC..376..249M,Buchhave2010ApJ...720.1118B,Mink2011ASPC..442..305M}. Given the lack of narrow flash features in the TRES and Hectoechelle spectra,\footnote{This is due to the combination of a low signal-to-noise ratio (the first TRES spectrum) and a late phase coverage (the later TRES and Hectoechelle spectra).} we only used them to measure the equivalent widths (EWs) of Na~{\sc i}~D lines at the M101 redshift, yielding ${\rm EW(D1)}=0.121\,\Ang$ and ${\rm EW(D2)}=0.180\,\Ang$. Using the \citet{Poznanski2012MNRAS.426.1465P} calibration, these EWs translate to an M101 extinction of $E(B-V)=0.032$\,mag, in agreement with \cite{Lundquist2023TNSAN.160....1L} and \cite{Smith2023arXiv230607964S}.

\section{Analysis}
\label{sec:ana}

\begin{figure*}
    \centering
    \includegraphics[width=1\textwidth]{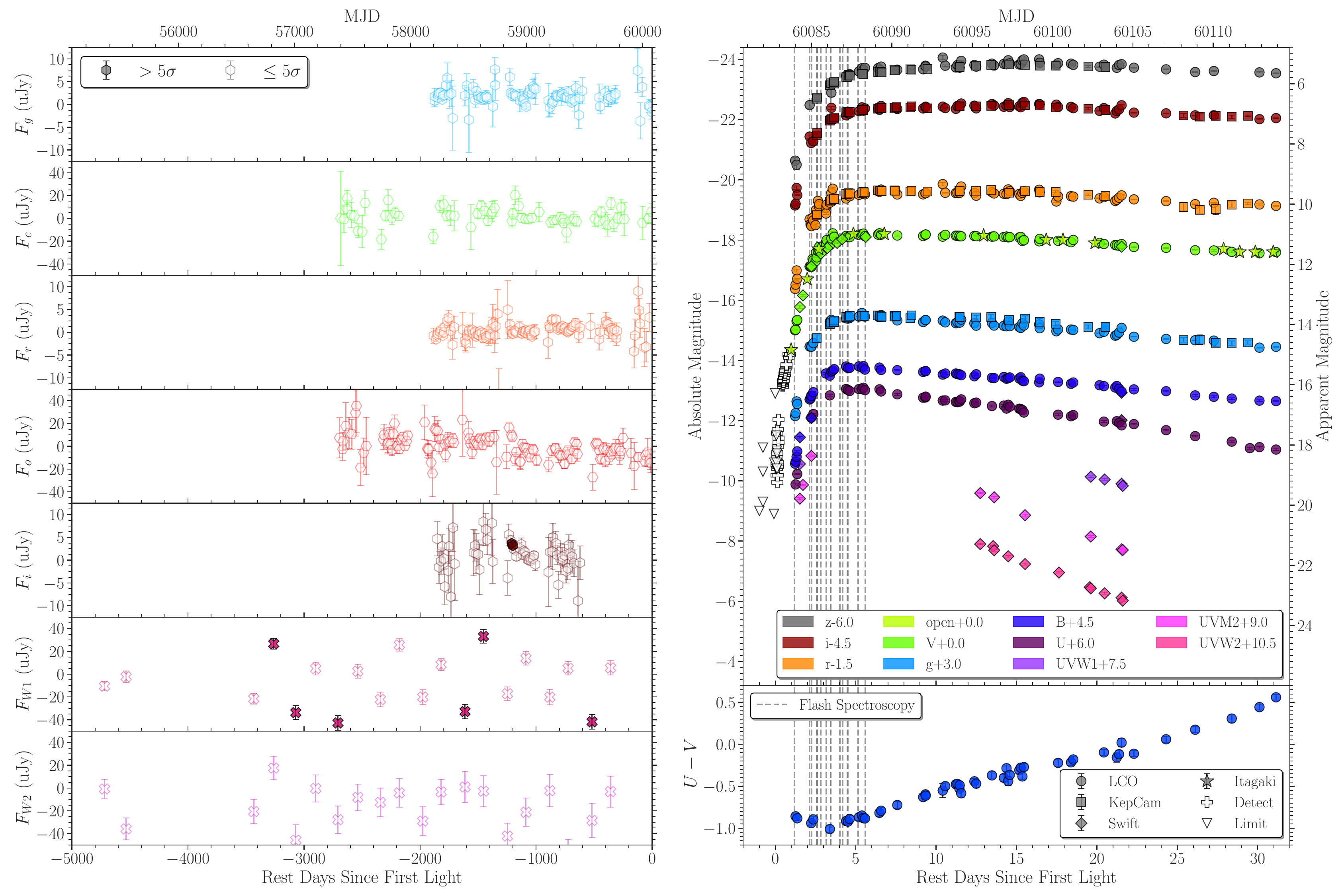}
\caption{\textit{Left}: pre-explosion differential flux measurements at the SN~2023ixf position in the ZTF \textit{g,r,i} bands; ATLAS \textit{c,o} bands; and WISE \textit{W1,W2} bands. Error bars denote $1\sigma$ uncertainties, and filled and open symbols are used for detections and $5\sigma$ limits, respectively. No significant variability ($>5\sigma$) is seen in the ZTF and ATLAS bands, except the two possible ZTF \textit{i}-band detections. The few sporadic WISE \textit{W1} detections are due to imperfect subtraction of the nearby H~{\sc ii} region (Figure~\ref{fig:m101}).
\textit{Right}: multiband light curve (\textit{top}) and $U-V$ color evolution (\textit{bottom}) of SN~2023ixf. Amateur detections and limits are shown as the open plus signs and downward-pointing triangles, respectively, with no vertical offset. The gray vertical dashed lines mark the epochs of flash spectroscopy (Figure~\ref{fig:spec}). SN~2023ixf is characterized by a rapid rise to the peak with blueward color evolution and a following bright plateau with redward color evolution. 
(The data used to create this figure will be available upon publication.)
}
\label{fig:phot}
\end{figure*}

\begin{figure*}
    \centering
    \includegraphics[width=1\textwidth]{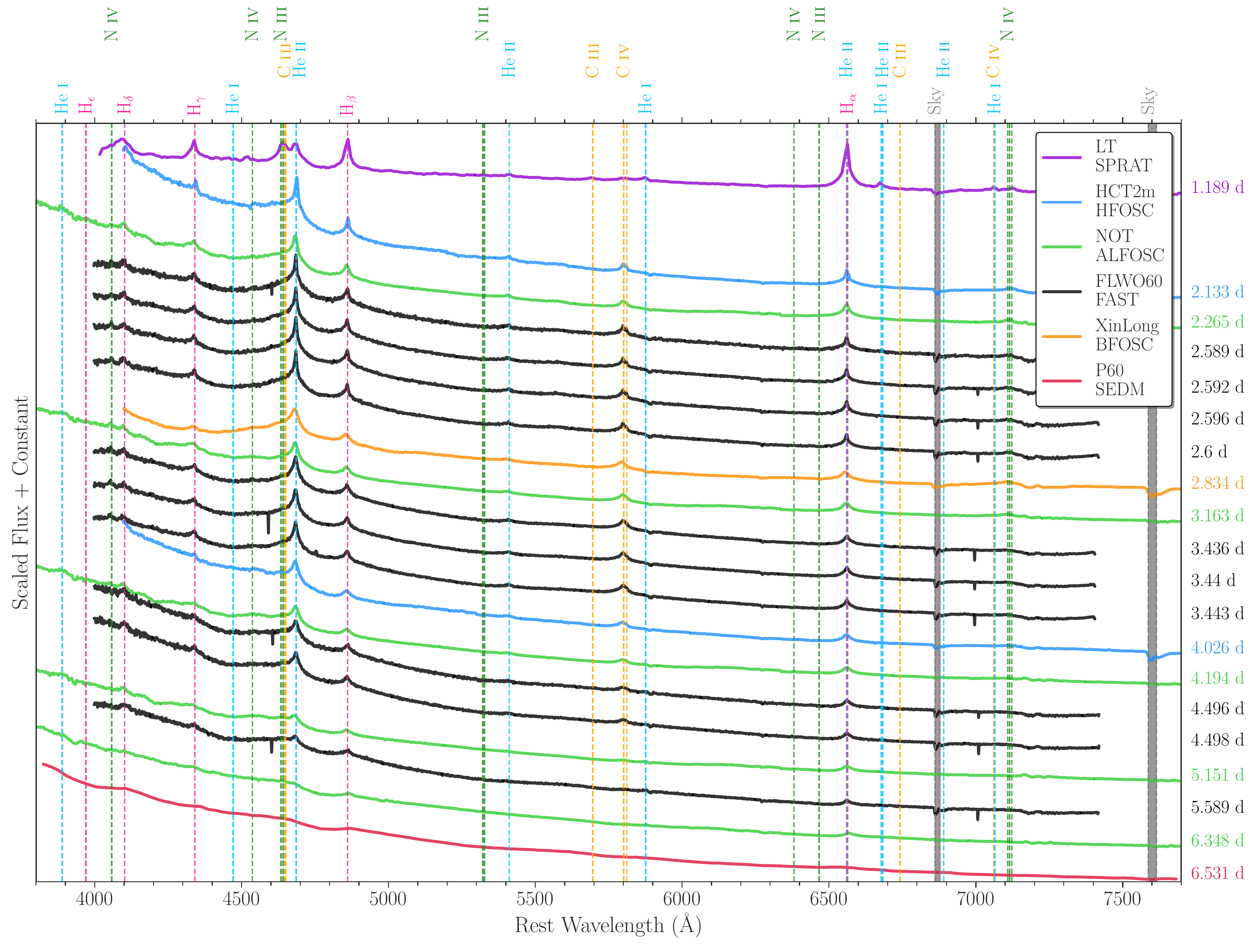}
\caption{Spectral time series of SN~2023ixf. The phase of each spectrum with respect to first light is noted on the right side. The prominent flash features of H, He~{\sc i}, He~{\sc ii}, C~{\sc iii}, C~{\sc iv}, N~{\sc iii}, and N~{\sc iv} persist up to 5.6 days, after which a blue featureless continuum dominates the spectra. A transition to a higher ionization state (i.e., C~{\sc iii} $\rightarrow$ C~{\sc iv}) happens during the first two spectra, in line with the blueward $U-V$ color evolution (Figure~\ref{fig:phot}).
(The data used to create this figure will be available upon publication.)
}
    \label{fig:spec}
\end{figure*}

All photometry and spectroscopy of SN~2023ixf are presented in Figures~\ref{fig:phot} and \ref{fig:spec}, respectively. 
We correct the photometry and spectroscopy for the Milky Way (MW) extinction of $E(B-V)=0.0089$\,mag \citep{Schlafly2011ApJ...737..103S},\footnote{Via the NASA/IPAC Infrared Science Archive: \url{https://irsa.ipac.caltech.edu/applications/DUST/}} as well as the M101 extinction of $E(B-V)=0.032$\,mag (measured in \S\ref{sec:obs}), assuming the \citet{Fitzpatrick1999PASP..111...63F} reddening law with $R_V=3.1$ and extended to the WISE bands with the relative optical to infrared extinction values from \citet{Wang2019ApJ...877..116W}.

\subsection{Light-curve Evolution}
\label{sec:lcevo}

We determine the detection significance ($\sigma$) of the ZTF, ATLAS, and WISE forced photometry from the ratio of measured flux ($f$) to its uncertainty ($f_{\rm err}$), i.e., $\sigma=|f|/f_{\rm err}$, where the absolute value of $f$ is taken to account for both positive and negative variability. As shown in Figure~\ref{fig:phot}, no significant pre-explosion variability ($\sigma>5$) is seen at the SN~2023ixf position in ZTF ($m_{g,r,i}\gtrsim21.8,22.0,21.3$) and ATLAS ($m_{c,o}\gtrsim20.2,20.3$) within $\approx2700$ days before first light (except the two possible ZTF \textit{i}-band detections at $22.5\pm0.2$\,mag ($\sigma=6.1$) and $22.6\pm0.2$\,mag ($\sigma=5.3$) from $\approx1205$ to $1195$ days before first light, which may allow short-term variability). This is consistent with the lack of long-term pre-explosion variability reported by \citet{Neustadt2023arXiv230606162N} in the deep Large Binocular Telescope (LBT) \textit{UBVR}-band observations ($m\gtrsim25-26$) from $\approx5600$ to $400$ days before first light and extends such a coverage to $0.32$ days before first light, albeit with the shallower limits. 
The few sporadic (nonperiodic) detections in the WISE $W1$ band ($\sim\pm25$\,uJy) are much higher than the variability ($\sim\pm8$\,uJy) seen in the \textit{Spitzer}/IRAC Chanel 1 ($3.6\,\mu{\rm m}$) observations with a similar temporal coverage ($\sim11.3–3.6$ yr before explosion; \citealt{Kilpatrick2023ApJ...952L..23K,Jencson2023ApJ...952L..30J,Soraisam2023arXiv230610783S}), and visual inspections of the WISE difference images indicate imperfect image subtraction of the nearby H~{\sc ii} region (Figure~\ref{fig:m101}) due to its relatively large PSF size (${\rm FWHM}\sim6"$). Thus, we do not consider these as significant detections.

The light curve of SN~2023ixf is characterized by a rapid rise to a bright peak, from $M_V\approx-10$ to $-18.2$\,mag within $\approx5$ days of first light (Figure~\ref{fig:phot}), with a possible deviation from a single power-law rise in the first $\approx 1$ day as noted by \cite{Hosseinzadeh2023ApJ...953L..16H}. During this rising phase, the $U-V$ color shows blueward evolution\footnote{Similar blueward color evolution is also seen in the other bands (e.g., $g-r$), indicating a change in the continuum rather than in specific lines.} from $-0.86\pm0.03$ mag (1.26 days) to $-1.01\pm0.04$ mag (3.39 days). Assuming a blackbody spectral energy distribution (SED), this color evolution corresponds to a temperature rise from $\approx12$,$000$ to $16$,$000$\,K. Such color/temperature evolution is not expected from pure shock cooling, and instead suggests a possible delayed shock breakout through dense CSM, in which a photosphere initially forms inside the unshocked optically thick CSM (e.g., \citealt{Moriya2018MNRAS.476.2840M,Forster2018NatAs...2..808F,Haynie2021ApJ...910..128,Khatami2023arXiv230403360K}), as similarly seen in  SN~II~2018zd \citep{Hiramatsu2021NatAs...5..903H}.

Following the bright peak, the light curve of SN~2023ixf settles on a plateau ($M_V\approx-17.6$\,mag) extending to 30 days with a smooth decline rate of $\approx0.03$\,mag\,day$^{-1}$ from the peak (Figure~\ref{fig:phot}). This plateau is on the bright and fast-declining ends of the SN~II population (e.g., \citealt{Anderson2014ApJ...786...67A,Valenti2016MNRAS.459.3939V}; see also \citealt{Jacobson-Galan2023arXiv230604721J,Teja2023ApJ...954L..12T,Sgro2023RNAAS...7..141S}). Coinciding with the peak, the $U-V$ color shows a transition to redward evolution, reaching $\approx 0.5$ mag ($\approx6000$\,K) at 30 days. In contrast to the early blueward evolution, this redward evolution is typical of the SN~II population (e.g., \citealt{Valenti2016MNRAS.459.3939V}), likely suggesting a switch in the location of the photosphere to the SN ejecta as it overruns the dense CSM. If we assume a typical SN shock velocity of $\sim10$,$000-15$,$000$\,km\,s$^{-1}$ (e.g., \citealt{Chevalier1994ApJ...420..268C,Matzner1999ApJ...510..379M}), the $U-V$ color transition at $3.4-4.4$ days corresponds to a CSM radial extent of $\sim(3-6)\times10^{14}$\,cm. 

To extract CSM properties from the light-curve modeling in \S\ref{sec:mod}, we construct a pseudobolometric light curve of SN~2023ixf by fitting a blackbody SED to every epoch of photometry containing at least three filters obtained within 30 minutes of each other and then integrating the fitted blackbody SED over the optical coverage (\textit{UBVRI}: $3250-8900\Ang$). For the unfiltered Itagaki photometry with a similar wavelength coverage ($\approx3200-9900\Ang$), we estimate its pseudobolometric correction at each epoch by linearly extrapolating or interpolating the blackbody temperatures from the multiband fits and taking the ratio of the integrated blackbody to observed flux convolved with the CCD response function. For the pre-discovery amateur points, we crudely estimate their pseudobolometric luminosity assuming the same correction as the Itagaki discovery point, albeit with uncharacterized CCD/filter response functions.
We note that the observed SED peaks are bluer than the $U$ band in the first $\approx10$ days ($\gtrsim10$,$000$\,K), and without reliable \textit{Swift} UVOT coverage (\S\ref{sec:phot}) during this phase, the fitted blackbody temperatures may be underestimated by up to $\sim10$,$000$\,K (e.g., \citealt{Valenti2016MNRAS.459.3939V,Arcavi2022ApJ...937...75}), which translates to a factor of $\sim2$ in the pseudobolometric luminosity.

\subsection{Spectral Evolution}
\label{sec:specevo}

As shown in Figure~\ref{fig:spec}, the early spectral evolution of SN~2023ixf is characterized by the flash features of H, He~{\sc i}, He~{\sc ii}, C~{\sc iii}, C~{\sc iv}, N~{\sc iii}, and N~{\sc iv} on top of a blue continuum (see also \citealt{Yamanaka2023PASJ..tmp...66Y, Jacobson-Galan2023arXiv230604721J,Smith2023arXiv230607964S,Bostroem2023arXiv230610119B,Teja2023ApJ...954L..12T,Vasylyev2023arXiv230701268V}). The N~{\sc iii} $\lambda\lambda\,4634,4641$ and C~{\sc iii} $\lambda\lambda\,4647,4650$ complex is comparable in line strength to He~{\sc ii} $\lambda\,4686$ in the first spectrum at 1.19 days but then quickly fades in the second spectrum at 2.13 days. Similarly, C~{\sc iii} $\lambda\,5696$ is comparable to C~{\sc iv} $\lambda\lambda\,5801,5812$ at 1.19 days but fades at 2.13 days. This transition to a higher ionization state (C~{\sc iii} $\rightarrow$ C~{\sc iv}) likely indicates a rise in temperature if we assume a smooth CSM density profile with constant elemental abundance \citep{Boian2019A&A...621A.109B,Boian2020MNRAS.496.1325B}, as similarly seen in SN~II~2018zd \citep{Hiramatsu2021NatAs...5..903H}. This is consistent with the possible delayed shock breakout seen in the $U-V$
color evolution (Figure~\ref{fig:phot}).

The prominent flash features of He~{\sc ii} $\lambda\,4686$ and C~{\sc iv} $\lambda\lambda\,5801,5812$ persist until $\approx 5.6$ and $4.5$ days, respectively (Figure~\ref{fig:spec}). Using these timescales as an alternate proxy to the color evolution\footnote{These two estimates do not agree perfectly, likely due to the difference in the optical depths they probe (i.e., optically thick photosphere and optically thin line emission; e.g., \citealt{Morozova2020ApJ...891L..32M}).} for the SN ejecta overrunning the dense CSM and assuming the same SN shock velocity as in \S\ref{sec:lcevo}, a radial CSM extent can be estimated as $\sim(4-7)\times10^{14}$\,cm. Together with the $U-V$ color evolution, the flash spectral series indicates a confined CSM, broadly in agreement with the estimates from \cite{Jacobson-Galan2023arXiv230604721J}, \cite{Smith2023arXiv230607964S}, \cite{Bostroem2023arXiv230610119B}, and \cite{Teja2023ApJ...954L..12T}.

\section{Light-curve Modeling with CSM Interaction}
\label{sec:mod}

Motivated by the early light curve (\S\ref{sec:lcevo}) and spectral evolution (\S\ref{sec:specevo}), we consider two different mass-loss scenarios for producing the confined CSM: continuous and eruptive.

\subsection{Continuous Mass Loss}
\label{sec:contmod}

As a first scenario, we explore ``superwind'' mass loss (e.g., \citealt{Moriya2018MNRAS.476.2840M,Forster2018NatAs...2..808F}) in which the continuous mass-loss rate is enhanced in the final years to decades before explosion by a few orders of magnitude ($\gtrsim10^{-3}\,M_\odot\,{\rm yr}^{-1}$) compared to a typical RSG mass-loss rate ($\sim10^{-5}\,M_\odot\,{\rm yr}^{-1}$; e.g., \citealt{Goldman2017MNRAS.465..403G,Beasor2020MNRAS.492.5994B}). We assume a constant wind velocity of $v_{\rm wind}=115$\,km\,s$^{-1}$ inferred by \cite{Smith2023arXiv230607964S} from the narrow H$\alpha$ line profile in a high-resolution spectrum taken at 2.62 days. As noted by \cite{Smith2023arXiv230607964S}, this velocity is higher than a typical RSG wind velocity of $\sim 20$\,km\,s$^{-1}$, and likely requires radiative acceleration \citep{Moriya2018MNRAS.476.2840M,Tsuna2023ApJ...952..115T} or eruptive mass loss (\S\ref{sec:erupmod}).
With the estimated CSM radial extent of $\sim(3-7)\times10^{14}$\,cm from the $U-V$ color and flash spectral evolution (\S\ref{sec:ana}), the velocity gives a mass-loss duration of $\sim0.8-1.9$ yr before explosion, which we explore here.

A broad range of possible masses, $8-20\,M_{\odot}$, has been reported for the progenitor of SN~2023ixf from pre-explosion imaging \citep{Pledger2023ApJ...953L..14P,Kilpatrick2023ApJ...952L..23K,Jencson2023ApJ...952L..30J,Soraisam2023arXiv230610783S}. As a base model, we use an RSG model with a low mass ($M_{\rm ZAMS}=12\,M_\odot$ and $M_{\rm final}=11\,M_\odot$) and large radius ($907\,R_\odot$) from \cite{Goldberg2020ApJ...895L..45G} given the observed bright light-curve plateau (\S\ref{sec:lcevo}). We note that according to the scaling relations for SN~II light-curve plateaus (e.g., \citealt{Popov1993ApJ...414..712P,Kasen2009ApJ...703.2205K,Sukhbold2016ApJ...821...38S,Goldberg2019ApJ...879....3G}), these progenitor properties are degenerate with the explosion energy, which we vary to account for the uncertainties in the progenitor properties. A better characterization of the progenitor properties from light-curve modeling should be explored when the light curve is sampled to the radioactive tail.
We also note that CSM estimates are less sensitive to a particular choice of degenerate parameters that result in a comparable CSM-free light curve (e.g., \citealt{Goldberg2020ApJ...895L..45G,Hiramatsu2021NatAs...5..903H,Khatami2023arXiv230403360K}).

\begin{figure*}
    \centering
    \includegraphics[width=1\textwidth]{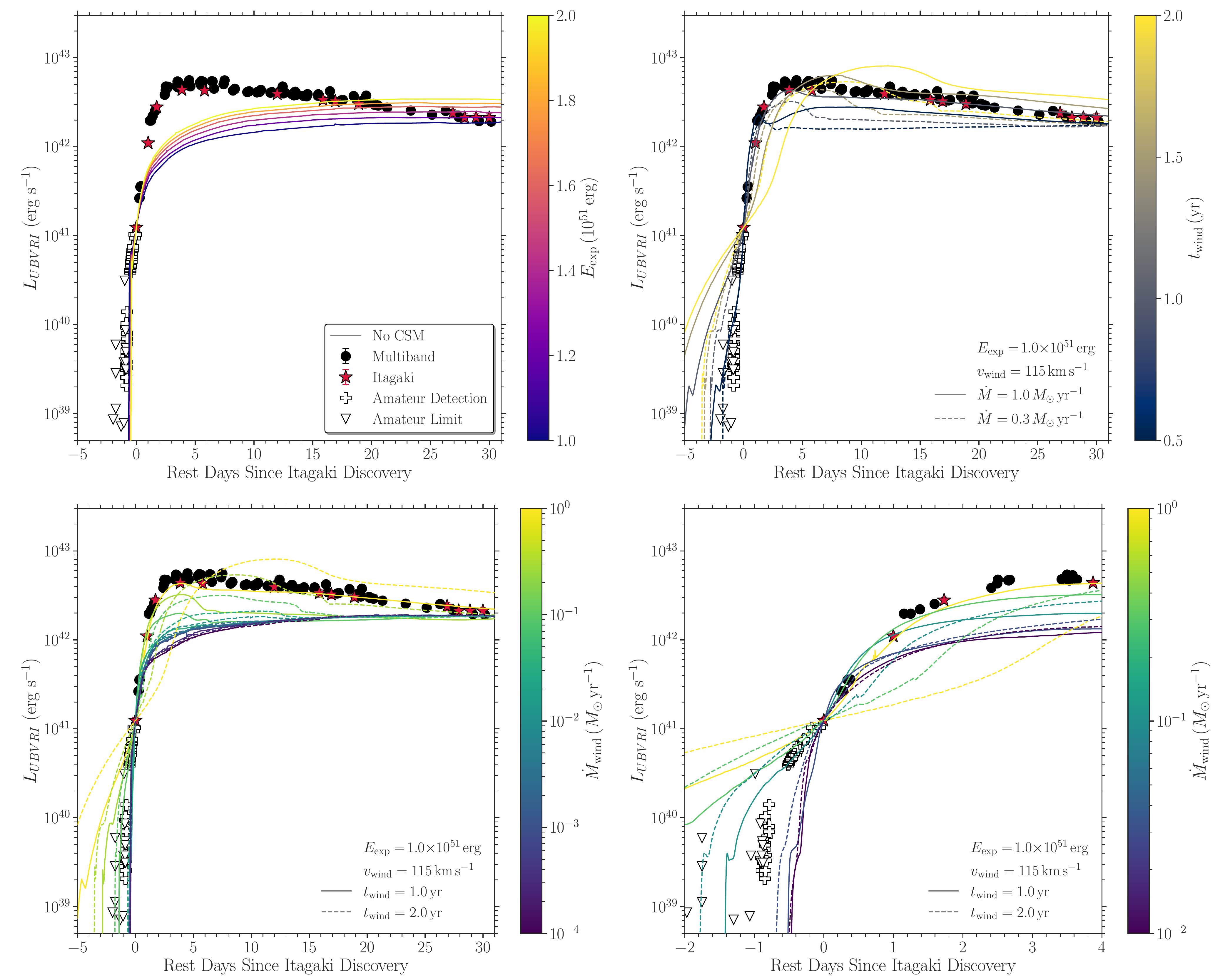}
\caption{Comparison of the pseudobolometric light curve of SN~2023ixf with CSM-free light-curve models with varying explosion energies ($E_{\rm exp}$; \textit{top left}), and continuous mass-loss light-curve models with a single $E_{\rm exp}$ ($1.0\times10^{51}$\,erg) and CSM velocity ($v_{\rm wind}=115$\,km\,s$^{-1}$): two fixed mass-loss rates ($\dot{M}_{\rm wind}$) with varying mass-loss durations ($t_{\rm wind}$; \textit{top right}), two fixed $t_{\rm wind}$ with varying $\dot{M}_{\rm wind}$ (\textit{bottom left}), and a zoomed-in version to the early phase (\textit{bottom right}). The phase is shifted with respect to the Itagaki discovery. CSM-free and minimal-CSM models are unable to reproduce the luminous peak/plateau up to $20$ days, and the luminosity excess requires $\dot{M}_{\rm wind}=0.3-1\,M_\odot\,{\rm yr}^{-1}$ for $t_{\rm wind}=2-1$ yr before explosion. The mismatch in the prediscovery excess may suggest a decreasing $\dot{M}_{\rm wind}$ toward the explosion.
}
    \label{fig:contmod}
\end{figure*}

We use a combination of \texttt{MESA} \citep{Paxton2011ApJS..192....3P,Paxton2013ApJS..208....4P,Paxton2015ApJS..220...15P,Paxton2018ApJS..234...34P,Paxton2019ApJS..243...10P} and \texttt{STELLA} \citep{Blinnikov1998ApJ...496..454B,Blinnikov2000ApJ...532.1132B,Blinnikov2004Ap&SS.290...13B,Baklanov2005AstL...31..429B,Blinnikov2006A&A...453..229B} for SN explosion and light-curve calculations, respectively. 
For the \texttt{MESA} explosion models, we vary the explosion energy ($E_{\rm exp}=(1.0-2.0) \times10^{51}$\,erg with $0.2\times10^{51}$\,erg increments), with a single value for the synthesized $^{56}$Ni mass ($M_{\rm Ni}=0.1\,M_{\odot}$), as it has no significant effect on the early light-curve evolution (e.g., \citealt{Kasen2009ApJ...703.2205K,Goldberg2019ApJ...879....3G,Kozyreva2019MNRAS.483.1211K,Hiramatsu2021ApJ...913...55H}). We then use the output of these explosion models as an input to \texttt{STELLA} by adding a wind density profile, $\rho_\mathrm{wind}(r) = \dot{M}_\mathrm{wind}/4\pi r^2 v_\mathrm{wind}$, with varying mass-loss rates ($\dot{M}_\mathrm{wind}=10^{-4}-1\,M_{\odot}$\,yr$^{-1}$ with $0.5$ dex increments) and durations ($t_\mathrm{wind}=0.5-2.0$\,yr with $0.5$ yr increments); the CSM mass is given by $M_{\rm CSM} = \dot{M}_\mathrm{wind} t_\mathrm{wind}$.  We use 700 and 100 spatial zones for the SN ejecta and CSM, respectively, and 100 frequency bins. A more detailed description of the \texttt{MESA}+\texttt{STELLA} workflow can be found in previous works (e.g, \citealt{Paxton2018ApJS..234...34P,Goldberg2019ApJ...879....3G,Hiramatsu2021ApJ...913...55H}).

In Figure~\ref{fig:contmod}, the pseudobolometric light curves of both CSM-free models and a subset of CSM models are compared with that of SN~2023ixf. In these model comparisons, we reference the phase with respect to the Itagaki discovery (the first point with the known CCD characteristic; \S\ref{sec:lcevo}) and shift the model phase to match the discovery point since the ``first light'' is not well defined in the models. 
The CSM-free models fail to reproduce the first $\approx20$ days of light-curve evolution (Figure~\ref{fig:contmod}, top left), which highlights the need for CSM interaction.

In the following CSM model comparison, we choose an explosion energy of $E_{\rm exp}=1.0\times10^{51}$\,erg, as its CSM-free model provides the best fit at 30 days after discovery when the effect of CSM interaction is expected to become less significant (but we do not intend to constrain $E_{\rm exp}$ given its degeneracy with the progenitor properties). For a given mass-loss rate of $\dot{M}_\mathrm{wind}=1.0\,M_\odot\,{\rm yr}^{-1}$, the peak luminosity is reasonably well reproduced by a mass-loss duration of $t_{\rm wind}=1.0$\,yr (Figure~\ref{fig:contmod}, top right). A longer (shorter) $t_{\rm wind}$ results in a peak that is too wide (narrow). For $\dot{M}_\mathrm{wind}=0.3\,M_\odot\,{\rm yr}^{-1}$, $t_{\rm wind}=2.0$\,yr better reproduces the peak luminosity. With a higher explosion energy of $E_{\rm exp}=2.0\times10^{51}$\,erg, these mass-loss estimates are reduced by $\approx0.5$\,dex (i.e., $\dot{M}_\mathrm{wind}\approx0.1-0.3\,M_\odot\,{\rm yr}^{-1}$ for $t_{\rm wind}=2.0-1.0$\,yr), albeit with an overshoot of the plateau luminosity at $>20$ days (as already seen in Figure~\ref{fig:contmod}, top left).
Below $\dot{M}_\mathrm{wind}=0.1\,M_\odot\,{\rm yr}^{-1}$, we do not find any models with a peak luminosity $\gtrsim 3\times10^{42}$\,erg\,s$^{-1}$ (even with $E_{\rm exp}=2.0\times10^{51}$\,erg) in this confined CSM configuration with $v_{\rm wind}=115$\,km\,s$^{-1}$. Thus, we estimate a range of mass-loss history to be $\dot{M}_{\rm wind}=0.1-1.0\,M_\odot\,{\rm yr}^{-1}$ for $t_{\rm wind}=2.0-1.0$ yr to reproduce the peak luminosity.

For $t_{\rm wind}=1.0$ and $2.0$\,yr, the effect of varying $\dot{M}_\mathrm{wind}$ is shown in Figure~\ref{fig:contmod} (bottom left). As discussed, $\dot{M}_\mathrm{wind}\geq0.3\,M_\odot\,{\rm yr}^{-1}$ is required to fit the peak luminosity well, and $\dot{M}_\mathrm{wind}\lesssim10^{-3}\,M_\odot\,{\rm yr}^{-1}$ has negligible effects on the peak luminosity. However, one difficulty with $\dot{M}_\mathrm{wind}=0.3-1.0\,M_\odot\,{\rm yr}^{-1}$ for $t_{\rm wind}=2.0-1.0$\,yr is that it overproduces the luminosity in prediscovery phase. The zoomed-in version to this phase is shown in Figure~\ref{fig:contmod} (bottom right). For $t_{\rm wind}=1.0$\,yr, $\dot{M}_\mathrm{wind}=0.1\,M_\odot\,{\rm yr}^{-1}$ captures the early excess (whose presence is first noted by \citealt{Hosseinzadeh2023ApJ...953L..16H}) down to $\approx0.6$ days before discovery and the following rise up to $\approx1$ day after discovery, albeit with an overshoot of the deep upper limit at $\approx8\times10^{38}$\,erg\,s$^{-1}$ by $\approx0.3$ days. Therefore, we speculate a time-variable mass-loss rate in the final $2-1$ years before explosion in which a lower mass-loss rate ($\approx0.01-0.1\,M_\odot\,{\rm yr}^{-1}$) follows a higher one ($\approx0.1-1\,M_\odot\,{\rm yr}^{-1}$) within the explosion properties explored here.\footnote{This may also be possible with an increasing $v_{\rm wind}$, as $\rho_\mathrm{wind} \propto \dot{M}_\mathrm{wind}/v_\mathrm{wind}$. But it would create wind-to-wind collision, which should be explored in future work.} Assuming that the lower mass-loss rate is responsible for the first $\sim2$ days of light-curve evolution after first light as the SN--CSM shock layer propagates through, this transition might have happened at $\sim0.7-0.4$ yr before explosion (with an SN shock velocity of $\sim10$,$000-15$,$000$\,km\,s$^{-1}$). Then, the total CSM mass could be estimated as $\sim0.1-0.7\,M_\odot$.

\subsection{Eruptive Mass Loss}

Another scenario usually considered for a dense CSM is eruptive mass loss months to years before core collapse. Such outbursts are observed for a significant fraction of SNe~IIn (strongly interacting SNe with narrow H emission lines throughout their evolution; \citealt{Ofek2013Natur.494...65O, Ofek2014ApJ...789..104O, Margutti2014ApJ...780...21M,Strotjohann2021ApJ...907...99S,Hiramatsu2023arXiv230511168}). While such detections are limited for more standard SNe~II (\citealt{Kochanek2017MNRAS.467.3347,Johnson2018MNRAS.480.1696}; but see \citealt{Jacobson-Galan2022ApJ...924...15}), this scenario is recently found to be favored over the superwind model from modeling of SN~II progenitors \citep{Davies2022MNRAS.517.1483}.

For this scenario, we follow and extend the methods of \cite{Tsuna2023ApJ...952..115T}, which used the open-source code \texttt{CHIPS} \citep{Takei2022ApJ...929..177} to simulate mass eruption, followed by \texttt{SNEC} \citep{Morozova2015ApJ...814...63} for calculating the bolometric light curve. The mass loss is assumed to be triggered by energy injection at the base of the envelope, which forms a shock that propagates to the surface and expels a part of the envelope. The model is characterized by two parameters: the injected energy and the amount of time from injection to core collapse. These control the mass, velocity, and extent of the CSM at core collapse. 

We extend the model grid from \cite{Tsuna2023ApJ...952..115T} while using the same RSG progenitor as the earlier work evolved by \texttt{MESA} of 15\ $M_\odot$ at ZAMS, which is within the mass range inferred for the SN~2023ixf progenitor. The progenitor has a mass of $12.8\,M_\odot$ and a radius of $670\,R_\odot$ at the time of energy injection, with modeling of mixing length and stellar wind outlined in \cite{Tsuna2023ApJ...952..115T}. This progenitor choice is different from that in \S\ref{sec:contmod}; however, as noted, a variation in explosion energies should account for the difference given the light-curve scaling relations. For the injected energy, the \texttt{CHIPS} code uses the injected energy scaled to the binding energy of the envelope ($\approx 4.86\times 10^{47}$ erg for this progenitor star), defined in the code as $f_{\rm inj}$. A fraction of the envelope is unbound for $f_{\rm inj}\gtrsim 0.2$, and we thus consider a range of values $f_{\rm inj}=[0.2, 0.25, 0.3, 0.4, 0.5, 0.6, 0.7]$.

\begin{figure*}
      \centering
    \begin{tabular}{cc}
     \begin{minipage}[t]{0.5\hsize}
    \centering
    \includegraphics[width=\linewidth]{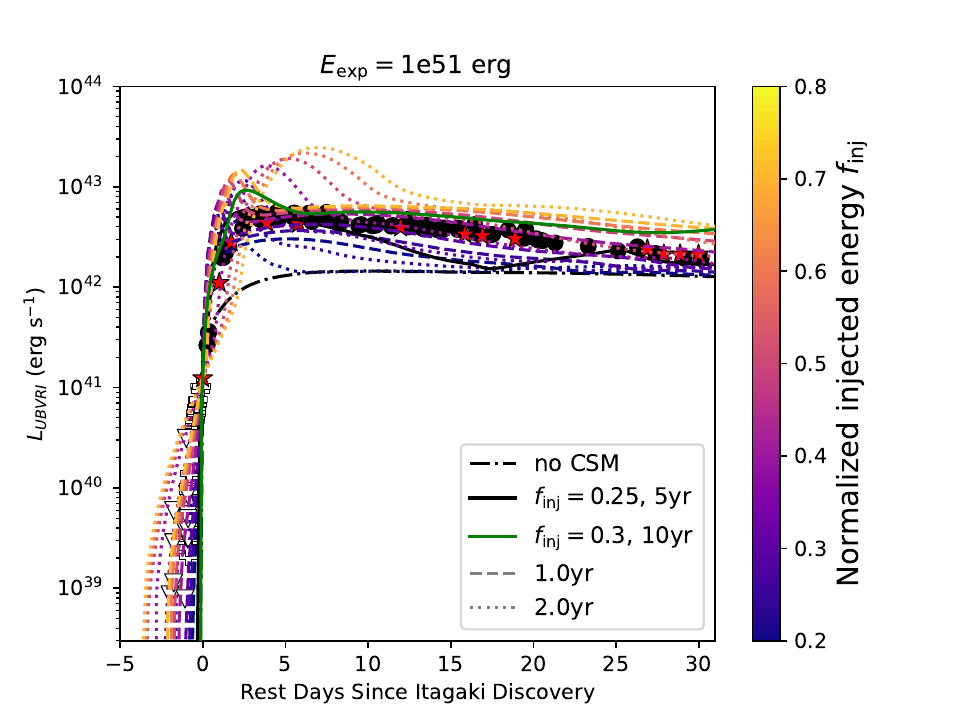}
    \end{minipage}
     \begin{minipage}[t]{0.5\hsize}
   \centering
    \includegraphics[width=\linewidth]{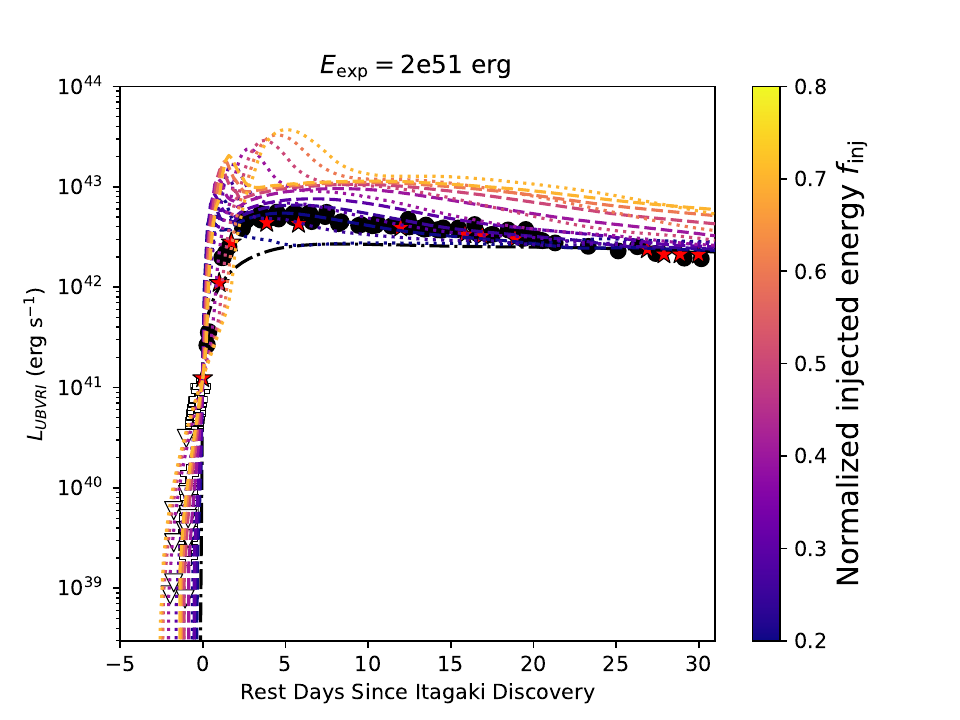} 
    \end{minipage} \\
     \begin{minipage}[t]{0.5\hsize}
    \centering
    \includegraphics[width=\linewidth]{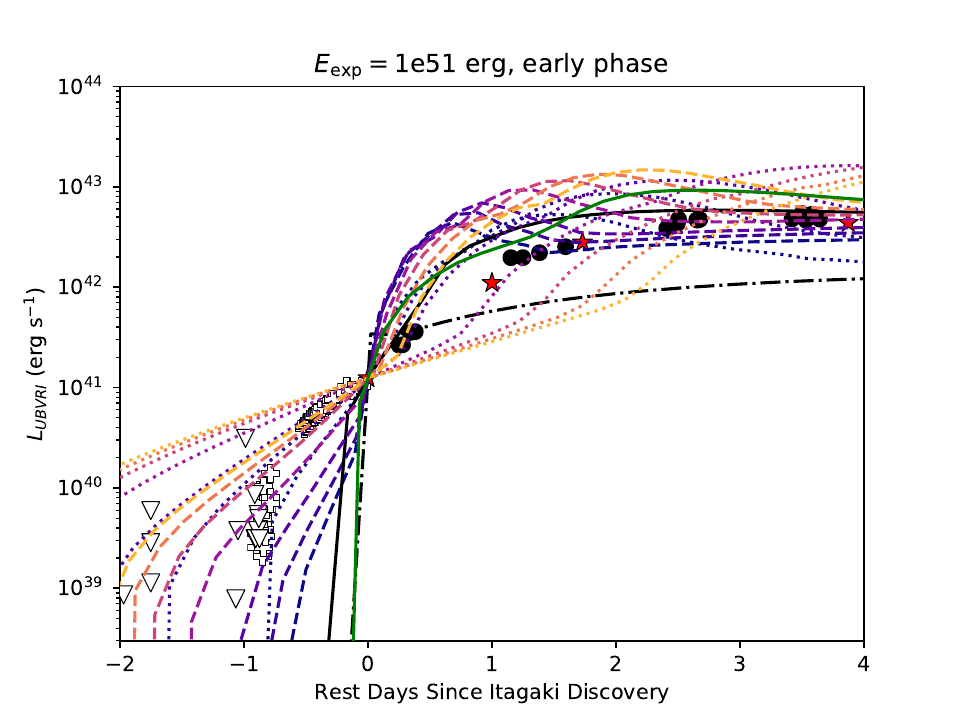}
    \end{minipage}
     \begin{minipage}[t]{0.5\hsize}
   \centering
    \includegraphics[width=\linewidth]{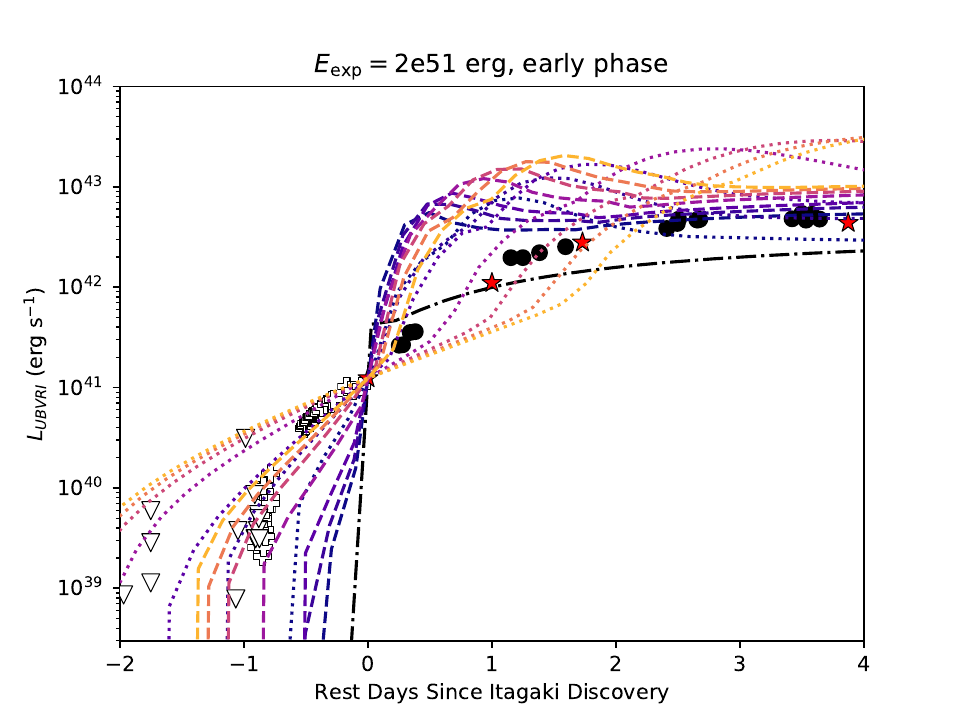} 
    \end{minipage} 
    \end{tabular}
\caption{Optical light curves of interacting SNe for the eruptive mass-loss model, color-coded by the normalized injected energy, $f_{\rm inj}$. The dashed and dotted lines are explosions with CSM formed by energy injections 1 and 2 yr before the SN, respectively. The black dashed-dotted lines show the case for the progenitor with no CSM. The green (black) solid lines in the left panels are models with an extended CSM expelled 10 (5) yr before the SN with $f_{\rm inj}=0.3\ (0.25)$. The data points from observations are the same as in Figure~\ref{fig:contmod}.}
 \label{fig:lc_eruption}
\end{figure*}
\label{sec:erupmod}

We first simulate explosions with CSM models of 1 and 2 yr from energy injection, motivated by the deep LBT limits on apparent pre-SN outbursts of the progenitor up to about 400 days prior to the SN \citep{Neustadt2023arXiv230606162N}.\footnote{We note that it takes several months from energy injection for the shock to propagate to the surface and be seen as an outburst.} For the range of $f_{\rm inj}$, the dense CSM expands in these 2 yr to radii of $\approx (2-7)\times 10^{14}$ cm, with respective masses of $\approx 0.04-1.7\,M_\odot$. As in \cite{Tsuna2023ApJ...952..115T}, we then attach the evolved envelope to the core of the \texttt{MESA} model and simulate the explosion using \texttt{SNEC}. We excise the inner $1.8\,M_\odot$ of the star and inject energy at the inner $0.1\,M_\odot$ as a thermal bomb. We consider the final explosion energy of $E_{\rm exp}=10^{51}$ and $2\times 10^{51}\ {\rm erg}$. 

Using the bolometric luminosity output by \texttt{SNEC}, we assume that the emission is thermalized at the photosphere and calculate the luminosities in the \textit{UBVRI} bands ($3250-8900$\,\AA).
The resulting light curves for injections 1 and 2 yr before the SN are shown as dashed and dotted lines in Figure~\ref{fig:lc_eruption}, respectively; a reference model with no CSM is shown with a dashed-dotted line. As in \S\ref{sec:contmod}, the model with no CSM has a monotonic rise and thus significantly underestimates around the peak. The bright plateau of SN 2023ixf after the peak is well reproduced by the cooling emission of the shocked CSM \citep[e.g.,][]{Piro2015ApJ...808L..51,Piro2021ApJ...909..209}, for a range of $f_{\rm inj}\approx 0.25$--$0.4$.
While no perfect match to observations is found, the 1 yr model with $E_{\rm exp}=10^{51}$ erg appears to best reproduce both the rise timescale and the plateau luminosity.\footnote{For a fixed $f_{\rm inj}$, the CSM extent generally scales with the escape velocity of the RSG progenitor. Thus, the required timescale can change for different progenitor models, but this change is expected to be of order unity.} The masses of the CSM in these models are $M_{\rm CSM}\approx 0.3$--$1\ M_\odot$, which agrees with the masses estimated in \S\ref{sec:contmod}.

A notable problem for the models is that they generally overpredict the luminosity around the peak. One possible reason for this is a departure from thermal equilibrium in the postshock region around breakout. 
In our CSM model, the density sharply drops around the breakout radius, which can also accelerate the shock. A low CSM density and, more importantly, a large shock velocity ($\gtrsim 10^4$ km s$^{-1}$) can both lead to deviation from thermal equilibrium \citep{Nakar2010ApJ...725..904,Svirski2012ApJ...759..108,Haynie2021ApJ...910..128}, which reduces the optical luminosity.

Another possibility for the eruptive case is a much earlier mass ejection, where breakout from a less dense CSM powers the peak and continued interaction powers the plateau. We explored models with longer intervals between the eruption and SN and found that two models, with energy injection 10 (5) yr before the SN with $f_{\rm inj}=0.3$ (0.25), reproduce the light curves better around the peak than the $1$--$2$ yr models for $E_{\rm exp}=10^{51}$ erg. These light curves are shown with solid lines in Figure~\ref{fig:lc_eruption}. In these models, the CSM extends to $\approx 10^{15} \ (5\times 10^{14})$ cm with mass of $\approx 0.06\ (0.04)\,M_\odot$, which is consistent with the results of \cite{Jacobson-Galan2023arXiv230604721J}.
However, as we explain in the next section, this model of early eruption is less preferred, as it may face two challenges: the nearly periodic pre-explosion variability \citep{Kilpatrick2023ApJ...952L..23K,Jencson2023ApJ...952L..30J,Soraisam2023arXiv230610783S} and the early X-ray detection \citep{Grefenstette2023ApJ...952L...3G}.

\subsection{Comparison to Other Observational Constraints}
\label{sec:comp}

In both the continuous and eruptive mass-loss cases, we find that the dense CSM ($0.1-1\,M_\odot$) confined within $(3-7)\times10^{14}$\,cm is able to reproduce the early light-curve evolution. Here we discuss other observational constraints obtained thus far and their implications for our mass-loss models.

The infrared light curve of the progenitor from 2012 is consistent with a nearly regular periodic variability, with no evidence of an additional precursor emission \citep{Kilpatrick2023ApJ...952L..23K,Jencson2023ApJ...952L..30J,Soraisam2023arXiv230610783S}. The absence of change in dust optical depth between $5600$ and $400$ days before explosion has independently constrained the potential outburst to be $\lesssim 3-5$ times brighter than the progenitor's original luminosity of $\sim10^{5}L_\odot=4\times10^{38}$\,erg\,s$^{-1}$ \citep{Neustadt2023arXiv230606162N}.

The continuous mass-loss scenario may result in precursor emission owing to the sustained energy injection that generates the outflow. With the inferred CSM properties, we estimate the luminosity of a possible precursor to be $\lesssim 2\times10^{39}$\,erg\,s$^{-1}$ from \cite{Matsumoto2022ApJ...936..114} in a dust-free environment. This is within the allowed luminosity range without altering the dust optical depth too much ($\tau_V=10-13$; \citealt{Neustadt2023arXiv230606162N}), and such an obscured precursor is consistent with the pre-explosion limits presented in Figure~\ref{fig:phot}, as well as the deeper LBT limits \citep{Neustadt2023arXiv230606162N} if the mass loss happens within $\approx400$ days before explosion (i.e., outside the LBT coverage). 
If the mass loss is instead driven by pulsation (e.g., \citealt{Yoon2010ApJ...717L..62Y}), a less luminous precursor is likely expected, which may provide a more natural explanation for the periodic progenitor variability.

For the eruptive mass-loss scenario, 
the precursor models of \cite{Tsuna2023ApJ...945..104} are dust-free and cannot be directly constrained from the pre-explosion observations of SN 2023ixf. However, the precursor emission for models with $f_{\rm inj}=0.25$ and $0.3$, which lasts for $\lesssim 1$ yr from eruption, is only $\lesssim 1$ mag brighter \citep{Tsuna2023ApJ...945..104} and is thus within the allowed luminosity range to avoid significant change of the dust optical depth \citep{Neustadt2023arXiv230606162N}. A caveat is that for eruptions much earlier than 2 yr before the SN, this scenario does not explain the observed periodic variability after the eruption. An eruption perturbs and oscillates the leftover envelope, which may cause the variability, but there is no evident reason that both its timescale and amplitude would be similar to the observed one. It is worthwhile to study the precursor emission in dusty environments, as well as its later oscillation, to quantitatively constrain this scenario.

The X-ray data obtained $4$ days after explosion suggest a rather tenuous CSM, with a mass-loss rate of $\approx 3\times 10^{-4}\ \,M_\odot\,{\rm yr}^{-1}$ for an assumed CSM velocity of $50\ {\rm km\ s^{-1}}$ \citep{Grefenstette2023ApJ...952L...3G}. Since the radius of the CSM probed is $\gtrsim(4-6)\times10^{14}$\,cm for an SN shock velocity of $\sim10$,$000-15$,$000$\,km\,s$^{-1}$, this does not contradict our preferred models (both continuous and eruptive) of the compact, dense CSM. However, this is in tension with an eruption $5-10$ yr before explosion, whose density is higher than this by about an order of magnitude. An asymmetric CSM may alleviate this problem, if the line of sight is not subtended by the dense CSM inferred from the optical light curve. An asymmetric CSM is also suggested from early spectra \citep{Smith2023arXiv230607964S,Vasylyev2023arXiv230701268V} and millimeter nondetections ($\lesssim$\,mJy at 230\,GHz; \citealt{Berger2023ApJ...951L..31B}), although the latter require the line of sight to be in the dense part of the CSM assuming a single wind density profile. Alternatively, the millimeter nondetections may be explained with our confined CSM models in that the early nondetections ($\lesssim4$ days) are due to free-free absorption by the confined CSM, while the later nondetections are due to weaker synchrotron emission from the underlying extended CSM ($\sim10^{-4}$ to $10^{-5}\,M_\odot\,{\rm yr}^{-1}$). This merits further investigations, especially with a recent fainter radio detection ($\approx40\,\mu{\rm Jy}$ at 10\,GHz) at 29 days after first light \citep{Matthews2023ATel16091....1M}.

Estimates based on spectra \citep{Jacobson-Galan2023arXiv230604721J,Bostroem2023arXiv230610119B} find a CSM of $M_{\rm CSM}\approx10^{-2}-0.1M_\odot$, much smaller than our results of $0.1-1M_\odot$. This may be due to the light curve and spectra probing different parts of the CSM (see e.g. \citealt{Morozova2020ApJ...891L..32M} for a similar discussion on SN~II 2013fs). Indeed, the light-curve model from \cite{Jacobson-Galan2023arXiv230604721J} with such a lower $M_{\rm CSM}$ underproduces the luminosity in the first week of its evolution (see also \citealt{Teja2023ApJ...954L..12T} for a similar discrepancy even with extremely large explosion energies of $(2-5)\times10^{51}$\,erg). 
Furthermore, the density profiles of our preferred CSM models are different from a wind profile assumed in spectral modeling. As in \S\ref{sec:contmod}, a single wind profile has difficulties explaining both the light-curve rise and plateau. The CSM from mass eruption is also expected to have a profile much different from that of a wind \citep{Kuriyama2020A&A...635A.127,Morozova2020ApJ...891L..32M,Tsuna2021PASJ...73.1128,Tsang2022ApJ...936...28T}.

\section{Summary}
\label{sec:sum}

We have presented our discovery of SN~2023ixf in M101 at $6.9$ Mpc and follow-up photometric and spectroscopic observations, respectively, in the first month and week of its evolution. Our observations have revealed the following:

\begin{itemize}
\vspace{-5pt}
  \item A rapid rise ($\approx5$ days) to a luminous light-curve peak ($M_V\approx-18.2$\,mag) and plateau ($M_V\approx-17.6$\,mag) extending to $30$ days after first light with a fast decline rate of $\approx0.03$\,mag\,day$^{-1}$.
\vspace{-5pt}
  \item Blueward $U-V$ color evolution in the first $4.4$ days toward the light-curve peak followed by redward color evolution during the light-curve plateau.
\vspace{-5pt}
  \item Prominent flash spectral features of H, He~{\sc i}, He~{\sc ii}, C~{\sc iii}, C~{\sc iv}, N~{\sc iii}, and N~{\sc iv} persisting up to $4.5-5.6$ days after first light.
\vspace{-5pt}
  \item A transition to a higher spectral ionization state (i.e., C~{\sc iii}\,$\rightarrow$\,C~{\sc iv}) in the first $2.1$ days after first light.
\vspace{-5pt}
\end{itemize}

These observed properties are not expected from pure shock-cooling emission, and instead suggest a delayed shock breakout from a dense CSM confined within $\sim(3-7)\times10^{14}$\,cm. 
Motivated by this, we have constructed numerical light-curve models based on both continuous and eruptive mass-loss scenarios shortly before explosion. Our key findings are as follows:

\begin{itemize}
\vspace{-5pt}
  \item For the continuous mass-loss scenario, we infer a range of mass-loss history with $0.1-1.0\,M_\odot\,{\rm yr}^{-1}$ in the final $2.0-1.0$ yr before explosion, with a potentially decreasing mass loss of $0.01-0.1\,M_\odot\,{\rm yr}^{-1}$ toward the explosion ($\lesssim0.7-0.4$ years) to reproduce the rapid rise and luminous peak.
\vspace{-5pt}
  \item For the eruptive mass-loss scenario, we favor energy injections of $\sim 10^{47}$ erg at the base of the envelope in the final year before explosion, which releases $0.3$--$1\ M_\odot$ of the envelope as CSM. The mass and extent of the CSM broadly agree with the continuous scenario.
\vspace{-5pt}
\end{itemize}

\noindent These confined, dense CSM models are compatible with the early X-ray detections, millimeter nondetections, and possibly the progenitor variability without clear outbursts before the final $\sim1$ yr of explosion.

Given its proximity, SN~2023ixf will remain as a primarily SN target for multiwavelength follow-up observations in the coming years. In ultraviolet to infrared, a fully sampled light curve up to a radioactive tail will allow better characterizations of the progenitor and explosion properties. Nebular-phase spectra have potential to constrain the progenitor mass, possibly better than the current pre-explosion estimates. Deviations from a smooth light-curve plateau and/or tail, or continued X-ray and/or radio/millimeter emission, will provide unique opportunities to probe the mass-loss history in the final years to decades of its progenitor evolution.

\section{acknowledgments}

We are grateful to Anthony Piro, Jim Fuller, Ken’ichi Nomoto, Takashi Moriya, and Charles Kilpatrick for useful discussions; to Warren Brown, Pascal Fortin, Murdock Hart, David Latham, and Jessica Mink for scheduling and reducing FLWO KeplerCam, FAST, and TRES observations; and to Nelson Caldwell, Daniel Fabricant, and Sean Moran for scheduling the MMT Hectochelle observations. 

The Berger Time-Domain research group at Harvard is supported by the NSF and NASA. 
The LCO supernova group is supported by NSF grants AST-1911151 and AST-1911225. 
D.T. is supported by the Sherman Fairchild Postdoctoral Fellowship at the California Institute of Technology. 
The Flatiron Institute is supported by the Simons Foundation.
This publication was made possible through the support of an LSSTC Catalyst Fellowship to K.A.B., funded through grant 62192 from the John Templeton Foundation to LSST Corporation. The opinions expressed in this publication are those of the authors and do not necessarily reflect the views of LSSTC or the John Templeton Foundation.
I.A. acknowledges support from the European Research Council (ERC) under the European Union’s Horizon 2020 research and innovation program (grant agreement No. 852097), from the Israel Science Foundation (grant No. 2752/19), from the United States - Israel Binational Science Foundation (BSF), and from the Israeli Council for Higher Education Alon Fellowship.
J.C.W. and J.V. are supported by NSF grant AST1813825. J.V. is also supported by by OTKA grant K-142534 of the National Research, Development and
Innovation Office, Hungary.

This work makes use of observations from KeplerCam on the 1.2 m telescope and FAST and TRES on the 1.5 m telescope at the Fred Lawrence Whipple Observatory.
Observations reported here were obtained at the MMT Observatory, a joint facility of the Smithsonian Institution and the University of Arizona. 
This paper uses data products produced by the OIR Telescope Data Center, supported by the Smithsonian Astrophysical Observatory.

This work makes use of observations from the Las Cumbres Observatory global telescope network. The authors wish to recognize and acknowledge the very significant cultural role and reverence that the summit of Haleakal$\bar{\text{a}}$ has always had within the indigenous Hawaiian community. We are most fortunate to have the opportunity to conduct observations from the mountain. 

We thank the support of the staffs at the Neil Gehrels \textit{Swift} Observatory.


This work has made use of data from the Zwicky Transient Facility (ZTF).  ZTF is supported by NSF grant No. AST-1440341 and a collaboration including Caltech, IPAC, the Weizmann Institute for Science, the Oskar Klein Center at Stockholm University, the University of Maryland, the University of Washington, Deutsches Elektronen-Synchrotron and Humboldt University, Los Alamos National Laboratories, the TANGO Consortium of Taiwan, the University of Wisconsin at Milwaukee, and Lawrence Berkeley National Laboratories. Operations are conducted by COO, IPAC, and UW. The ZTF forced-photometry service was funded under the Heising-Simons Foundation grant
No. 12540303 (PI: Graham).


This work has made use of data from the Asteroid Terrestrial-impact Last Alert System (ATLAS) project. ATLAS is primarily funded to search for near-Earth asteroids through NASA grant Nos. NN12AR55G, 80NSSC18K0284, and 80NSSC18K1575; byproducts of the NEO search include images and catalogs from the survey area. This work was partially funded by Kepler/K2 grant No. J1944/80NSSC19K0112 and HST grant No. GO-15889, and STFC grant Nos. ST/T000198/1 and ST/S006109/1. The ATLAS science products have been made possible through the contributions of the University of Hawaii Institute for Astronomy, the Queen’s University Belfast, the Space Telescope Science Institute, the South African Astronomical Observatory, and The Millennium Institute of Astrophysics (MAS), Chile.

The PS1 and the PS1 public science archives have been made possible through contributions by the Institute for Astronomy, the University of Hawaii, the Pan-STARRS Project Office, the Max-Planck Society and its participating institutes, the Max Planck Institute for Astronomy, Heidelberg and the Max Planck Institute for Extraterrestrial Physics, Garching, The Johns Hopkins University, Durham University, the University of Edinburgh, the Queen's University Belfast, the Harvard-Smithsonian Center for Astrophysics, the Las Cumbres Observatory Global Telescope Network Incorporated, the National Central University of Taiwan, the Space Telescope Science Institute, NASA under grant No. NNX08AR22G issued through the Planetary Science Division of the NASA Science Mission Directorate, NSF grant No. AST-1238877, the University of Maryland, Eotvos Lorand University, the Los Alamos National Laboratory, and the Gordon and Betty Moore Foundation.

This publication makes use of data products from the Wide-field Infrared Survey Explorer, which is a joint project of the University of California, Los Angeles, and the Jet Propulsion Laboratory/California Institute of Technology, funded by the National Aeronautics and Space Administration.

This research has made use of the NASA Astrophysics Data System (ADS), the NASA/IPAC Extragalactic Database (NED), and NASA/IPAC Infrared Science Archive (IRSA, which is funded by NASA and operated by the California Institute of Technology), and IRAF (which is distributed by the National Optical Astronomy Observatory, NOAO, operated by the Association of Universities for Research in Astronomy, AURA, Inc., under cooperative agreement with the NSF).

TNS is supported by funding from the Weizmann Institute of Science, as well as grants from the Israeli Institute for Advanced Studies and the European Union via ERC grant No. 725161.



\vspace{5mm}
\facilities{ADS, ATLAS, FLWO:1.2m (KeplerCam), FLWO:1.5m (FAST, TRES), IRSA, LCOGT (SBIG, Sinistro), MMT (Hectoechelle), NED, NEOWISE, PS1, \textit{Swift} (UVOT), WISE, ZTF}.


\defcitealias{Astropy2018AJ....156..123A}{Astropy Collaboration 2018}
\software{Astropy \citepalias{Astropy2018AJ....156..123A}, 
\texttt{atlas-fp} (\url{https://gist.github.com/thespacedoctor/86777fa5a9567b7939e8d84fd8cf6a76}), 
BANZAI \citep{curtis_mccully_2018_1257560},
\texttt{CHIPS} \citep{Takei2022ApJ...929..177},
\texttt{emcee} \citep{Foreman-Mackey2013PASP..125..306F},
\texttt{lcogtsnpipe} \citep{Valenti2016MNRAS.459.3939V}, 
Matplotlib \citep{Hunter2007CSE.....9...90H}, 
\texttt{MESA} \citep{Paxton2011ApJS..192....3P,Paxton2013ApJS..208....4P,Paxton2015ApJS..220...15P,Paxton2018ApJS..234...34P,Paxton2019ApJS..243...10P},
NumPy \citep{Oliphant2006}, 
\texttt{photutils} \citep{photutils},
PyRAF \citep{PyRAF2012ascl.soft07011S},
SciPy \citep{SciPy2020NatMe..17..261V},
seaborn \citep{seaborn2020zndo...3629446W},
\texttt{SExtractor} \citep{Bertin1996A&AS..117..393B},
\texttt{STELLA} \citep{Blinnikov1998ApJ...496..454B,Blinnikov2000ApJ...532.1132B,Blinnikov2004Ap&SS.290...13B,Baklanov2005AstL...31..429B,Blinnikov2006A&A...453..229B},
\texttt{SNEC}
\citep{Morozova2015ApJ...814...63}
}.

\bibliography{main}

\begin{thebibliography}{}\footnotesize
\expandafter\ifx\csname natexlab\endcsname\relax\def\natexlab#1{#1}\fi
\providecommand{\url}[1]{\href{#1}{#1}}
\providecommand{\dodoi}[1]{doi:~\href{http://doi.org/#1}{\nolinkurl{#1}}}

\bibitem[{{Anderson} {et~al.}(2014){Anderson}, {Gonz{\'a}lez-Gait{\'a}n},
  {Hamuy}, {Guti{\'e}rrez}, {Stritzinger}, {Olivares E.}, {Phillips},
  {Schulze}, {Antezana}, {Bolt}, {Campillay}, {Castell{\'o}n}, {Contreras}, {de
  Jaeger}, {Folatelli}, {F{\"o}rster}, {Freedman}, {Gonz{\'a}lez}, {Hsiao},
  {Krzemi{\'n}ski}, {Krisciunas}, {Maza}, {McCarthy}, {Morrell}, {Persson},
  {Roth}, {Salgado}, {Suntzeff}, \&
  {Thomas-Osip}}]{Anderson2014ApJ...786...67A}
{Anderson}, J.~P., {Gonz{\'a}lez-Gait{\'a}n}, S., {Hamuy}, M., {et~al.} 2014,
  \hypersetup{urlcolor=magenta}\href{https://doi.org/10.1088/0004-637X/786/1/67}{\apj},
  \hypersetup{urlcolor=blue}\href{https://ui.adsabs.harvard.edu/abs/2014ApJ...786...67A}{786,
  67}

\bibitem[{{Arcavi}(2017)}]{Arcavi2017hsn..book..239A}
{Arcavi}, I. 2017, in Handbook of Supernovae, ed. A.~W. {Alsabti} \&
  P.~{Murdin}

\bibitem[{{Arcavi}(2022)}]{Arcavi2022ApJ...937...75}
{Arcavi}, I. 2022,
  \hypersetup{urlcolor=magenta}\href{https://doi.org/10.3847/1538-4357/ac90c0}{\apj},
  \hypersetup{urlcolor=blue}\href{https://ui.adsabs.harvard.edu/abs/2022ApJ...937...75A}{937,
  75}

\bibitem[{{Astropy Collaboration} {et~al.}(2018){Astropy Collaboration},
  {Price-Whelan}, {Sip{\H{o}}cz}, {G{\"u}nther}, {Lim}, {Crawford}, {Conseil},
  {Shupe}, {Craig}, {Dencheva}, {Ginsburg}, {VanderPlas}, {Bradley},
  {P{\'e}rez-Su{\'a}rez}, {de Val-Borro}, {Aldcroft}, {Cruz}, {Robitaille},
  {Tollerud}, {Ardelean}, {Babej}, {Bach}, {Bachetti}, {Bakanov}, {Bamford},
  {Barentsen}, {Barmby}, {Baumbach}, {Berry}, {Biscani}, {Boquien}, {Bostroem},
  {Bouma}, {Brammer}, {Bray}, {Breytenbach}, {Buddelmeijer}, {Burke},
  {Calderone}, {Cano Rodr{\'\i}guez}, {Cara}, {Cardoso}, {Cheedella}, {Copin},
  {Corrales}, {Crichton}, {D'Avella}, {Deil}, {Depagne}, {Dietrich}, {Donath},
  {Droettboom}, {Earl}, {Erben}, {Fabbro}, {Ferreira}, {Finethy}, {Fox},
  {Garrison}, {Gibbons}, {Goldstein}, {Gommers}, {Greco}, {Greenfield},
  {Groener}, {Grollier}, {Hagen}, {Hirst}, {Homeier}, {Horton}, {Hosseinzadeh},
  {Hu}, {Hunkeler}, {Ivezi{\'c}}, {Jain}, {Jenness}, {Kanarek}, {Kendrew},
  {Kern}, {Kerzendorf}, {Khvalko}, {King}, {Kirkby}, {Kulkarni}, {Kumar},
  {Lee}, {Lenz}, {Littlefair}, {Ma}, {Macleod}, {Mastropietro}, {McCully},
  {Montagnac}, {Morris}, {Mueller}, {Mumford}, {Muna}, {Murphy}, {Nelson},
  {Nguyen}, {Ninan}, {N{\"o}the}, {Ogaz}, {Oh}, {Parejko}, {Parley}, {Pascual},
  {Patil}, {Patil}, {Plunkett}, {Prochaska}, {Rastogi}, {Reddy Janga},
  {Sabater}, {Sakurikar}, {Seifert}, {Sherbert}, {Sherwood-Taylor}, {Shih},
  {Sick}, {Silbiger}, {Singanamalla}, {Singer}, {Sladen}, {Sooley},
  {Sornarajah}, {Streicher}, {Teuben}, {Thomas}, {Tremblay}, {Turner},
  {Terr{\'o}n}, {van Kerkwijk}, {de la Vega}, {Watkins}, {Weaver}, {Whitmore},
  {Woillez}, {Zabalza}, \& {Astropy Contributors}}]{Astropy2018AJ....156..123A}
{Astropy Collaboration}, {Price-Whelan}, A.~M., {Sip{\H{o}}cz}, B.~M., {et~al.}
  2018,
  \hypersetup{urlcolor=magenta}\href{https://doi.org/10.3847/1538-3881/aabc4f}{\aj},
  \hypersetup{urlcolor=blue}\href{https://ui.adsabs.harvard.edu/abs/2018AJ....156..123A}{156,
  123}

\bibitem[{{Baklanov} {et~al.}(2005){Baklanov}, {Blinnikov}, \&
  {Pavlyuk}}]{Baklanov2005AstL...31..429B}
{Baklanov}, P.~V., {Blinnikov}, S.~I., \& {Pavlyuk}, N.~N. 2005,
  \hypersetup{urlcolor=magenta}\href{https://doi.org/10.1134/1.1958107}{Astronomy
  Letters},
  \hypersetup{urlcolor=blue}\href{https://ui.adsabs.harvard.edu/abs/2005AstL...31..429B}{31,
  429}

\bibitem[{{Beasor} {et~al.}(2020){Beasor}, {Davies}, {Smith}, {van Loon},
  {Gehrz}, \& {Figer}}]{Beasor2020MNRAS.492.5994B}
{Beasor}, E.~R., {Davies}, B., {Smith}, N., {et~al.} 2020,
  \hypersetup{urlcolor=magenta}\href{https://doi.org/10.1093/mnras/staa255}{\mnras},
  \hypersetup{urlcolor=blue}\href{https://ui.adsabs.harvard.edu/abs/2020MNRAS.492.5994B}{492,
  5994}

\bibitem[{{Bellm} {et~al.}(2019){Bellm}, {Kulkarni}, {Graham}, {Dekany},
  {Smith}, {Riddle}, {Masci}, {Helou}, {Prince}, {Adams}, {Barbarino},
  {Barlow}, {Bauer}, {Beck}, {Belicki}, {Biswas}, {Blagorodnova}, {Bodewits},
  {Bolin}, {Brinnel}, {Brooke}, {Bue}, {Bulla}, {Burruss}, {Cenko}, {Chang},
  {Connolly}, {Coughlin}, {Cromer}, {Cunningham}, {De}, {Delacroix}, {Desai},
  {Duev}, {Eadie}, {Farnham}, {Feeney}, {Feindt}, {Flynn}, {Franckowiak},
  {Frederick}, {Fremling}, {Gal-Yam}, {Gezari}, {Giomi}, {Goldstein},
  {Golkhou}, {Goobar}, {Groom}, {Hacopians}, {Hale}, {Henning}, {Ho}, {Hover},
  {Howell}, {Hung}, {Huppenkothen}, {Imel}, {Ip}, {Ivezi{\'c}}, {Jackson},
  {Jones}, {Juric}, {Kasliwal}, {Kaspi}, {Kaye}, {Kelley}, {Kowalski},
  {Kramer}, {Kupfer}, {Landry}, {Laher}, {Lee}, {Lin}, {Lin}, {Lunnan},
  {Giomi}, {Mahabal}, {Mao}, {Miller}, {Monkewitz}, {Murphy}, {Ngeow},
  {Nordin}, {Nugent}, {Ofek}, {Patterson}, {Penprase}, {Porter}, {Rauch},
  {Rebbapragada}, {Reiley}, {Rigault}, {Rodriguez}, {van Roestel}, {Rusholme},
  {van Santen}, {Schulze}, {Shupe}, {Singer}, {Soumagnac}, {Stein}, {Surace},
  {Sollerman}, {Szkody}, {Taddia}, {Terek}, {Van Sistine}, {van Velzen},
  {Vestrand}, {Walters}, {Ward}, {Ye}, {Yu}, {Yan}, \&
  {Zolkower}}]{Bellm2019PASP..131a8002B}
{Bellm}, E.~C., {Kulkarni}, S.~R., {Graham}, M.~J., {et~al.} 2019,
  \hypersetup{urlcolor=magenta}\href{https://doi.org/10.1088/1538-3873/aaecbe}{\pasp},
  \hypersetup{urlcolor=blue}\href{https://ui.adsabs.harvard.edu/abs/2019PASP..131a8002B}{131,
  018002}

\bibitem[{{Berger} {et~al.}(2023){Berger}, {Keating}, {Margutti}, {Maeda},
  {Alexander}, {Cendes}, {Eftekhari}, {Gurwell}, {Hiramatsu}, {Ho}, {Laskar},
  {Rao}, \& {Williams}}]{Berger2023ApJ...951L..31B}
{Berger}, E., {Keating}, G.~K., {Margutti}, R., {et~al.} 2023,
  \hypersetup{urlcolor=magenta}\href{https://doi.org/10.3847/2041-8213/ace0c4}{\apjl},
  \hypersetup{urlcolor=blue}\href{https://ui.adsabs.harvard.edu/abs/2023ApJ...951L..31B}{951,
  L31}

\bibitem[{{Bertin} \& {Arnouts}(1996)}]{Bertin1996A&AS..117..393B}
{Bertin}, E., \& {Arnouts}, S. 1996,
  \hypersetup{urlcolor=magenta}\href{https://doi.org/10.1051/aas:1996164}{\aaps},
  \hypersetup{urlcolor=blue}\href{https://ui.adsabs.harvard.edu/abs/1996A&AS..117..393B}{117,
  393}

\bibitem[{{Blagorodnova} {et~al.}(2018){Blagorodnova}, {Neill}, {Walters},
  {Kulkarni}, {Fremling}, {Ben-Ami}, {Dekany}, {Fucik}, {Konidaris}, {Nash},
  {Ngeow}, {Ofek}, {O' Sullivan}, {Quimby}, {Ritter}, \&
  {Vyhmeister}}]{Blagorodnova2018PASP..130c5003B}
{Blagorodnova}, N., {Neill}, J.~D., {Walters}, R., {et~al.} 2018,
  \hypersetup{urlcolor=magenta}\href{https://doi.org/10.1088/1538-3873/aaa53f}{\pasp},
  \hypersetup{urlcolor=blue}\href{https://ui.adsabs.harvard.edu/abs/2018PASP..130c5003B}{130,
  035003}

\bibitem[{{Blinnikov} {et~al.}(2000){Blinnikov}, {Lundqvist}, {Bartunov},
  {Nomoto}, \& {Iwamoto}}]{Blinnikov2000ApJ...532.1132B}
{Blinnikov}, S., {Lundqvist}, P., {Bartunov}, O., {Nomoto}, K., \& {Iwamoto},
  K. 2000,
  \hypersetup{urlcolor=magenta}\href{https://doi.org/10.1086/308588}{\apj},
  \hypersetup{urlcolor=blue}\href{https://ui.adsabs.harvard.edu/abs/2000ApJ...532.1132B}{532,
  1132}

\bibitem[{{Blinnikov} \& {Sorokina}(2004)}]{Blinnikov2004Ap&SS.290...13B}
{Blinnikov}, S., \& {Sorokina}, E. 2004,
  \hypersetup{urlcolor=magenta}\href{https://doi.org/10.1023/B:ASTR.0000022161.03559.42}{\apss},
  \hypersetup{urlcolor=blue}\href{https://ui.adsabs.harvard.edu/abs/2004Ap&SS.290...13B}{290,
  13}

\bibitem[{{Blinnikov} {et~al.}(1998){Blinnikov}, {Eastman}, {Bartunov},
  {Popolitov}, \& {Woosley}}]{Blinnikov1998ApJ...496..454B}
{Blinnikov}, S.~I., {Eastman}, R., {Bartunov}, O.~S., {Popolitov}, V.~A., \&
  {Woosley}, S.~E. 1998,
  \hypersetup{urlcolor=magenta}\href{https://doi.org/10.1086/305375}{\apj},
  \hypersetup{urlcolor=blue}\href{https://ui.adsabs.harvard.edu/abs/1998ApJ...496..454B}{496,
  454}

\bibitem[{{Blinnikov} {et~al.}(2006){Blinnikov}, {R{\"o}pke}, {Sorokina},
  {Gieseler}, {Reinecke}, {Travaglio}, {Hillebrandt}, \&
  {Stritzinger}}]{Blinnikov2006A&A...453..229B}
{Blinnikov}, S.~I., {R{\"o}pke}, F.~K., {Sorokina}, E.~I., {et~al.} 2006,
  \hypersetup{urlcolor=magenta}\href{https://doi.org/10.1051/0004-6361:20054594}{\aap},
  \hypersetup{urlcolor=blue}\href{https://ui.adsabs.harvard.edu/abs/2006A&A...453..229B}{453,
  229}

\bibitem[{{Boian} \& {Groh}(2019)}]{Boian2019A&A...621A.109B}
{Boian}, I., \& {Groh}, J.~H. 2019,
  \hypersetup{urlcolor=magenta}\href{https://doi.org/10.1051/0004-6361/201833779}{\aap},
  \hypersetup{urlcolor=blue}\href{https://ui.adsabs.harvard.edu/abs/2019A&A...621A.109B}{621,
  A109}

\bibitem[{{Boian} \& {Groh}(2020)}]{Boian2020MNRAS.496.1325B}
{Boian}, I., \& {Groh}, J.~H. 2020,
  \hypersetup{urlcolor=magenta}\href{https://doi.org/10.1093/mnras/staa1540}{\mnras},
  \hypersetup{urlcolor=blue}\href{https://ui.adsabs.harvard.edu/abs/2020MNRAS.496.1325B}{496,
  1325}

\bibitem[{{Bostroem} {et~al.}(2023){Bostroem}, {Pearson}, {Shrestha}, {Sand},
  {Valenti}, {Jha}, {Andrews}, {Smith}, {Terreran}, {Green}, {Dong},
  {Lundquist}, {Haislip}, {Hoang}, {Hosseinzadeh}, {Janzen}, {Jencson},
  {Kouprianov}, {Paraskeva}, {Meza Retamal}, {Reichart}, {Arcavi}, {Bonanos},
  {Coughlin}, {Farah}, {Hawley}, {Hebb}, {Hiramatsu}, {Howell}, {Iijima},
  {Ilyin}, {McCully}, {Moran}, {Morris}, {Mura}, {Newsome}, {Pabst}, {Ochner},
  {Padilla Gonzalez}, {Pastorello}, {Pellegrino}, {Ravi}, {Reguitti}, {Salo},
  {Vinko}, {Wheeler}, {Williams}, \& {Wyatt}}]{Bostroem2023arXiv230610119B}
{Bostroem}, K.~A., {Pearson}, J., {Shrestha}, M., {et~al.} 2023,
  \hypersetup{urlcolor=magenta}\href{https://doi.org/10.48550/arXiv.2306.10119}{arXiv
  e-prints},
  \hypersetup{urlcolor=magenta}\href{https://doi.org/10.48550/arXiv.2306.10119}{arXiv}{:}\hypersetup{urlcolor=blue}\href{https://ui.adsabs.harvard.edu/abs/2023arXiv230610119B}{2306.10119}

\bibitem[{Bradley {et~al.}(2022)Bradley, Sipőcz, Robitaille, Tollerud,
  Vinícius, Deil, Barbary, Wilson, Busko, Donath, Günther, Cara, Lim,
  Meßlinger, Conseil, Bostroem, Droettboom, Bray, Bratholm, Barentsen, Craig,
  Rathi, Pascual, Perren, Georgiev, de~Val-Borro, Kerzendorf, Bach, Quint, \&
  Souchereau}]{photutils}
Bradley, L., Sipőcz, B., Robitaille, T., {et~al.} 2022, astropy/photutils:
  1.5.0, 1.5.0,  Zenodo,
  \hypersetup{urlcolor=magenta}doi:\href{https://doi.org/10.5281/zenodo.6825092}{10.5281/zenodo.6825092}

\bibitem[{{Breeveld} {et~al.}(2011){Breeveld}, {Landsman}, {Holland}, {Roming},
  {Kuin}, \& {Page}}]{Breeveld2011AIPC.1358..373B}
{Breeveld}, A.~A., {Landsman}, W., {Holland}, S.~T., {et~al.} 2011, in American
  Institute of Physics Conference Series, Vol. 1358, Gamma Ray Bursts 2010, ed.
  J.~E. {McEnery}, J.~L. {Racusin}, \& N.~{Gehrels}\hypersetup{urlcolor=blue},
  \href{https://ui.adsabs.harvard.edu/abs/2011AIPC.1358..373B}{373--376}

\bibitem[{{Brown} {et~al.}(2014){Brown}, {Breeveld}, {Holland}, {Kuin}, \&
  {Pritchard}}]{Brown2014Ap&SS.354...89B}
{Brown}, P.~J., {Breeveld}, A.~A., {Holland}, S., {Kuin}, P., \& {Pritchard},
  T. 2014,
  \hypersetup{urlcolor=magenta}\href{https://doi.org/10.1007/s10509-014-2059-8}{\apss},
  \hypersetup{urlcolor=blue}\href{https://ui.adsabs.harvard.edu/abs/2014Ap&SS.354...89B}{354,
  89}

\bibitem[{{Brown} {et~al.}(2013){Brown}, {Baliber}, {Bianco}, {Bowman},
  {Burleson}, {Conway}, {Crellin}, {Depagne}, {De Vera}, {Dilday}, {Dragomir},
  {Dubberley}, {Eastman}, {Elphick}, {Falarski}, {Foale}, {Ford}, {Fulton},
  {Garza}, {Gomez}, {Graham}, {Greene}, {Haldeman}, {Hawkins}, {Haworth},
  {Haynes}, {Hidas}, {Hjelstrom}, {Howell}, {Hygelund}, {Lister}, {Lobdill},
  {Martinez}, {Mullins}, {Norbury}, {Parrent}, {Paulson}, {Petry}, {Pickles},
  {Posner}, {Rosing}, {Ross}, {Sand}, {Saunders}, {Shobbrook}, {Shporer},
  {Street}, {Thomas}, {Tsapras}, {Tufts}, {Valenti}, {Vander Horst}, {Walker},
  {White}, \& {Willis}}]{Brown2013PASP..125.1031B}
{Brown}, T.~M., {Baliber}, N., {Bianco}, F.~B., {et~al.} 2013,
  \hypersetup{urlcolor=magenta}\href{https://doi.org/10.1086/673168}{\pasp},
  \hypersetup{urlcolor=blue}\href{https://ui.adsabs.harvard.edu/abs/2013PASP..125.1031B}{125,
  1031}

\bibitem[{{Buchhave} {et~al.}(2010){Buchhave}, {Bakos}, {Hartman}, {Torres},
  {Kov{\'a}cs}, {Latham}, {Noyes}, {Esquerdo}, {Everett}, {Howard}, {Marcy},
  {Fischer}, {Johnson}, {Andersen}, {F{\H{u}}r{\'e}sz}, {Perumpilly},
  {Sasselov}, {Stefanik}, {B{\'e}ky}, {L{\'a}z{\'a}r}, {Papp}, \&
  {S{\'a}ri}}]{Buchhave2010ApJ...720.1118B}
{Buchhave}, L.~A., {Bakos}, G.~{\'A}., {Hartman}, J.~D., {et~al.} 2010,
  \hypersetup{urlcolor=magenta}\href{https://doi.org/10.1088/0004-637X/720/2/1118}{\apj},
  \hypersetup{urlcolor=blue}\href{https://ui.adsabs.harvard.edu/abs/2010ApJ...720.1118B}{720,
  1118}

\bibitem[{{Chambers} {et~al.}(2016){Chambers}, {Magnier}, {Metcalfe},
  {Flewelling}, {Huber}, {Waters}, {Denneau}, {Draper}, {Farrow}, {Finkbeiner},
  {Holmberg}, {Koppenhoefer}, {Price}, {Rest}, {Saglia}, {Schlafly}, {Smartt},
  {Sweeney}, {Wainscoat}, {Burgett}, {Chastel}, {Grav}, {Heasley}, {Hodapp},
  {Jedicke}, {Kaiser}, {Kudritzki}, {Luppino}, {Lupton}, {Monet}, {Morgan},
  {Onaka}, {Shiao}, {Stubbs}, {Tonry}, {White}, {Ba{\~n}ados}, {Bell},
  {Bender}, {Bernard}, {Boegner}, {Boffi}, {Botticella}, {Calamida},
  {Casertano}, {Chen}, {Chen}, {Cole}, {Deacon}, {Frenk}, {Fitzsimmons},
  {Gezari}, {Gibbs}, {Goessl}, {Goggia}, {Gourgue}, {Goldman}, {Grant},
  {Grebel}, {Hambly}, {Hasinger}, {Heavens}, {Heckman}, {Henderson}, {Henning},
  {Holman}, {Hopp}, {Ip}, {Isani}, {Jackson}, {Keyes}, {Koekemoer}, {Kotak},
  {Le}, {Liska}, {Long}, {Lucey}, {Liu}, {Martin}, {Masci}, {McLean}, {Mindel},
  {Misra}, {Morganson}, {Murphy}, {Obaika}, {Narayan}, {Nieto-Santisteban},
  {Norberg}, {Peacock}, {Pier}, {Postman}, {Primak}, {Rae}, {Rai}, {Riess},
  {Riffeser}, {Rix}, {R{\"o}ser}, {Russel}, {Rutz}, {Schilbach}, {Schultz},
  {Scolnic}, {Strolger}, {Szalay}, {Seitz}, {Small}, {Smith}, {Soderblom},
  {Taylor}, {Thomson}, {Taylor}, {Thakar}, {Thiel}, {Thilker}, {Unger},
  {Urata}, {Valenti}, {Wagner}, {Walder}, {Walter}, {Watters}, {Werner},
  {Wood-Vasey}, \& {Wyse}}]{Chambers2016arXiv161205560C}
{Chambers}, K.~C., {Magnier}, E.~A., {Metcalfe}, N., {et~al.} 2016, arXiv
  e-prints,
  \hypersetup{urlcolor=magenta}\href{https://arxiv.org/abs/1612.05560}{arXiv}{:}\hypersetup{urlcolor=blue}\href{https://ui.adsabs.harvard.edu/abs/2016arXiv161205560C}{1612.05560}

\bibitem[{{Chevalier} \& {Fransson}(1994)}]{Chevalier1994ApJ...420..268C}
{Chevalier}, R.~A., \& {Fransson}, C. 1994,
  \hypersetup{urlcolor=magenta}\href{https://doi.org/10.1086/173557}{\apj},
  \hypersetup{urlcolor=blue}\href{https://ui.adsabs.harvard.edu/abs/1994ApJ...420..268C}{420,
  268}

\bibitem[{{Chevalier} \& {Fransson}(2017)}]{Chevalier2017hsn..book..875C}
{Chevalier}, R.~A., \& {Fransson}, C. 2017, in Handbook of Supernovae, ed.
  A.~W. {Alsabti} \& P.~{Murdin}

\bibitem[{{Chufarin} {et~al.}(2023){Chufarin}, {Potapov}, {Ionov}, {Korotkiy},
  {Nazarov}, \& {Sokolovsky}}]{Chufarin2023TNSAN.150....1C}
{Chufarin}, V., {Potapov}, N., {Ionov}, I., {et~al.} 2023, Transient Name
  Server AstroNote,
  \hypersetup{urlcolor=blue}\href{https://ui.adsabs.harvard.edu/abs/2023TNSAN.150....1C}{150,
  1}

\bibitem[{{Davies} {et~al.}(2022){Davies}, {Plez}, \&
  {Petrault}}]{Davies2022MNRAS.517.1483}
{Davies}, B., {Plez}, B., \& {Petrault}, M. 2022,
  \hypersetup{urlcolor=magenta}\href{https://doi.org/10.1093/mnras/stac2427}{\mnras},
  \hypersetup{urlcolor=blue}\href{https://ui.adsabs.harvard.edu/abs/2022MNRAS.517.1483D}{517,
  1483}

\bibitem[{{De} {et~al.}(2020){De}, {Hankins}, {Kasliwal}, {Moore}, {Ofek},
  {Adams}, {Ashley}, {Babul}, {Bagdasaryan}, {Burdge}, {Burnham}, {Dekany},
  {Declacroix}, {Galla}, {Greffe}, {Hale}, {Jencson}, {Lau}, {Mahabal},
  {McKenna}, {Sharma}, {Shopbell}, {Smith}, {Soon}, {Sokoloski}, {Soria}, \&
  {Travouillon}}]{De2020PASP..132b5001D}
{De}, K., {Hankins}, M.~J., {Kasliwal}, M.~M., {et~al.} 2020,
  \hypersetup{urlcolor=magenta}\href{https://doi.org/10.1088/1538-3873/ab6069}{\pasp},
  \hypersetup{urlcolor=blue}\href{https://ui.adsabs.harvard.edu/abs/2020PASP..132b5001D}{132,
  025001}

\bibitem[{{de Vaucouleurs} {et~al.}(1991){de Vaucouleurs}, {de Vaucouleurs},
  {Corwin}, {Buta}, {Paturel}, \& {Fouque}}]{deVaucouleurs1991rc3..book.....D}
{de Vaucouleurs}, G., {de Vaucouleurs}, A., {Corwin}, Herold~G., J., {et~al.}
  1991, {Third Reference Catalogue of Bright Galaxies}

\bibitem[{{Dessart} {et~al.}(2017){Dessart}, {Hillier}, \&
  {Audit}}]{Dessart2017A&A...605A..83D}
{Dessart}, L., {Hillier}, D.~J., \& {Audit}, E. 2017,
  \hypersetup{urlcolor=magenta}\href{https://doi.org/10.1051/0004-6361/201730942}{\aap},
  \hypersetup{urlcolor=blue}\href{https://ui.adsabs.harvard.edu/abs/2017A&A...605A..83D}{605,
  A83}

\bibitem[{{Fabricant} {et~al.}(1998){Fabricant}, {Cheimets}, {Caldwell}, \&
  {Geary}}]{Fabricant1998PASP..110...79F}
{Fabricant}, D., {Cheimets}, P., {Caldwell}, N., \& {Geary}, J. 1998,
  \hypersetup{urlcolor=magenta}\href{https://doi.org/10.1086/316111}{\pasp},
  \hypersetup{urlcolor=blue}\href{https://ui.adsabs.harvard.edu/abs/1998PASP..110...79F}{110,
  79}

\bibitem[{{Fan} {et~al.}(2016){Fan}, {Wang}, {Jiang}, {Wu}, {Li}, {Huang},
  {Xu}, {Hu}, {Zhu}, {Wang}, {Komossa}, \& {Zhang}}]{Fan2016PASP..128k5005F}
{Fan}, Z., {Wang}, H., {Jiang}, X., {et~al.} 2016,
  \hypersetup{urlcolor=magenta}\href{https://doi.org/10.1088/1538-3873/128/969/115005}{\pasp},
  \hypersetup{urlcolor=blue}\href{https://ui.adsabs.harvard.edu/abs/2016PASP..128k5005F}{128,
  115005}

\bibitem[{{Filippenko} {et~al.}(2023){Filippenko}, {Zheng}, \&
  {Yang}}]{Filippenko2023TNSAN.123....1F}
{Filippenko}, A.~V., {Zheng}, W., \& {Yang}, Y. 2023, Transient Name Server
  AstroNote,
  \hypersetup{urlcolor=blue}\href{https://ui.adsabs.harvard.edu/abs/2023TNSAN.123....1F}{123,
  1}

\bibitem[{{Fitzpatrick}(1999)}]{Fitzpatrick1999PASP..111...63F}
{Fitzpatrick}, E.~L. 1999,
  \hypersetup{urlcolor=magenta}\href{https://doi.org/10.1086/316293}{\pasp},
  \hypersetup{urlcolor=blue}\href{https://ui.adsabs.harvard.edu/abs/1999PASP..111...63F}{111,
  63}

\bibitem[{{Flewelling} {et~al.}(2020){Flewelling}, {Magnier}, {Chambers},
  {Heasley}, {Holmberg}, {Huber}, {Sweeney}, {Waters}, {Calamida}, {Casertano},
  {Chen}, {Farrow}, {Hasinger}, {Henderson}, {Long}, {Metcalfe}, {Narayan},
  {Nieto-Santisteban}, {Norberg}, {Rest}, {Saglia}, {Szalay}, {Thakar},
  {Tonry}, {Valenti}, {Werner}, {White}, {Denneau}, {Draper}, {Hodapp},
  {Jedicke}, {Kaiser}, {Kudritzki}, {Price}, {Wainscoat}, {Chastel}, {McLean},
  {Postman}, \& {Shiao}}]{Flewelling2020ApJS..251....7F}
{Flewelling}, H.~A., {Magnier}, E.~A., {Chambers}, K.~C., {et~al.} 2020,
  \hypersetup{urlcolor=magenta}\href{https://doi.org/10.3847/1538-4365/abb82d}{\apjs},
  \hypersetup{urlcolor=blue}\href{https://ui.adsabs.harvard.edu/abs/2020ApJS..251....7F}{251,
  7}

\bibitem[{{Foreman-Mackey} {et~al.}(2013){Foreman-Mackey}, {Hogg}, {Lang}, \&
  {Goodman}}]{Foreman-Mackey2013PASP..125..306F}
{Foreman-Mackey}, D., {Hogg}, D.~W., {Lang}, D., \& {Goodman}, J. 2013,
  \hypersetup{urlcolor=magenta}\href{https://doi.org/10.1086/670067}{\pasp},
  \hypersetup{urlcolor=blue}\href{https://ui.adsabs.harvard.edu/abs/2013PASP..125..306F}{125,
  306}

\bibitem[{{F{\"o}rster} {et~al.}(2018){F{\"o}rster}, {Moriya}, {Maureira},
  {Anderson}, {Blinnikov}, {Bufano}, {Cabrera-Vives}, {Clocchiatti}, {de
  Jaeger}, {Est{\'e}vez}, {Galbany}, {Gonz{\'a}lez-Gait{\'a}n}, {Gr{\"a}fener},
  {Hamuy}, {Hsiao}, {Huentelemu}, {Huijse}, {Kuncarayakti}, {Mart{\'\i}nez},
  {Medina}, {Olivares E.}, {Pignata}, {Razza}, {Reyes}, {San Mart{\'\i}n},
  {Smith}, {Vera}, {Vivas}, {de Ugarte Postigo}, {Yoon}, {Ashall}, {Fraser},
  {Gal-Yam}, {Kankare}, {Le Guillou}, {Mazzali}, {Walton}, \&
  {Young}}]{Forster2018NatAs...2..808F}
{F{\"o}rster}, F., {Moriya}, T.~J., {Maureira}, J.~C., {et~al.} 2018,
  \hypersetup{urlcolor=magenta}\href{https://doi.org/10.1038/s41550-018-0563-4}{Nature
  Astronomy},
  \hypersetup{urlcolor=blue}\href{https://ui.adsabs.harvard.edu/abs/2018NatAs...2..808F}{2,
  808}

\bibitem[{{Fulton} {et~al.}(2023){Fulton}, {Nicholl}, {Smith}, {Srivastav},
  {Young}, {McCollum}, {Moore}, {Sim}, {Weston}, {Sheng}, {Shingles}, {Sommer},
  {Aamer}, {Smartt}, {Stevance}, {Rhodes}, {Andersson}, {Denneau}, {Tonry},
  {Weiland}, {Lawrence}, {Siverd}, {Erasmus}, {Koorts}, {Anderson}, {Jordan},
  {Suc}, {Rest}, {Chen}, \& {Stubbs}}]{Fulton2023TNSAN.124....1F}
{Fulton}, M.~D., {Nicholl}, M., {Smith}, K.~W., {et~al.} 2023, Transient Name
  Server AstroNote,
  \hypersetup{urlcolor=blue}\href{https://ui.adsabs.harvard.edu/abs/2023TNSAN.124....1F}{124,
  1}

\bibitem[{{Gal-Yam} {et~al.}(2014){Gal-Yam}, {Arcavi}, {Ofek}, {Ben-Ami},
  {Cenko}, {Kasliwal}, {Cao}, {Yaron}, {Tal}, {Silverman}, {Horesh}, {De Cia},
  {Taddia}, {Sollerman}, {Perley}, {Vreeswijk}, {Kulkarni}, {Nugent},
  {Filippenko}, \& {Wheeler}}]{Gal-Yam2014Natur.509..471G}
{Gal-Yam}, A., {Arcavi}, I., {Ofek}, E.~O., {et~al.} 2014,
  \hypersetup{urlcolor=magenta}\href{https://doi.org/10.1038/nature13304}{\nat},
  \hypersetup{urlcolor=blue}\href{https://ui.adsabs.harvard.edu/abs/2014Natur.509..471G}{509,
  471}

\bibitem[{{Goldberg} \& {Bildsten}(2020)}]{Goldberg2020ApJ...895L..45G}
{Goldberg}, J.~A., \& {Bildsten}, L. 2020,
  \hypersetup{urlcolor=magenta}\href{https://doi.org/10.3847/2041-8213/ab9300}{\apjl},
  \hypersetup{urlcolor=blue}\href{https://ui.adsabs.harvard.edu/abs/2020ApJ...895L..45G}{895,
  L45}

\bibitem[{{Goldberg} {et~al.}(2019){Goldberg}, {Bildsten}, \&
  {Paxton}}]{Goldberg2019ApJ...879....3G}
{Goldberg}, J.~A., {Bildsten}, L., \& {Paxton}, B. 2019,
  \hypersetup{urlcolor=magenta}\href{https://doi.org/10.3847/1538-4357/ab22b6}{\apj},
  \hypersetup{urlcolor=blue}\href{https://ui.adsabs.harvard.edu/abs/2019ApJ...879....3G}{879,
  3}

\bibitem[{{Goldman} {et~al.}(2017){Goldman}, {van Loon}, {Zijlstra}, {Green},
  {Wood}, {Nanni}, {Imai}, {Whitelock}, {Matsuura}, {Groenewegen}, \&
  {G{\'o}mez}}]{Goldman2017MNRAS.465..403G}
{Goldman}, S.~R., {van Loon}, J.~T., {Zijlstra}, A.~A., {et~al.} 2017,
  \hypersetup{urlcolor=magenta}\href{https://doi.org/10.1093/mnras/stw2708}{\mnras},
  \hypersetup{urlcolor=blue}\href{https://ui.adsabs.harvard.edu/abs/2017MNRAS.465..403G}{465,
  403}

\bibitem[{{Gomez} {et~al.}(2023){Gomez}, {Berger}, {Blanchard}, {Hosseinzadeh},
  {Nicholl}, {Hiramatsu}, {Villar}, \& {Yin}}]{Gomez2023ApJ...949..114G}
{Gomez}, S., {Berger}, E., {Blanchard}, P.~K., {et~al.} 2023,
  \hypersetup{urlcolor=magenta}\href{https://doi.org/10.3847/1538-4357/acc536}{\apj},
  \hypersetup{urlcolor=blue}\href{https://ui.adsabs.harvard.edu/abs/2023ApJ...949..114G}{949,
  114}

\bibitem[{{Gomez} {et~al.}(2020){Gomez}, {Berger}, {Blanchard}, {Hosseinzadeh},
  {Nicholl}, {Villar}, \& {Yin}}]{Gomez2020ApJ...904...74G}
{Gomez}, S., {Berger}, E., {Blanchard}, P.~K., {et~al.} 2020,
  \hypersetup{urlcolor=magenta}\href{https://doi.org/10.3847/1538-4357/abbf49}{\apj},
  \hypersetup{urlcolor=blue}\href{https://ui.adsabs.harvard.edu/abs/2020ApJ...904...74G}{904,
  74}

\bibitem[{{Graham} {et~al.}(2019){Graham}, {Kulkarni}, {Bellm}, {Adams},
  {Barbarino}, {Blagorodnova}, {Bodewits}, {Bolin}, {Brady}, {Cenko}, {Chang},
  {Coughlin}, {De}, {Eadie}, {Farnham}, {Feindt}, {Franckowiak}, {Fremling},
  {Gezari}, {Ghosh}, {Goldstein}, {Golkhou}, {Goobar}, {Ho}, {Huppenkothen},
  {Ivezi{\'c}}, {Jones}, {Juric}, {Kaplan}, {Kasliwal}, {Kelley}, {Kupfer},
  {Lee}, {Lin}, {Lunnan}, {Mahabal}, {Miller}, {Ngeow}, {Nugent}, {Ofek},
  {Prince}, {Rauch}, {van Roestel}, {Schulze}, {Singer}, {Sollerman}, {Taddia},
  {Yan}, {Ye}, {Yu}, {Barlow}, {Bauer}, {Beck}, {Belicki}, {Biswas}, {Brinnel},
  {Brooke}, {Bue}, {Bulla}, {Burruss}, {Connolly}, {Cromer}, {Cunningham},
  {Dekany}, {Delacroix}, {Desai}, {Duev}, {Feeney}, {Flynn}, {Frederick},
  {Gal-Yam}, {Giomi}, {Groom}, {Hacopians}, {Hale}, {Helou}, {Henning},
  {Hover}, {Hillenbrand}, {Howell}, {Hung}, {Imel}, {Ip}, {Jackson}, {Kaspi},
  {Kaye}, {Kowalski}, {Kramer}, {Kuhn}, {Landry}, {Laher}, {Mao}, {Masci},
  {Monkewitz}, {Murphy}, {Nordin}, {Patterson}, {Penprase}, {Porter},
  {Rebbapragada}, {Reiley}, {Riddle}, {Rigault}, {Rodriguez}, {Rusholme}, {van
  Santen}, {Shupe}, {Smith}, {Soumagnac}, {Stein}, {Surace}, {Szkody}, {Terek},
  {Van Sistine}, {van Velzen}, {Vestrand}, {Walters}, {Ward}, {Zhang}, \&
  {Zolkower}}]{Graham2019PASP..131g8001G}
{Graham}, M.~J., {Kulkarni}, S.~R., {Bellm}, E.~C., {et~al.} 2019,
  \hypersetup{urlcolor=magenta}\href{https://doi.org/10.1088/1538-3873/ab006c}{\pasp},
  \hypersetup{urlcolor=blue}\href{https://ui.adsabs.harvard.edu/abs/2019PASP..131g8001G}{131,
  078001}

\bibitem[{{Grefenstette} {et~al.}(2023){Grefenstette}, {Brightman}, {Earnshaw},
  {Harrison}, \& {Margutti}}]{Grefenstette2023ApJ...952L...3G}
{Grefenstette}, B.~W., {Brightman}, M., {Earnshaw}, H.~P., {Harrison}, F.~A.,
  \& {Margutti}, R. 2023,
  \hypersetup{urlcolor=magenta}\href{https://doi.org/10.3847/2041-8213/acdf4e}{\apjl},
  \hypersetup{urlcolor=blue}\href{https://ui.adsabs.harvard.edu/abs/2023ApJ...952L...3G}{952,
  L3}

\bibitem[{{Haynie} \& {Piro}(2021)}]{Haynie2021ApJ...910..128}
{Haynie}, A., \& {Piro}, A.~L. 2021,
  \hypersetup{urlcolor=magenta}\href{https://doi.org/10.3847/1538-4357/abe938}{\apj},
  \hypersetup{urlcolor=blue}\href{https://ui.adsabs.harvard.edu/abs/2021ApJ...910..128H}{910,
  128}

\bibitem[{{Hiramatsu} {et~al.}(2021{\natexlab{\hspace{0pt}a}}){Hiramatsu},
  {Howell}, {Moriya}, {Goldberg}, {Hosseinzadeh}, {Arcavi}, {Anderson},
  {Guti{\'e}rrez}, {Burke}, {McCully}, {Valenti}, {Galbany}, {Fang}, {Maeda},
  {Folatelli}, {Hsiao}, {Morrell}, {Phillips}, {Stritzinger}, {Suntzeff},
  {Gromadzki}, {Maguire}, {M{\"u}ller-Bravo}, \&
  {Young}}]{Hiramatsu2021ApJ...913...55H}
{Hiramatsu}, D., {Howell}, D.~A., {Moriya}, T.~J., {et~al.}
  2021{\natexlab{\hspace{0pt}a}},
  \hypersetup{urlcolor=magenta}\href{https://doi.org/10.3847/1538-4357/abf6d6}{\apj},
  \hypersetup{urlcolor=blue}\href{https://ui.adsabs.harvard.edu/abs/2021ApJ...913...55H}{913,
  55}

\bibitem[{{Hiramatsu} {et~al.}(2021{\natexlab{\hspace{0pt}b}}){Hiramatsu},
  {Howell}, {Van Dyk}, {Goldberg}, {Maeda}, {Moriya}, {Tominaga}, {Nomoto},
  {Hosseinzadeh}, {Arcavi}, {McCully}, {Burke}, {Bostroem}, {Valenti}, {Dong},
  {Brown}, {Andrews}, {Bilinski}, {Williams}, {Smith}, {Smith}, {Sand},
  {Anand}, {Xu}, {Filippenko}, {Bersten}, {Folatelli}, {Kelly}, {Noguchi}, \&
  {Itagaki}}]{Hiramatsu2021NatAs...5..903H}
{Hiramatsu}, D., {Howell}, D.~A., {Van Dyk}, S.~D., {et~al.}
  2021{\natexlab{\hspace{0pt}b}},
  \hypersetup{urlcolor=magenta}\href{https://doi.org/10.1038/s41550-021-01384-2}{Nature
  Astronomy},
  \hypersetup{urlcolor=blue}\href{https://ui.adsabs.harvard.edu/abs/2021NatAs...5..903H}{5,
  903}

\bibitem[{{Hiramatsu} {et~al.}(2023){Hiramatsu}, {Matsumoto}, {Berger},
  {Ransome}, {Villar}, {Gomez}, {Cendes}, {De}, {Farah}, {Howell}, {McCully},
  {Newsome}, {Padilla Gonzalez}, {Pellegrino}, {Suzuki}, \&
  {Terreran}}]{Hiramatsu2023arXiv230511168}
{Hiramatsu}, D., {Matsumoto}, T., {Berger}, E., {et~al.} 2023,
  \hypersetup{urlcolor=magenta}\href{https://doi.org/10.48550/arXiv.2305.11168}{arXiv
  e-prints},
  \hypersetup{urlcolor=magenta}\href{https://doi.org/10.48550/arXiv.2305.11168}{arXiv}{:}\hypersetup{urlcolor=blue}\href{https://ui.adsabs.harvard.edu/abs/2023arXiv230511168H}{2305.11168}

\bibitem[{{Hosseinzadeh} {et~al.}(2023){Hosseinzadeh}, {Farah}, {Shrestha},
  {Sand}, {Dong}, {Brown}, {Bostroem}, {Valenti}, {Jha}, {Andrews}, {Arcavi},
  {Haislip}, {Hiramatsu}, {Hoang}, {Howell}, {Janzen}, {Jencson}, {Kouprianov},
  {Lundquist}, {McCully}, {Meza Retamal}, {Modjaz}, {Newsome}, {Gonzalez},
  {Pearson}, {Pellegrino}, {Ravi}, {Reichart}, {Smith}, {Terreran}, \&
  {Vink{\'o}}}]{Hosseinzadeh2023ApJ...953L..16H}
{Hosseinzadeh}, G., {Farah}, J., {Shrestha}, M., {et~al.} 2023,
  \hypersetup{urlcolor=magenta}\href{https://doi.org/10.3847/2041-8213/ace4c4}{\apjl},
  \hypersetup{urlcolor=blue}\href{https://ui.adsabs.harvard.edu/abs/2023ApJ...953L..16H}{953,
  L16}

\bibitem[{{Howell} \& {Global Supernova
  Project}(2017)}]{Howell2017AAS...23031803H}
{Howell}, D.~A., \& {Global Supernova Project}. 2017, in American Astronomical
  Society Meeting Abstracts, Vol. 230, American Astronomical Society Meeting
  Abstracts \#230\hypersetup{urlcolor=blue},
  \href{https://ui.adsabs.harvard.edu/abs/2017AAS...23031803H}{318.03}

\bibitem[{{Hu} {et~al.}(2018){Hu}, {Wang}, {Lin}, {Kong}, {Cheng}, {Fan},
  {Fang}, {Lin}, {Mao}, {Wang}, {Zhou}, {Zhou}, {Zhu}, \&
  {Zou}}]{Hu2018ApJ...854...68H}
{Hu}, N., {Wang}, E., {Lin}, Z., {et~al.} 2018,
  \hypersetup{urlcolor=magenta}\href{https://doi.org/10.3847/1538-4357/aaa6ca}{\apj},
  \hypersetup{urlcolor=blue}\href{https://ui.adsabs.harvard.edu/abs/2018ApJ...854...68H}{854,
  68}

\bibitem[{{Hunter}(2007)}]{Hunter2007CSE.....9...90H}
{Hunter}, J.~D. 2007,
  \hypersetup{urlcolor=magenta}\href{https://doi.org/10.1109/MCSE.2007.55}{Computing
  in Science and Engineering},
  \hypersetup{urlcolor=blue}\href{https://ui.adsabs.harvard.edu/abs/2007CSE.....9...90H}{9,
  90}

\bibitem[{{Itagaki}(2023)}]{Itagaki2023TNSTR1158....1I}
{Itagaki}, K. 2023, Transient Name Server Discovery Report,
  \hypersetup{urlcolor=blue}\href{https://ui.adsabs.harvard.edu/abs/2023TNSTR1158....1I}{2023-1158,
  1}

\bibitem[{{Jacobson-Gal{\'a}n} {et~al.}(2022){Jacobson-Gal{\'a}n}, {Dessart},
  {Jones}, {Margutti}, {Coppejans}, {Dimitriadis}, {Foley}, {Kilpatrick},
  {Matthews}, {Rest}, {Terreran}, {Aleo}, {Auchettl}, {Blanchard}, {Coulter},
  {Davis}, {de Boer}, {DeMarchi}, {Drout}, {Earl}, {Gagliano}, {Gall},
  {Hjorth}, {Huber}, {Ibik}, {Milisavljevic}, {Pan}, {Rest}, {Ridden-Harper},
  {Rojas-Bravo}, {Siebert}, {Smith}, {Taggart}, {Tinyanont}, {Wang}, \&
  {Zenati}}]{Jacobson-Galan2022ApJ...924...15}
{Jacobson-Gal{\'a}n}, W.~V., {Dessart}, L., {Jones}, D.~O., {et~al.} 2022,
  \hypersetup{urlcolor=magenta}\href{https://doi.org/10.3847/1538-4357/ac3f3a}{\apj},
  \hypersetup{urlcolor=blue}\href{https://ui.adsabs.harvard.edu/abs/2022ApJ...924...15J}{924,
  15}

\bibitem[{{Jacobson-Galan} {et~al.}(2023){Jacobson-Galan}, {Dessart},
  {Margutti}, {Chornock}, {Foley}, {Kilpatrick}, {Jones}, {Taggart}, {Angus},
  {Bhattacharjee}, {Braff}, {Brethauer}, {Burgasser}, {Cao}, {Carlile},
  {Chambers}, {Coulter}, {Dominguez-Ruiz}, {Dickinson}, {de Boer}, {Gagliano},
  {Gall}, {Gao}, {Gates}, {Gomez}, {Guolo}, {Halford}, {Hjorth}, {Huber},
  {Johnson}, {Karpoor}, {Laskar}, {LeBaron}, {Li}, {Lin}, {Loch}, {Lynam},
  {Magnier}, {Maloney}, {Matthews}, {McDonald}, {Miao}, {Milisavljevic}, {Pan},
  {Pradyumna}, {Ransome}, {Rees}, {Rest}, {Rojas-Bravo}, {Sandford}, {Sandoval
  Ascencio}, {Sanjaripour}, {Savino}, {Sears}, {Sharei}, {Smartt}, {Softich},
  {Theissen}, {Tinyanont}, {Tohfa}, {Villar}, {Wang}, {Wainscoat},
  {Westerling}, {Wiston}, {Wozniak}, {Yadavalli}, \&
  {Zenati}}]{Jacobson-Galan2023arXiv230604721J}
{Jacobson-Galan}, W.~V., {Dessart}, L., {Margutti}, R., {et~al.} 2023, arXiv
  e-prints,
  \hypersetup{urlcolor=magenta}\href{https://arxiv.org/abs/2306.04721}{arXiv}{:}\hypersetup{urlcolor=blue}\href{https://ui.adsabs.harvard.edu/abs/2023arXiv230604721J}{2306.04721}

\bibitem[{{Jencson} {et~al.}(2023){Jencson}, {Pearson}, {Beasor}, {Lau},
  {Andrews}, {Bostroem}, {Dong}, {Engesser}, {Gomez}, {Guolo}, {Hoang},
  {Hosseinzadeh}, {Jha}, {Karambelkar}, {Kasliwal}, {Lundquist}, {Meza
  Retamal}, {Rest}, {Sand}, {Shahbandeh}, {Shrestha}, {Smith}, {Strader},
  {Valenti}, {Wang}, \& {Zenati}}]{Jencson2023ApJ...952L..30J}
{Jencson}, J.~E., {Pearson}, J., {Beasor}, E.~R., {et~al.} 2023,
  \hypersetup{urlcolor=magenta}\href{https://doi.org/10.3847/2041-8213/ace618}{\apjl},
  \hypersetup{urlcolor=blue}\href{https://ui.adsabs.harvard.edu/abs/2023ApJ...952L..30J}{952,
  L30}

\bibitem[{{Johnson} {et~al.}(2018){Johnson}, {Kochanek}, \&
  {Adams}}]{Johnson2018MNRAS.480.1696}
{Johnson}, S.~A., {Kochanek}, C.~S., \& {Adams}, S.~M. 2018,
  \hypersetup{urlcolor=magenta}\href{https://doi.org/10.1093/mnras/sty1966}{\mnras},
  \hypersetup{urlcolor=blue}\href{https://ui.adsabs.harvard.edu/abs/2018MNRAS.480.1696J}{480,
  1696}

\bibitem[{{Kasen} \& {Woosley}(2009)}]{Kasen2009ApJ...703.2205K}
{Kasen}, D., \& {Woosley}, S.~E. 2009,
  \hypersetup{urlcolor=magenta}\href{https://doi.org/10.1088/0004-637X/703/2/2205}{\apj},
  \hypersetup{urlcolor=blue}\href{https://ui.adsabs.harvard.edu/abs/2009ApJ...703.2205K}{703,
  2205}

\bibitem[{{Khatami} \& {Kasen}(2023)}]{Khatami2023arXiv230403360K}
{Khatami}, D., \& {Kasen}, D. 2023,
  \hypersetup{urlcolor=magenta}\href{https://doi.org/10.48550/arXiv.2304.03360}{arXiv
  e-prints},
  \hypersetup{urlcolor=magenta}\href{https://doi.org/10.48550/arXiv.2304.03360}{arXiv}{:}\hypersetup{urlcolor=blue}\href{https://ui.adsabs.harvard.edu/abs/2023arXiv230403360K}{2304.03360}

\bibitem[{{Kilpatrick} {et~al.}(2023){Kilpatrick}, {Foley},
  {Jacobson-Gal{\'a}n}, {Piro}, {Smartt}, {Drout}, {Gagliano}, {Gall},
  {Hjorth}, {Jones}, {Mandel}, {Margutti}, {Ramirez-Ruiz}, {Ransome}, {Villar},
  {Coulter}, {Gao}, {Matthews}, {Taggart}, \&
  {Zenati}}]{Kilpatrick2023ApJ...952L..23K}
{Kilpatrick}, C.~D., {Foley}, R.~J., {Jacobson-Gal{\'a}n}, W.~V., {et~al.}
  2023,
  \hypersetup{urlcolor=magenta}\href{https://doi.org/10.3847/2041-8213/ace4ca}{\apjl},
  \hypersetup{urlcolor=blue}\href{https://ui.adsabs.harvard.edu/abs/2023ApJ...952L..23K}{952,
  L23}

\bibitem[{{Kochanek} {et~al.}(2017){Kochanek}, {Fraser}, {Adams}, {Sukhbold},
  {Prieto}, {M{\"u}ller}, {Bock}, {Brown}, {Dong}, {Holoien}, {Khan},
  {Shappee}, \& {Stanek}}]{Kochanek2017MNRAS.467.3347}
{Kochanek}, C.~S., {Fraser}, M., {Adams}, S.~M., {et~al.} 2017,
  \hypersetup{urlcolor=magenta}\href{https://doi.org/10.1093/mnras/stx291}{\mnras},
  \hypersetup{urlcolor=blue}\href{https://ui.adsabs.harvard.edu/abs/2017MNRAS.467.3347K}{467,
  3347}

\bibitem[{{Koltenbah}(2023)}]{Koltenbah2023TNSAN.144....1K}
{Koltenbah}, B. 2023, Transient Name Server AstroNote,
  \hypersetup{urlcolor=blue}\href{https://ui.adsabs.harvard.edu/abs/2023TNSAN.144....1K}{144,
  1}

\bibitem[{{Kozyreva} {et~al.}(2019){Kozyreva}, {Nakar}, \&
  {Waldman}}]{Kozyreva2019MNRAS.483.1211K}
{Kozyreva}, A., {Nakar}, E., \& {Waldman}, R. 2019,
  \hypersetup{urlcolor=magenta}\href{https://doi.org/10.1093/mnras/sty3185}{\mnras},
  \hypersetup{urlcolor=blue}\href{https://ui.adsabs.harvard.edu/abs/2019MNRAS.483.1211K}{483,
  1211}

\bibitem[{{Kuriyama} \& {Shigeyama}(2020)}]{Kuriyama2020A&A...635A.127}
{Kuriyama}, N., \& {Shigeyama}, T. 2020,
  \hypersetup{urlcolor=magenta}\href{https://doi.org/10.1051/0004-6361/201937226}{\aap},
  \hypersetup{urlcolor=blue}\href{https://ui.adsabs.harvard.edu/abs/2020A&A...635A.127K}{635,
  A127}

\bibitem[{{Landolt}(1983)}]{Landolt1983AJ.....88..439L}
{Landolt}, A.~U. 1983,
  \hypersetup{urlcolor=magenta}\href{https://doi.org/10.1086/113329}{\aj},
  \hypersetup{urlcolor=blue}\href{https://ui.adsabs.harvard.edu/abs/1983AJ.....88..439L}{88,
  439}

\bibitem[{{Landolt}(1992)}]{Landolt1992AJ....104..340L}
{Landolt}, A.~U. 1992,
  \hypersetup{urlcolor=magenta}\href{https://doi.org/10.1086/116242}{\aj},
  \hypersetup{urlcolor=blue}\href{https://ui.adsabs.harvard.edu/abs/1992AJ....104..340L}{104,
  340}

\bibitem[{{Lang}(2014)}]{Lang2014AJ....147..108L}
{Lang}, D. 2014,
  \hypersetup{urlcolor=magenta}\href{https://doi.org/10.1088/0004-6256/147/5/108}{\aj},
  \hypersetup{urlcolor=blue}\href{https://ui.adsabs.harvard.edu/abs/2014AJ....147..108L}{147,
  108}

\bibitem[{{Lundquist} {et~al.}(2023){Lundquist}, {O'Meara}, \&
  {Walawender}}]{Lundquist2023TNSAN.160....1L}
{Lundquist}, M., {O'Meara}, J., \& {Walawender}, J. 2023, Transient Name Server
  AstroNote,
  \hypersetup{urlcolor=blue}\href{https://ui.adsabs.harvard.edu/abs/2023TNSAN.160....1L}{160,
  1}

\bibitem[{{Mainzer} {et~al.}(2014){Mainzer}, {Bauer}, {Cutri}, {Grav},
  {Masiero}, {Beck}, {Clarkson}, {Conrow}, {Dailey}, {Eisenhardt}, {Fabinsky},
  {Fajardo-Acosta}, {Fowler}, {Gelino}, {Grillmair}, {Heinrichsen}, {Kendall},
  {Kirkpatrick}, {Liu}, {Masci}, {McCallon}, {Nugent}, {Papin}, {Rice},
  {Royer}, {Ryan}, {Sevilla}, {Sonnett}, {Stevenson}, {Thompson}, {Wheelock},
  {Wiemer}, {Wittman}, {Wright}, \& {Yan}}]{Mainzer2014ApJ...792...30M}
{Mainzer}, A., {Bauer}, J., {Cutri}, R.~M., {et~al.} 2014,
  \hypersetup{urlcolor=magenta}\href{https://doi.org/10.1088/0004-637X/792/1/30}{\apj},
  \hypersetup{urlcolor=blue}\href{https://ui.adsabs.harvard.edu/abs/2014ApJ...792...30M}{792,
  30}

\bibitem[{{Mao} {et~al.}(2023){Mao}, {Zhang}, {Cai}, {Chen}, {Chen}, {Gao},
  {Li}, {Lyu}, {Qin}, {Sun}, {Xu}, {Zhang}, {Zhang}, {Zhao}, {Zheng}, {Zhou},
  \& {Ye}}]{Mao2023TNSAN.130....1M}
{Mao}, Y., {Zhang}, M., {Cai}, G., {et~al.} 2023, Transient Name Server
  AstroNote,
  \hypersetup{urlcolor=blue}\href{https://ui.adsabs.harvard.edu/abs/2023TNSAN.130....1M}{130,
  1}

\bibitem[{{Margutti} {et~al.}(2014){Margutti}, {Milisavljevic}, {Soderberg},
  {Chornock}, {Zauderer}, {Murase}, {Guidorzi}, {Sanders}, {Kuin}, {Fransson},
  {Levesque}, {Chandra}, {Berger}, {Bianco}, {Brown}, {Challis},
  {Chatzopoulos}, {Cheung}, {Choi}, {Chomiuk}, {Chugai}, {Contreras}, {Drout},
  {Fesen}, {Foley}, {Fong}, {Friedman}, {Gall}, {Gehrels}, {Hjorth}, {Hsiao},
  {Kirshner}, {Im}, {Leloudas}, {Lunnan}, {Marion}, {Martin}, {Morrell},
  {Neugent}, {Omodei}, {Phillips}, {Rest}, {Silverman}, {Strader},
  {Stritzinger}, {Szalai}, {Utterback}, {Vinko}, {Wheeler}, {Arnett},
  {Campana}, {Chevalier}, {Ginsburg}, {Kamble}, {Roming}, {Pritchard}, \&
  {Stringfellow}}]{Margutti2014ApJ...780...21M}
{Margutti}, R., {Milisavljevic}, D., {Soderberg}, A.~M., {et~al.} 2014,
  \hypersetup{urlcolor=magenta}\href{https://doi.org/10.1088/0004-637X/780/1/21}{\apj},
  \hypersetup{urlcolor=blue}\href{https://ui.adsabs.harvard.edu/abs/2014ApJ...780...21M}{780,
  21}

\bibitem[{{Masci} {et~al.}(2023){Masci}, {Laher}, {Rusholme}, {Shupe},
  {Paladini}, {Groom}, \& {Wold}}]{Masci2023arXiv230516279M}
{Masci}, F.~J., {Laher}, R.~R., {Rusholme}, B., {et~al.} 2023,
  \hypersetup{urlcolor=magenta}\href{https://doi.org/10.48550/arXiv.2305.16279}{arXiv
  e-prints},
  \hypersetup{urlcolor=magenta}\href{https://doi.org/10.48550/arXiv.2305.16279}{arXiv}{:}\hypersetup{urlcolor=blue}\href{https://ui.adsabs.harvard.edu/abs/2023arXiv230516279M}{2305.16279}

\bibitem[{{Masci} {et~al.}(2019){Masci}, {Laher}, {Rusholme}, {Shupe}, {Groom},
  {Surace}, {Jackson}, {Monkewitz}, {Beck}, {Flynn}, {Terek}, {Landry},
  {Hacopians}, {Desai}, {Howell}, {Brooke}, {Imel}, {Wachter}, {Ye}, {Lin},
  {Cenko}, {Cunningham}, {Rebbapragada}, {Bue}, {Miller}, {Mahabal}, {Bellm},
  {Patterson}, {Juri{\'c}}, {Golkhou}, {Ofek}, {Walters}, {Graham}, {Kasliwal},
  {Dekany}, {Kupfer}, {Burdge}, {Cannella}, {Barlow}, {Van Sistine}, {Giomi},
  {Fremling}, {Blagorodnova}, {Levitan}, {Riddle}, {Smith}, {Helou}, {Prince},
  \& {Kulkarni}}]{Masci2019PASP..131a8003M}
{Masci}, F.~J., {Laher}, R.~R., {Rusholme}, B., {et~al.} 2019,
  \hypersetup{urlcolor=magenta}\href{https://doi.org/10.1088/1538-3873/aae8ac}{\pasp},
  \hypersetup{urlcolor=blue}\href{https://ui.adsabs.harvard.edu/abs/2019PASP..131a8003M}{131,
  018003}

\bibitem[{{Matsumoto} \& {Metzger}(2022)}]{Matsumoto2022ApJ...936..114}
{Matsumoto}, T., \& {Metzger}, B.~D. 2022,
  \hypersetup{urlcolor=magenta}\href{https://doi.org/10.3847/1538-4357/ac892c}{\apj},
  \hypersetup{urlcolor=blue}\href{https://ui.adsabs.harvard.edu/abs/2022ApJ...936..114M}{936,
  114}

\bibitem[{{Matthews} {et~al.}(2023){Matthews}, {Margutti}, {AJ},
  {Jacobson-Galan}, {Chornock}, {Alexander}, {Laskar}, {Cendes}, {Berger},
  {Drout}, \& {Milisavljevic}}]{Matthews2023ATel16091....1M}
{Matthews}, D., {Margutti}, R., {AJ}, N., {et~al.} 2023, The Astronomer's
  Telegram,
  \hypersetup{urlcolor=blue}\href{https://ui.adsabs.harvard.edu/abs/2023ATel16091....1M}{16091,
  1}

\bibitem[{{Matzner} \& {McKee}(1999)}]{Matzner1999ApJ...510..379M}
{Matzner}, C.~D., \& {McKee}, C.~F. 1999,
  \hypersetup{urlcolor=magenta}\href{https://doi.org/10.1086/306571}{\apj},
  \hypersetup{urlcolor=blue}\href{https://ui.adsabs.harvard.edu/abs/1999ApJ...510..379M}{510,
  379}

\bibitem[{McCully {et~al.}(2018)McCully, Turner, Volgenau, Harbeck, Valenti,
  Riba, Bachelet, Snyder, Kurczynski, Norbury, \&
  Street}]{curtis_mccully_2018_1257560}
McCully, C., Turner, M., Volgenau, N., {et~al.} 2018, LCOGT/banzai: Initial
  Release, 0.9.4,  Zenodo,
  \hypersetup{urlcolor=magenta}doi:\href{https://doi.org/10.5281/zenodo.1257560}{10.5281/zenodo.1257560}

\bibitem[{{Meisner} {et~al.}(2018){Meisner}, {Lang}, \&
  {Schlegel}}]{Meisner2018AJ....156...69M}
{Meisner}, A.~M., {Lang}, D., \& {Schlegel}, D.~J. 2018,
  \hypersetup{urlcolor=magenta}\href{https://doi.org/10.3847/1538-3881/aacbcd}{\aj},
  \hypersetup{urlcolor=blue}\href{https://ui.adsabs.harvard.edu/abs/2018AJ....156...69M}{156,
  69}

\bibitem[{{Mink}(2011)}]{Mink2011ASPC..442..305M}
{Mink}, D.~J. 2011, in Astronomical Society of the Pacific Conference Series,
  Vol. 442, Astronomical Data Analysis Software and Systems XX, ed. I.~N.
  {Evans}, A.~{Accomazzi}, D.~J. {Mink}, \& A.~H.
  {Rots}\hypersetup{urlcolor=blue},
  \href{https://ui.adsabs.harvard.edu/abs/2011ASPC..442..305M}{305}

\bibitem[{{Mink} {et~al.}(2007){Mink}, {Wyatt}, {Caldwell}, {Conroy}, {Furesz},
  \& {Tokarz}}]{Mink2007ASPC..376..249M}
{Mink}, D.~J., {Wyatt}, W.~F., {Caldwell}, N., {et~al.} 2007, in Astronomical
  Society of the Pacific Conference Series, Vol. 376, Astronomical Data
  Analysis Software and Systems XVI, ed. R.~A. {Shaw}, F.~{Hill}, \& D.~J.
  {Bell}\hypersetup{urlcolor=blue},
  \href{https://ui.adsabs.harvard.edu/abs/2007ASPC..376..249M}{249}

\bibitem[{{Mink} {et~al.}(2005){Mink}, {Wyatt}, {Roll}, {Tokarz}, {Conroy},
  {Caldwell}, {Kurtz}, \& {Geller}}]{Mink2005ASPC..347..228M}
{Mink}, D.~J., {Wyatt}, W.~F., {Roll}, J.~B., {et~al.} 2005, in Astronomical
  Society of the Pacific Conference Series, Vol. 347, Astronomical Data
  Analysis Software and Systems XIV, ed. P.~{Shopbell}, M.~{Britton}, \&
  R.~{Ebert}\hypersetup{urlcolor=blue},
  \href{https://ui.adsabs.harvard.edu/abs/2005ASPC..347..228M}{228}

\bibitem[{{Moriya} {et~al.}(2018){Moriya}, {F{\"o}rster}, {Yoon},
  {Gr{\"a}fener}, \& {Blinnikov}}]{Moriya2018MNRAS.476.2840M}
{Moriya}, T.~J., {F{\"o}rster}, F., {Yoon}, S.-C., {Gr{\"a}fener}, G., \&
  {Blinnikov}, S.~I. 2018,
  \hypersetup{urlcolor=magenta}\href{https://doi.org/10.1093/mnras/sty475}{\mnras},
  \hypersetup{urlcolor=blue}\href{https://ui.adsabs.harvard.edu/abs/2018MNRAS.476.2840M}{476,
  2840}

\bibitem[{{Morozova} {et~al.}(2020){Morozova}, {Piro}, {Fuller}, \& {Van
  Dyk}}]{Morozova2020ApJ...891L..32M}
{Morozova}, V., {Piro}, A.~L., {Fuller}, J., \& {Van Dyk}, S.~D. 2020,
  \hypersetup{urlcolor=magenta}\href{https://doi.org/10.3847/2041-8213/ab77c8}{\apjl},
  \hypersetup{urlcolor=blue}\href{https://ui.adsabs.harvard.edu/abs/2020ApJ...891L..32M}{891,
  L32}

\bibitem[{{Morozova} {et~al.}(2015){Morozova}, {Piro}, {Renzo}, {Ott},
  {Clausen}, {Couch}, {Ellis}, \& {Roberts}}]{Morozova2015ApJ...814...63}
{Morozova}, V., {Piro}, A.~L., {Renzo}, M., {et~al.} 2015,
  \hypersetup{urlcolor=magenta}\href{https://doi.org/10.1088/0004-637X/814/1/63}{\apj},
  \hypersetup{urlcolor=blue}\href{https://ui.adsabs.harvard.edu/abs/2015ApJ...814...63M}{814,
  63}

\bibitem[{{Morozova} {et~al.}(2018){Morozova}, {Piro}, \&
  {Valenti}}]{Morozova2018ApJ...858...15}
{Morozova}, V., {Piro}, A.~L., \& {Valenti}, S. 2018,
  \hypersetup{urlcolor=magenta}\href{https://doi.org/10.3847/1538-4357/aab9a6}{\apj},
  \hypersetup{urlcolor=blue}\href{https://ui.adsabs.harvard.edu/abs/2018ApJ...858...15M}{858,
  15}

\bibitem[{{Nakar} \& {Sari}(2010)}]{Nakar2010ApJ...725..904}
{Nakar}, E., \& {Sari}, R. 2010,
  \hypersetup{urlcolor=magenta}\href{https://doi.org/10.1088/0004-637X/725/1/904}{\apj},
  \hypersetup{urlcolor=blue}\href{https://ui.adsabs.harvard.edu/abs/2010ApJ...725..904N}{725,
  904}

\bibitem[{{Neustadt} {et~al.}(2023){Neustadt}, {Kochanek}, \& {Rizzo
  Smith}}]{Neustadt2023arXiv230606162N}
{Neustadt}, J.~M.~M., {Kochanek}, C.~S., \& {Rizzo Smith}, M. 2023,
  \hypersetup{urlcolor=magenta}\href{https://doi.org/10.48550/arXiv.2306.06162}{arXiv
  e-prints},
  \hypersetup{urlcolor=magenta}\href{https://doi.org/10.48550/arXiv.2306.06162}{arXiv}{:}\hypersetup{urlcolor=blue}\href{https://ui.adsabs.harvard.edu/abs/2023arXiv230606162N}{2306.06162}

\bibitem[{{Ofek} {et~al.}(2013){Ofek}, {Sullivan}, {Cenko}, {Kasliwal},
  {Gal-Yam}, {Kulkarni}, {Arcavi}, {Bildsten}, {Bloom}, {Horesh}, {Howell},
  {Filippenko}, {Laher}, {Murray}, {Nakar}, {Nugent}, {Silverman}, {Shaviv},
  {Surace}, \& {Yaron}}]{Ofek2013Natur.494...65O}
{Ofek}, E.~O., {Sullivan}, M., {Cenko}, S.~B., {et~al.} 2013,
  \hypersetup{urlcolor=magenta}\href{https://doi.org/10.1038/nature11877}{\nat},
  \hypersetup{urlcolor=blue}\href{https://ui.adsabs.harvard.edu/abs/2013Natur.494...65O}{494,
  65}

\bibitem[{{Ofek} {et~al.}(2014){Ofek}, {Sullivan}, {Shaviv}, {Steinbok},
  {Arcavi}, {Gal-Yam}, {Tal}, {Kulkarni}, {Nugent}, {Ben-Ami}, {Kasliwal},
  {Cenko}, {Laher}, {Surace}, {Bloom}, {Filippenko}, {Silverman}, \&
  {Yaron}}]{Ofek2014ApJ...789..104O}
{Ofek}, E.~O., {Sullivan}, M., {Shaviv}, N.~J., {et~al.} 2014,
  \hypersetup{urlcolor=magenta}\href{https://doi.org/10.1088/0004-637X/789/2/104}{\apj},
  \hypersetup{urlcolor=blue}\href{https://ui.adsabs.harvard.edu/abs/2014ApJ...789..104O}{789,
  104}

\bibitem[{Oliphant(2006)}]{Oliphant2006}
Oliphant, T.~E. 2006, A Guide to NumPy
  (\hypersetup{urlcolor=magenta}\href{https://web.mit.edu/dvp/Public/numpybook.pdf}{USA:
  Trelgol Publishing})

\bibitem[{{Paxton} {et~al.}(2011){Paxton}, {Bildsten}, {Dotter}, {Herwig},
  {Lesaffre}, \& {Timmes}}]{Paxton2011ApJS..192....3P}
{Paxton}, B., {Bildsten}, L., {Dotter}, A., {et~al.} 2011,
  \hypersetup{urlcolor=magenta}\href{https://doi.org/10.1088/0067-0049/192/1/3}{\apjs},
  \hypersetup{urlcolor=blue}\href{https://ui.adsabs.harvard.edu/abs/2011ApJS..192....3P}{192,
  3}

\bibitem[{{Paxton} {et~al.}(2013){Paxton}, {Cantiello}, {Arras}, {Bildsten},
  {Brown}, {Dotter}, {Mankovich}, {Montgomery}, {Stello}, {Timmes}, \&
  {Townsend}}]{Paxton2013ApJS..208....4P}
{Paxton}, B., {Cantiello}, M., {Arras}, P., {et~al.} 2013,
  \hypersetup{urlcolor=magenta}\href{https://doi.org/10.1088/0067-0049/208/1/4}{\apjs},
  \hypersetup{urlcolor=blue}\href{https://ui.adsabs.harvard.edu/abs/2013ApJS..208....4P}{208,
  4}

\bibitem[{{Paxton} {et~al.}(2015){Paxton}, {Marchant}, {Schwab}, {Bauer},
  {Bildsten}, {Cantiello}, {Dessart}, {Farmer}, {Hu}, {Langer}, {Townsend},
  {Townsley}, \& {Timmes}}]{Paxton2015ApJS..220...15P}
{Paxton}, B., {Marchant}, P., {Schwab}, J., {et~al.} 2015,
  \hypersetup{urlcolor=magenta}\href{https://doi.org/10.1088/0067-0049/220/1/15}{\apjs},
  \hypersetup{urlcolor=blue}\href{https://ui.adsabs.harvard.edu/abs/2015ApJS..220...15P}{220,
  15}

\bibitem[{{Paxton} {et~al.}(2018){Paxton}, {Schwab}, {Bauer}, {Bildsten},
  {Blinnikov}, {Duffell}, {Farmer}, {Goldberg}, {Marchant}, {Sorokina},
  {Thoul}, {Townsend}, \& {Timmes}}]{Paxton2018ApJS..234...34P}
{Paxton}, B., {Schwab}, J., {Bauer}, E.~B., {et~al.} 2018,
  \hypersetup{urlcolor=magenta}\href{https://doi.org/10.3847/1538-4365/aaa5a8}{\apjs},
  \hypersetup{urlcolor=blue}\href{https://ui.adsabs.harvard.edu/abs/2018ApJS..234...34P}{234,
  34}

\bibitem[{{Paxton} {et~al.}(2019){Paxton}, {Smolec}, {Schwab}, {Gautschy},
  {Bildsten}, {Cantiello}, {Dotter}, {Farmer}, {Goldberg}, {Jermyn}, {Kanbur},
  {Marchant}, {Thoul}, {Townsend}, {Wolf}, {Zhang}, \&
  {Timmes}}]{Paxton2019ApJS..243...10P}
{Paxton}, B., {Smolec}, R., {Schwab}, J., {et~al.} 2019,
  \hypersetup{urlcolor=magenta}\href{https://doi.org/10.3847/1538-4365/ab2241}{\apjs},
  \hypersetup{urlcolor=blue}\href{https://ui.adsabs.harvard.edu/abs/2019ApJS..243...10P}{243,
  10}

\bibitem[{{Perley} {et~al.}(2023){Perley}, {Gal-Yam}, {Irani}, \&
  {Zimmerman}}]{Perley2023TNSAN.119....1P}
{Perley}, D.~A., {Gal-Yam}, A., {Irani}, I., \& {Zimmerman}, E. 2023, Transient
  Name Server AstroNote,
  \hypersetup{urlcolor=blue}\href{https://ui.adsabs.harvard.edu/abs/2023TNSAN.119....1P}{119,
  1}

\bibitem[{{Perley} \& {Irani}(2023)}]{Perley2023TNSAN.120....1P}
{Perley}, D.~A., \& {Irani}, I. 2023, Transient Name Server AstroNote,
  \hypersetup{urlcolor=blue}\href{https://ui.adsabs.harvard.edu/abs/2023TNSAN.120....1P}{120,
  1}

\bibitem[{{Piascik} {et~al.}(2014){Piascik}, {Steele}, {Bates}, {Mottram},
  {Smith}, {Barnsley}, \& {Bolton}}]{Piascik2014SPIE.9147E..8HP}
{Piascik}, A.~S., {Steele}, I.~A., {Bates}, S.~D., {et~al.} 2014, in Society of
  Photo-Optical Instrumentation Engineers (SPIE) Conference Series, Vol. 9147,
  Ground-based and Airborne Instrumentation for Astronomy V, ed. S.~K.
  {Ramsay}, I.~S. {McLean}, \& H.~{Takami}\hypersetup{urlcolor=blue},
  \href{https://ui.adsabs.harvard.edu/abs/2014SPIE.9147E..8HP}{91478H}

\bibitem[{{Piro}(2015)}]{Piro2015ApJ...808L..51}
{Piro}, A.~L. 2015,
  \hypersetup{urlcolor=magenta}\href{https://doi.org/10.1088/2041-8205/808/2/L51}{\apjl},
  \hypersetup{urlcolor=blue}\href{https://ui.adsabs.harvard.edu/abs/2015ApJ...808L..51P}{808,
  L51}

\bibitem[{{Piro} {et~al.}(2021){Piro}, {Haynie}, \&
  {Yao}}]{Piro2021ApJ...909..209}
{Piro}, A.~L., {Haynie}, A., \& {Yao}, Y. 2021,
  \hypersetup{urlcolor=magenta}\href{https://doi.org/10.3847/1538-4357/abe2b1}{\apj},
  \hypersetup{urlcolor=blue}\href{https://ui.adsabs.harvard.edu/abs/2021ApJ...909..209P}{909,
  209}

\bibitem[{{Pledger} \& {Shara}(2023)}]{Pledger2023ApJ...953L..14P}
{Pledger}, J.~L., \& {Shara}, M.~M. 2023,
  \hypersetup{urlcolor=magenta}\href{https://doi.org/10.3847/2041-8213/ace88b}{\apjl},
  \hypersetup{urlcolor=blue}\href{https://ui.adsabs.harvard.edu/abs/2023ApJ...953L..14P}{953,
  L14}

\bibitem[{{Popov}(1993)}]{Popov1993ApJ...414..712P}
{Popov}, D.~V. 1993,
  \hypersetup{urlcolor=magenta}\href{https://doi.org/10.1086/173117}{\apj},
  \hypersetup{urlcolor=blue}\href{https://ui.adsabs.harvard.edu/abs/1993ApJ...414..712P}{414,
  712}

\bibitem[{{Poznanski} {et~al.}(2012){Poznanski}, {Prochaska}, \&
  {Bloom}}]{Poznanski2012MNRAS.426.1465P}
{Poznanski}, D., {Prochaska}, J.~X., \& {Bloom}, J.~S. 2012,
  \hypersetup{urlcolor=magenta}\href{https://doi.org/10.1111/j.1365-2966.2012.21796.x}{\mnras},
  \hypersetup{urlcolor=blue}\href{https://ui.adsabs.harvard.edu/abs/2012MNRAS.426.1465P}{426,
  1465}

\bibitem[{{Riess} {et~al.}(2022){Riess}, {Yuan}, {Macri}, {Scolnic}, {Brout},
  {Casertano}, {Jones}, {Murakami}, {Anand}, {Breuval}, {Brink}, {Filippenko},
  {Hoffmann}, {Jha}, {D'arcy Kenworthy}, {Mackenty}, {Stahl}, \&
  {Zheng}}]{Riess2022ApJ...934L...7R}
{Riess}, A.~G., {Yuan}, W., {Macri}, L.~M., {et~al.} 2022,
  \hypersetup{urlcolor=magenta}\href{https://doi.org/10.3847/2041-8213/ac5c5b}{\apjl},
  \hypersetup{urlcolor=blue}\href{https://ui.adsabs.harvard.edu/abs/2022ApJ...934L...7R}{934,
  L7}

\bibitem[{{Schlafly} \& {Finkbeiner}(2011)}]{Schlafly2011ApJ...737..103S}
{Schlafly}, E.~F., \& {Finkbeiner}, D.~P. 2011,
  \hypersetup{urlcolor=magenta}\href{https://doi.org/10.1088/0004-637X/737/2/103}{\apj},
  \hypersetup{urlcolor=blue}\href{https://ui.adsabs.harvard.edu/abs/2011ApJ...737..103S}{737,
  103}

\bibitem[{{Science Software Branch at STScI}(2012)}]{PyRAF2012ascl.soft07011S}
{Science Software Branch at STScI}. 2012, {PyRAF: Python alternative for IRAF},
  Astrophysics Source Code Library, record ascl:1207.011,
  \hypersetup{urlcolor=magenta}\href{https://ascl.net/1207.011}{ascl}:\hypersetup{urlcolor=blue}\href{https://ui.adsabs.harvard.edu/abs/2012ascl.soft07011S}{1207.011}

\bibitem[{{SDSS Collaboration} {et~al.}(2017){SDSS Collaboration}, {Albareti},
  {Allende Prieto}, {Almeida}, {Anders}, {Anderson}, {Andrews},
  {Arag{\'o}n-Salamanca}, {Argudo-Fern{\'a}ndez}, {Armengaud}, {Aubourg},
  {Avila-Reese}, {Badenes}, {Bailey}, {Barbuy}, {Barger},
  {Barrera-Ballesteros}, {Bartosz}, {Basu}, {Bates}, {Battaglia}, {Baumgarten},
  {Baur}, {Bautista}, {Beers}, {Belfiore}, {Bershady}, {Bertran de Lis},
  {Bird}, {Bizyaev}, {Blanc}, {Blanton}, {Blomqvist}, {Bolton}, {Borissova},
  {Bovy}, {Brandt}, {Brinkmann}, {Brownstein}, {Bundy}, {Burtin}, {Busca},
  {Camacho Chavez}, {Cano D{\'\i}az}, {Cappellari}, {Carrera}, {Chen},
  {Cherinka}, {Cheung}, {Chiappini}, {Chojnowski}, {Chuang}, {Chung},
  {Cirolini}, {Clerc}, {Cohen}, {Comerford}, {Comparat}, {Correa do
  Nascimento}, {Cousinou}, {Covey}, {Crane}, {Croft}, {Cunha}, {Darling},
  {Davidson}, {Dawson}, {Da Costa}, {Da Silva Ilha}, {Deconto Machado},
  {Delubac}, {De Lee}, {De la Macorra}, {De la Torre}, {Diamond-Stanic},
  {Donor}, {Downes}, {Drory}, {Du}, {Du Mas des Bourboux}, {Dwelly}, {Ebelke},
  {Eigenbrot}, {Eisenstein}, {Elsworth}, {Emsellem}, {Eracleous}, {Escoffier},
  {Evans}, {Falc{\'o}n-Barroso}, {Fan}, {Favole}, {Fernandez-Alvar},
  {Fernandez-Trincado}, {Feuillet}, {Fleming}, {Font-Ribera}, {Freischlad},
  {Frinchaboy}, {Fu}, {Gao}, {Garcia}, {Garcia-Dias}, {Garcia-Hern{\'a}ndez},
  {Garcia P{\'e}rez}, {Gaulme}, {Ge}, {Geisler}, {Gillespie}, {Gil Marin},
  {Girardi}, {Goddard}, {Gomez Maqueo Chew}, {Gonzalez-Perez}, {Grabowski},
  {Green}, {Grier}, {Grier}, {Guo}, {Guy}, {Hagen}, {Hall}, {Harding},
  {Harley}, {Hasselquist}, {Hawley}, {Hayes}, {Hearty}, {Hekker}, {Hernandez
  Toledo}, {Ho}, {Hogg}, {Holley-Bockelmann}, {Holtzman}, {Holzer}, {Hu},
  {Huber}, {Hutchinson}, {Hwang}, {Ibarra-Medel}, {Ivans}, {Ivory}, {Jaehnig},
  {Jensen}, {Johnson}, {Jones}, {Jullo}, {Kallinger}, {Kinemuchi}, {Kirkby},
  {Klaene}, {Kneib}, {Kollmeier}, {Lacerna}, {Lane}, {Lang}, {Laurent}, {Law},
  {Leauthaud}, {Le Goff}, {Li}, {Li}, {Li}, {Li}, {Liang}, {Liang}, {Lima},
  {Lin}, {Lin}, {Lin}, {Liu}, {Long}, {Lucatello}, {MacDonald}, {MacLeod},
  {Mackereth}, {Mahadevan}, {Maia}, {Maiolino}, {Majewski}, {Malanushenko},
  {Malanushenko}, {Mallmann}, {Manchado}, {Maraston}, {Marques-Chaves},
  {Martinez Valpuesta}, {Masters}, {Mathur}, {McGreer}, {Merloni},
  {Merrifield}, {M{\'e}sz{\'a}ros}, {Meza}, {Miglio}, {Minchev},
  {Molaverdikhani}, {Montero-Dorta}, {Mosser}, {Muna}, {Myers}, {Nair},
  {Nandra}, {Ness}, {Newman}, {Nichol}, {Nidever}, {Nitschelm}, {O'Connell},
  {Oravetz}, {Oravetz}, {Pace}, {Padilla}, {Palanque-Delabrouille}, {Pan},
  {Parejko}, {Paris}, {Park}, {Peacock}, {Peirani}, {Pellejero-Ibanez},
  {Penny}, {Percival}, {Percival}, {Perez-Fournon}, {Petitjean}, {Pieri},
  {Pinsonneault}, {Pisani}, {Prada}, {Prakash}, {Price-Jones}, {Raddick},
  {Rahman}, {Raichoor}, {Barboza Rembold}, {Reyna}, {Rich}, {Richstein},
  {Ridl}, {Riffel}, {Riffel}, {Rix}, {Robin}, {Rockosi},
  {Rodr{\'\i}guez-Torres}, {Rodrigues}, {Roe}, {Roman Lopes},
  {Rom{\'a}n-Z{\'u}{\~n}iga}, {Ross}, {Rossi}, {Ruan}, {Ruggeri}, {Runnoe},
  {Salazar-Albornoz}, {Salvato}, {Sanchez}, {Sanchez}, {Sanchez-Gallego},
  {Santiago}, {Schiavon}, {Schimoia}, {Schlafly}, {Schlegel}, {Schneider},
  {Sch{\"o}nrich}, {Schultheis}, {Schwope}, {Seo}, {Serenelli}, {Sesar},
  {Shao}, {Shetrone}, {Shull}, {Silva Aguirre}, {Skrutskie}, {Slosar}, {Smith},
  {Smith}, {Sobeck}, {Somers}, {Souto}, {Stark}, {Stassun}, {Steinmetz},
  {Stello}, {Storchi Bergmann}, {Strauss}, {Streblyanska}, {Stringfellow},
  {Suarez}, {Sun}, {Taghizadeh-Popp}, {Tang}, {Tao}, {Tayar}, {Tembe},
  {Thomas}, {Tinker}, {Tojeiro}, {Tremonti}, {Troup}, {Trump}, {Unda-Sanzana},
  {Valenzuela}, {Van den Bosch}, {Vargas-Maga{\~n}a}, {Vazquez}, {Villanova},
  {Vivek}, {Vogt}, {Wake}, {Walterbos}, {Wang}, {Wang}, {Weaver}, {Weijmans},
  {Weinberg}, {Westfall}, {Whelan}, {Wilcots}, {Wild}, {Williams}, {Wilson},
  {Wood-Vasey}, {Wylezalek}, {Xiao}, {Yan}, {Yang}, {Ybarra}, {Yeche}, {Yuan},
  {Zakamska}, {Zamora}, {Zasowski}, {Zhang}, {Zhao}, {Zhao}, {Zheng}, {Zheng},
  {Zhou}, {Zhu}, {Zinn}, \& {Zou}}]{Albareti2017ApJS..233...25A}
{SDSS Collaboration}, {Albareti}, F.~D., {Allende Prieto}, C., {et~al.} 2017,
  \hypersetup{urlcolor=magenta}\href{https://doi.org/10.3847/1538-4365/aa8992}{\apjs},
  \hypersetup{urlcolor=blue}\href{https://ui.adsabs.harvard.edu/abs/2017ApJS..233...25A}{233,
  25}

\bibitem[{{Sgro} {et~al.}(2023){Sgro}, {Esposito}, {Blaclard}, {Gomez},
  {Marchis}, {Filippenko}, {Peluso}, {Lawrence}, {Verveen}, {Wagner}, {Nardi},
  {Wiart}, {Mirwald}, {Christensen}, {Eramia}, {Parker}, {Guillet}, {Kim},
  {Logan}, {Kyba}, {Toulmin}, {Vantaggiato}, {Adhis}, {Gary}, {Goodey},
  {Dickinson}, {Koster}, {Martin}, {Bonilla}, {Chung}, {Miny}, {Mortecrette},
  {Saibi}, {Gagnon}, {Simard}, {Vacon}, {Simard}, {Dreise}, {Funakoshi},
  {Vacon}, {Yaniz}, {Le Tarnec}, {Laugier}, {Siders}, {Sweitzer}, {Dvoracek},
  {Archer}, {Deitz}, {Bradley}, {Fukui}, {Sibbernsen}, {Borrot}, {Cross},
  {Heider}, {Yamaguchi}, {Hirsch}, {Leroux}, {Billiani}, {Lorber}, {Smallen},
  {Shimizu}, {Nishimura}, {Ryno}, {Cunningham}, {Gagnon}, {Primm}, {Rushton},
  {Sibbernsen}, {Mitchell}, {Yoblonsky}, {Leroux}, {Clerget}, {Stojanovi{\'c}},
  {Unique}, {Huth}, {Ang}, {Santoni}, {Foster}, {Poggiali}, {Xu}, {Kukita},
  {{\v{S}}{\'c}epanovi{\'c}}, {Saibi}, {Will}, {Latour}, {Haythornthwaite},
  {Cadieux}, {M{\"u}ller}, {Chung}, {Watanabe}, \&
  {Arnaud}}]{Sgro2023RNAAS...7..141S}
{Sgro}, L.~A., {Esposito}, T.~M., {Blaclard}, G., {et~al.} 2023,
  \hypersetup{urlcolor=magenta}\href{https://doi.org/10.3847/2515-5172/ace41f}{Research
  Notes of the American Astronomical Society},
  \hypersetup{urlcolor=blue}\href{https://ui.adsabs.harvard.edu/abs/2023RNAAS...7..141S}{7,
  141}

\bibitem[{{Shingles} {et~al.}(2021){Shingles}, {Smith}, {Young}, {Smartt},
  {Tonry}, {Denneau}, {Heinze}, {Weiland}, {Flewelling}, {Stalder},
  {Clocchiatti}, {F{\"o}rster}, {Pignata}, {Rest}, {Anderson}, {Stubbs}, \&
  {Erasmus}}]{Shingles2021TNSAN...7....1S}
{Shingles}, L., {Smith}, K.~W., {Young}, D.~R., {et~al.} 2021, Transient Name
  Server AstroNote,
  \hypersetup{urlcolor=blue}\href{https://ui.adsabs.harvard.edu/abs/2021TNSAN...7....1S}{7,
  1}

\bibitem[{{Smartt}(2009)}]{Smartt2009ARA&A..47...63S}
{Smartt}, S.~J. 2009,
  \hypersetup{urlcolor=magenta}\href{https://doi.org/10.1146/annurev-astro-082708-101737}{\araa},
  \hypersetup{urlcolor=blue}\href{https://ui.adsabs.harvard.edu/abs/2009ARA&A..47...63S}{47,
  63}

\bibitem[{{Smartt}(2015)}]{Smartt2015PASA...32...16S}
{Smartt}, S.~J. 2015,
  \hypersetup{urlcolor=magenta}\href{https://doi.org/10.1017/pasa.2015.17}{\pasa},
  \hypersetup{urlcolor=blue}\href{https://ui.adsabs.harvard.edu/abs/2015PASA...32...16S}{32,
  e016}

\bibitem[{{Smith} {et~al.}(2020){Smith}, {Smartt}, {Young}, {Tonry}, {Denneau},
  {Flewelling}, {Heinze}, {Weiland}, {Stalder}, {Rest}, {Stubbs}, {Anderson},
  {Chen}, {Clark}, {Do}, {F{\"o}rster}, {Fulton}, {Gillanders}, {McBrien},
  {O'Neill}, {Srivastav}, \& {Wright}}]{Smith2020PASP..132h5002S}
{Smith}, K.~W., {Smartt}, S.~J., {Young}, D.~R., {et~al.} 2020,
  \hypersetup{urlcolor=magenta}\href{https://doi.org/10.1088/1538-3873/ab936e}{\pasp},
  \hypersetup{urlcolor=blue}\href{https://ui.adsabs.harvard.edu/abs/2020PASP..132h5002S}{132,
  085002}

\bibitem[{{Smith} {et~al.}(2023){Smith}, {Pearson}, {Sand}, {Ilyin},
  {Bostroem}, {Hosseinzadeh}, \& {Shrestha}}]{Smith2023arXiv230607964S}
{Smith}, N., {Pearson}, J., {Sand}, D.~J., {et~al.} 2023,
  \hypersetup{urlcolor=magenta}\href{https://doi.org/10.48550/arXiv.2306.07964}{arXiv
  e-prints},
  \hypersetup{urlcolor=magenta}\href{https://doi.org/10.48550/arXiv.2306.07964}{arXiv}{:}\hypersetup{urlcolor=blue}\href{https://ui.adsabs.harvard.edu/abs/2023arXiv230607964S}{2306.07964}

\bibitem[{{Soraisam} {et~al.}(2023){Soraisam}, {Szalai}, {Van Dyk}, {Andrews},
  {Srinivasan}, {Chun}, {Matheson}, {Scicluna}, \&
  {Vasquez-Torres}}]{Soraisam2023arXiv230610783S}
{Soraisam}, M.~D., {Szalai}, T., {Van Dyk}, S.~D., {et~al.} 2023,
  \hypersetup{urlcolor=magenta}\href{https://doi.org/10.48550/arXiv.2306.10783}{arXiv
  e-prints},
  \hypersetup{urlcolor=magenta}\href{https://doi.org/10.48550/arXiv.2306.10783}{arXiv}{:}\hypersetup{urlcolor=blue}\href{https://ui.adsabs.harvard.edu/abs/2023arXiv230610783S}{2306.10783}

\bibitem[{{Steele} {et~al.}(2004){Steele}, {Smith}, {Rees}, {Baker}, {Bates},
  {Bode}, {Bowman}, {Carter}, {Etherton}, {Ford}, {Fraser}, {Gomboc}, {Lett},
  {Mansfield}, {Marchant}, {Medrano-Cerda}, {Mottram}, {Raback}, {Scott},
  {Tomlinson}, \& {Zamanov}}]{Steele2004SPIE.5489..679S}
{Steele}, I.~A., {Smith}, R.~J., {Rees}, P.~C., {et~al.} 2004, in Society of
  Photo-Optical Instrumentation Engineers (SPIE) Conference Series, Vol. 5489,
  Ground-based Telescopes, ed. J.~{Oschmann},
  Jacobus~M.\hypersetup{urlcolor=blue},
  \href{https://ui.adsabs.harvard.edu/abs/2004SPIE.5489..679S}{679--692}

\bibitem[{{Stritzinger} {et~al.}(2023){Stritzinger}, {Valerin}, {Elias-Rosa},
  {Fraser}, {Galbany}, {Gutierrez}, {Kankare}, {Kotak}, {Moran}, {Lundqvist},
  {Matilainen}, {Reguitti}, {Reynolds}, {Salmaso}, \&
  {Shappee}}]{Stritzinger2023TNSAN.145....1S}
{Stritzinger}, M., {Valerin}, G., {Elias-Rosa}, N., {et~al.} 2023, Transient
  Name Server AstroNote,
  \hypersetup{urlcolor=blue}\href{https://ui.adsabs.harvard.edu/abs/2023TNSAN.145....1S}{145,
  1}

\bibitem[{{Strotjohann} {et~al.}(2021){Strotjohann}, {Ofek}, {Gal-Yam},
  {Bruch}, {Schulze}, {Shaviv}, {Sollerman}, {Filippenko}, {Yaron}, {Fremling},
  {Nordin}, {Kool}, {Perley}, {Ho}, {Yang}, {Yao}, {Soumagnac}, {Graham},
  {Barbarino}, {Tartaglia}, {De}, {Goldstein}, {Cook}, {Brink}, {Taggart},
  {Yan}, {Lunnan}, {Kasliwal}, {Kulkarni}, {Nugent}, {Masci}, {Rosnet},
  {Adams}, {Andreoni}, {Bagdasaryan}, {Bellm}, {Burdge}, {Duev}, {Dugas},
  {Frederick}, {Goldwasser}, {Hankins}, {Irani}, {Karambelkar}, {Kupfer},
  {Liang}, {Neill}, {Porter}, {Riddle}, {Sharma}, {Short}, {Taddia},
  {Tzanidakis}, {van Roestel}, {Walters}, \&
  {Zhuang}}]{Strotjohann2021ApJ...907...99S}
{Strotjohann}, N.~L., {Ofek}, E.~O., {Gal-Yam}, A., {et~al.} 2021,
  \hypersetup{urlcolor=magenta}\href{https://doi.org/10.3847/1538-4357/abd032}{\apj},
  \hypersetup{urlcolor=blue}\href{https://ui.adsabs.harvard.edu/abs/2021ApJ...907...99S}{907,
  99}

\bibitem[{{Sukhbold} {et~al.}(2016){Sukhbold}, {Ertl}, {Woosley}, {Brown}, \&
  {Janka}}]{Sukhbold2016ApJ...821...38S}
{Sukhbold}, T., {Ertl}, T., {Woosley}, S.~E., {Brown}, J.~M., \& {Janka}, H.~T.
  2016,
  \hypersetup{urlcolor=magenta}\href{https://doi.org/10.3847/0004-637X/821/1/38}{\apj},
  \hypersetup{urlcolor=blue}\href{https://ui.adsabs.harvard.edu/abs/2016ApJ...821...38S}{821,
  38}

\bibitem[{{Svirski} {et~al.}(2012){Svirski}, {Nakar}, \&
  {Sari}}]{Svirski2012ApJ...759..108}
{Svirski}, G., {Nakar}, E., \& {Sari}, R. 2012,
  \hypersetup{urlcolor=magenta}\href{https://doi.org/10.1088/0004-637X/759/2/108}{\apj},
  \hypersetup{urlcolor=blue}\href{https://ui.adsabs.harvard.edu/abs/2012ApJ...759..108S}{759,
  108}

\bibitem[{{Szentgyorgyi} {et~al.}(1998){Szentgyorgyi}, {Cheimets}, {Eng},
  {Fabricant}, {Geary}, {Hartmann}, {Pieri}, \&
  {Roll}}]{Szentgyorgyi1998SPIE.3355..242S}
{Szentgyorgyi}, A.~H., {Cheimets}, P., {Eng}, R., {et~al.} 1998, in Society of
  Photo-Optical Instrumentation Engineers (SPIE) Conference Series, Vol. 3355,
  Optical Astronomical Instrumentation, ed.
  S.~{D'Odorico}\hypersetup{urlcolor=blue},
  \href{https://ui.adsabs.harvard.edu/abs/1998SPIE.3355..242S}{242--252}

\bibitem[{{Szentgyorgyi} \&
  {Fur{\'e}sz}(2007)}]{Szentgyorgyi2007RMxAC..28..129S}
{Szentgyorgyi}, A.~H., \& {Fur{\'e}sz}, G. 2007, in Revista Mexicana de
  Astronomia y Astrofisica Conference Series, Vol.~28, Revista Mexicana de
  Astronomia y Astrofisica Conference Series, ed.
  S.~{Kurtz}\hypersetup{urlcolor=blue},
  \href{https://ui.adsabs.harvard.edu/abs/2007RMxAC..28..129S}{129--133}

\bibitem[{{Szentgyorgyi} {et~al.}(2005){Szentgyorgyi}, {Geary}, {Latham},
  {Groner}, {Amato}, {Bennett}, {Falco}, {Peters}, {Ordway}, \&
  {Fata}}]{Szentgyorgyi2005AAS...20711010S}
{Szentgyorgyi}, A.~H., {Geary}, J.~G., {Latham}, D.~W., {et~al.} 2005, in
  American Astronomical Society Meeting Abstracts, Vol. 207, American
  Astronomical Society Meeting Abstracts\hypersetup{urlcolor=blue},
  \href{https://ui.adsabs.harvard.edu/abs/2005AAS...20711010S}{110.10}

\bibitem[{{Takei} {et~al.}(2022){Takei}, {Tsuna}, {Kuriyama}, {Ko}, \&
  {Shigeyama}}]{Takei2022ApJ...929..177}
{Takei}, Y., {Tsuna}, D., {Kuriyama}, N., {Ko}, T., \& {Shigeyama}, T. 2022,
  \hypersetup{urlcolor=magenta}\href{https://doi.org/10.3847/1538-4357/ac60fe}{\apj},
  \hypersetup{urlcolor=blue}\href{https://ui.adsabs.harvard.edu/abs/2022ApJ...929..177T}{929,
  177}

\bibitem[{{Teja} {et~al.}(2023{\natexlab{\hspace{0pt}a}}){Teja}, {Anupama},
  {Sahu}, {Kurre}, \& {Pramod}}]{Teja2023TNSCR1233....1T}
{Teja}, R.~S., {Anupama}, G., {Sahu}, D., {Kurre}, M., \& {Pramod}.
  2023{\natexlab{\hspace{0pt}a}}, Transient Name Server Classification Report,
  \hypersetup{urlcolor=blue}\href{https://ui.adsabs.harvard.edu/abs/2023TNSCR1233....1T}{2023-1233,
  1}

\bibitem[{{Teja} {et~al.}(2023{\natexlab{\hspace{0pt}b}}){Teja}, {Singh},
  {Basu}, {Anupama}, {Sahu}, {Dutta}, {Swain}, {Nakaoka}, {Pathak}, {Bhalerao},
  {Barway}, {Kumar}, {A.~J.}, {Imazawa}, {Kumar}, \&
  {Kawabata}}]{Teja2023ApJ...954L..12T}
{Teja}, R.~S., {Singh}, A., {Basu}, J., {et~al.}
  2023{\natexlab{\hspace{0pt}b}},
  \hypersetup{urlcolor=magenta}\href{https://doi.org/10.3847/2041-8213/acef20}{\apjl},
  \hypersetup{urlcolor=blue}\href{https://ui.adsabs.harvard.edu/abs/2023ApJ...954L..12T}{954,
  L12}

\bibitem[{{Terreran} {et~al.}(2022){Terreran}, {Jacobson-Gal{\'a}n}, {Groh},
  {Margutti}, {Coppejans}, {Dimitriadis}, {Kilpatrick}, {Matthews}, {Siebert},
  {Angus}, {Brink}, {Filippenko}, {Foley}, {Jones}, {Tinyanont}, {Gall},
  {Pfister}, {Zenati}, {Ansari}, {Auchettl}, {El-Badry}, {Magnier}, \&
  {Zheng}}]{Terreran2022ApJ...926...20T}
{Terreran}, G., {Jacobson-Gal{\'a}n}, W.~V., {Groh}, J.~H., {et~al.} 2022,
  \hypersetup{urlcolor=magenta}\href{https://doi.org/10.3847/1538-4357/ac3820}{\apj},
  \hypersetup{urlcolor=blue}\href{https://ui.adsabs.harvard.edu/abs/2022ApJ...926...20T}{926,
  20}

\bibitem[{{Tokarz} \& {Roll}(1997)}]{Tokarz1997ASPC..125..140T}
{Tokarz}, S.~P., \& {Roll}, J. 1997, in Astronomical Society of the Pacific
  Conference Series, Vol. 125, Astronomical Data Analysis Software and Systems
  VI, ed. G.~{Hunt} \& H.~{Payne}\hypersetup{urlcolor=blue},
  \href{https://ui.adsabs.harvard.edu/abs/1997ASPC..125..140T}{140}

\bibitem[{{Tonry} {et~al.}(2018){Tonry}, {Denneau}, {Heinze}, {Stalder},
  {Smith}, {Smartt}, {Stubbs}, {Weiland}, \& {Rest}}]{Tonry2018PASP..130f4505T}
{Tonry}, J.~L., {Denneau}, L., {Heinze}, A.~N., {et~al.} 2018,
  \hypersetup{urlcolor=magenta}\href{https://doi.org/10.1088/1538-3873/aabadf}{\pasp},
  \hypersetup{urlcolor=blue}\href{https://ui.adsabs.harvard.edu/abs/2018PASP..130f4505T}{130,
  064505}

\bibitem[{{Tsang} {et~al.}(2022){Tsang}, {Kasen}, \&
  {Bildsten}}]{Tsang2022ApJ...936...28T}
{Tsang}, B. T.~H., {Kasen}, D., \& {Bildsten}, L. 2022,
  \hypersetup{urlcolor=magenta}\href{https://doi.org/10.3847/1538-4357/ac83bc}{\apj},
  \hypersetup{urlcolor=blue}\href{https://ui.adsabs.harvard.edu/abs/2022ApJ...936...28T}{936,
  28}

\bibitem[{{Tsuna} {et~al.}(2023{\natexlab{\hspace{0pt}a}}){Tsuna}, {Murase}, \&
  {Moriya}}]{Tsuna2023ApJ...952..115T}
{Tsuna}, D., {Murase}, K., \& {Moriya}, T.~J. 2023{\natexlab{\hspace{0pt}a}},
  \hypersetup{urlcolor=magenta}\href{https://doi.org/10.3847/1538-4357/acdb71}{\apj},
  \hypersetup{urlcolor=blue}\href{https://ui.adsabs.harvard.edu/abs/2023ApJ...952..115T}{952,
  115}

\bibitem[{{Tsuna} {et~al.}(2021){Tsuna}, {Takei}, {Kuriyama}, \&
  {Shigeyama}}]{Tsuna2021PASJ...73.1128}
{Tsuna}, D., {Takei}, Y., {Kuriyama}, N., \& {Shigeyama}, T. 2021,
  \hypersetup{urlcolor=magenta}\href{https://doi.org/10.1093/pasj/psab063}{\pasj},
  \hypersetup{urlcolor=blue}\href{https://ui.adsabs.harvard.edu/abs/2021PASJ...73.1128T}{73,
  1128}

\bibitem[{{Tsuna} {et~al.}(2023{\natexlab{\hspace{0pt}b}}){Tsuna}, {Takei}, \&
  {Shigeyama}}]{Tsuna2023ApJ...945..104}
{Tsuna}, D., {Takei}, Y., \& {Shigeyama}, T. 2023{\natexlab{\hspace{0pt}b}},
  \hypersetup{urlcolor=magenta}\href{https://doi.org/10.3847/1538-4357/acbbc6}{\apj},
  \hypersetup{urlcolor=blue}\href{https://ui.adsabs.harvard.edu/abs/2023ApJ...945..104T}{945,
  104}

\bibitem[{{UnWISE Team}(2021)}]{unWISE}
{UnWISE Team}. 2021, unWISE Catalog,  IPAC,
  \hypersetup{urlcolor=magenta}doi:\href{https://doi.org/10.26131/IRSA525}{10.26131/IRSA525}

\bibitem[{{Valenti} {et~al.}(2016){Valenti}, {Howell}, {Stritzinger}, {Graham},
  {Hosseinzadeh}, {Arcavi}, {Bildsten}, {Jerkstrand}, {McCully}, {Pastorello},
  {Piro}, {Sand}, {Smartt}, {Terreran}, {Baltay}, {Benetti}, {Brown},
  {Filippenko}, {Fraser}, {Rabinowitz}, {Sullivan}, \&
  {Yuan}}]{Valenti2016MNRAS.459.3939V}
{Valenti}, S., {Howell}, D.~A., {Stritzinger}, M.~D., {et~al.} 2016,
  \hypersetup{urlcolor=magenta}\href{https://doi.org/10.1093/mnras/stw870}{\mnras},
  \hypersetup{urlcolor=blue}\href{https://ui.adsabs.harvard.edu/abs/2016MNRAS.459.3939V}{459,
  3939}

\bibitem[{{Van Dyk}(2017)}]{VanDyk2017hsn..book..693V}
{Van Dyk}, S.~D. 2017, in Handbook of Supernovae, ed. A.~W. {Alsabti} \&
  P.~{Murdin}

\bibitem[{{Vasylyev} {et~al.}(2023){Vasylyev}, {Yang}, {Filippenko}, {Patra},
  {Brink}, {Wang}, {Chornock}, {Margutti}, {Gates}, {Burgasser}, {Karpoor},
  {LeBaron}, {Softich}, {Theissen}, {Wiston}, \&
  {Zheng}}]{Vasylyev2023arXiv230701268V}
{Vasylyev}, S.~S., {Yang}, Y., {Filippenko}, A.~V., {et~al.} 2023,
  \hypersetup{urlcolor=magenta}\href{https://doi.org/10.48550/arXiv.2307.01268}{arXiv
  e-prints},
  \hypersetup{urlcolor=magenta}\href{https://doi.org/10.48550/arXiv.2307.01268}{arXiv}{:}\hypersetup{urlcolor=blue}\href{https://ui.adsabs.harvard.edu/abs/2023arXiv230701268V}{2307.01268}

\bibitem[{{Virtanen} {et~al.}(2020){Virtanen}, {Gommers}, {Oliphant},
  {Haberland}, {Reddy}, {Cournapeau}, {Burovski}, {Peterson}, {Weckesser},
  {Bright}, {van der Walt}, {Brett}, {Wilson}, {Millman}, {Mayorov}, {Nelson},
  {Jones}, {Kern}, {Larson}, {Carey}, {Polat}, {Feng}, {Moore}, {VanderPlas},
  {Laxalde}, {Perktold}, {Cimrman}, {Henriksen}, {Quintero}, {Harris},
  {Archibald}, {Ribeiro}, {Pedregosa}, {van Mulbregt}, \& {SciPy 1. 0
  Contributors}}]{SciPy2020NatMe..17..261V}
{Virtanen}, P., {Gommers}, R., {Oliphant}, T.~E., {et~al.} 2020,
  \hypersetup{urlcolor=magenta}\href{https://doi.org/10.1038/s41592-019-0686-2}{Nature
  Methods},
  \hypersetup{urlcolor=blue}\href{https://ui.adsabs.harvard.edu/abs/2020NatMe..17..261V}{17,
  261}

\bibitem[{{Wang} \& {Chen}(2019)}]{Wang2019ApJ...877..116W}
{Wang}, S., \& {Chen}, X. 2019,
  \hypersetup{urlcolor=magenta}\href{https://doi.org/10.3847/1538-4357/ab1c61}{\apj},
  \hypersetup{urlcolor=blue}\href{https://ui.adsabs.harvard.edu/abs/2019ApJ...877..116W}{877,
  116}

\bibitem[{{Waskom} {et~al.}(2020){Waskom}, {Botvinnik}, {Ostblom}, {Lukauskas},
  {Hobson}, {MaozGelbart}, {Gemperline}, {Augspurger}, {Halchenko}, {Cole},
  {Warmenhoven}, {De Ruiter}, {Pye}, {Hoyer}, {Vanderplas}, {Villalba},
  {Kunter}, {Quintero}, {Bachant}, {Martin}, {Meyer}, {Swain}, {Miles},
  {Brunner}, {O'Kane}, {Yarkoni}, {Williams}, \&
  {Evans}}]{seaborn2020zndo...3629446W}
{Waskom}, M., {Botvinnik}, O., {Ostblom}, J., {et~al.} 2020, {mwaskom/seaborn:
  v0.10.0 (January 2020)}, v0.10.0, Zenodo,  Zenodo,
  \hypersetup{urlcolor=magenta}doi:\href{https://doi.org/10.5281/zenodo.3629446}{10.5281/zenodo.3629446}

\bibitem[{{Wright} {et~al.}(2010){Wright}, {Eisenhardt}, {Mainzer}, {Ressler},
  {Cutri}, {Jarrett}, {Kirkpatrick}, {Padgett}, {McMillan}, {Skrutskie},
  {Stanford}, {Cohen}, {Walker}, {Mather}, {Leisawitz}, {Gautier}, {McLean},
  {Benford}, {Lonsdale}, {Blain}, {Mendez}, {Irace}, {Duval}, {Liu}, {Royer},
  {Heinrichsen}, {Howard}, {Shannon}, {Kendall}, {Walsh}, {Larsen}, {Cardon},
  {Schick}, {Schwalm}, {Abid}, {Fabinsky}, {Naes}, \&
  {Tsai}}]{Wright2010AJ....140.1868W}
{Wright}, E.~L., {Eisenhardt}, P. R.~M., {Mainzer}, A.~K., {et~al.} 2010,
  \hypersetup{urlcolor=magenta}\href{https://doi.org/10.1088/0004-6256/140/6/1868}{\aj},
  \hypersetup{urlcolor=blue}\href{https://ui.adsabs.harvard.edu/abs/2010AJ....140.1868W}{140,
  1868}

\bibitem[{{Yamanaka} {et~al.}(2023){Yamanaka}, {Fujii}, \&
  {Nagayama}}]{Yamanaka2023PASJ..tmp...66Y}
{Yamanaka}, M., {Fujii}, M., \& {Nagayama}, T. 2023,
  \hypersetup{urlcolor=magenta}\href{https://doi.org/10.1093/pasj/psad051}{\pasj},
  \hypersetup{urlcolor=magenta}\href{https://doi.org/10.1093/pasj/psad051}{arXiv}{:}\hypersetup{urlcolor=blue}\href{https://ui.adsabs.harvard.edu/abs/2023PASJ..tmp...66Y}{2306.00263}

\bibitem[{{Yaron}(2023)}]{Yaron2023TNSCR1233....1T}
{Yaron}, O. 2023, Transient Name Server Classification Report,
  \hypersetup{urlcolor=blue}\href{https://ui.adsabs.harvard.edu/abs/2023TNSCR1267....1Y}{2023-1267,
  1}

\bibitem[{{Yaron} {et~al.}(2017){Yaron}, {Perley}, {Gal-Yam}, {Groh}, {Horesh},
  {Ofek}, {Kulkarni}, {Sollerman}, {Fransson}, {Rubin}, {Szabo}, {Sapir},
  {Taddia}, {Cenko}, {Valenti}, {Arcavi}, {Howell}, {Kasliwal}, {Vreeswijk},
  {Khazov}, {Fox}, {Cao}, {Gnat}, {Kelly}, {Nugent}, {Filippenko}, {Laher},
  {Wozniak}, {Lee}, {Rebbapragada}, {Maguire}, {Sullivan}, \&
  {Soumagnac}}]{Yaron2017NatPh..13..510Y}
{Yaron}, O., {Perley}, D.~A., {Gal-Yam}, A., {et~al.} 2017,
  \hypersetup{urlcolor=magenta}\href{https://doi.org/10.1038/nphys4025}{Nature
  Physics},
  \hypersetup{urlcolor=blue}\href{https://ui.adsabs.harvard.edu/abs/2017NatPh..13..510Y}{13,
  510}

\bibitem[{{Yoon} \& {Cantiello}(2010)}]{Yoon2010ApJ...717L..62Y}
{Yoon}, S.-C., \& {Cantiello}, M. 2010,
  \hypersetup{urlcolor=magenta}\href{https://doi.org/10.1088/2041-8205/717/1/L62}{\apjl},
  \hypersetup{urlcolor=blue}\href{https://ui.adsabs.harvard.edu/abs/2010ApJ...717L..62Y}{717,
  L62}

\bibitem[{{Zacharias} {et~al.}(2013){Zacharias}, {Finch}, {Girard}, {Henden},
  {Bartlett}, {Monet}, \& {Zacharias}}]{Zacharias2013AJ....145...44Z}
{Zacharias}, N., {Finch}, C.~T., {Girard}, T.~M., {et~al.} 2013,
  \hypersetup{urlcolor=magenta}\href{https://doi.org/10.1088/0004-6256/145/2/44}{\aj},
  \hypersetup{urlcolor=blue}\href{https://ui.adsabs.harvard.edu/abs/2013AJ....145...44Z}{145,
  44}

\bibitem[{{Zackay} {et~al.}(2016){Zackay}, {Ofek}, \&
  {Gal-Yam}}]{Zackay2016ApJ...830...27Z}
{Zackay}, B., {Ofek}, E.~O., \& {Gal-Yam}, A. 2016,
  \hypersetup{urlcolor=magenta}\href{https://doi.org/10.3847/0004-637X/830/1/27}{\apj},
  \hypersetup{urlcolor=blue}\href{https://ui.adsabs.harvard.edu/abs/2016ApJ...830...27Z}{830,
  27}

\bibitem[{{Zhang} {et~al.}(2023){Zhang}, {Fan}, {Zheng}, {Zhang}, \&
  {He}}]{Zhang2023TNSAN.132....1Z}
{Zhang}, Y., {Fan}, Z., {Zheng}, J., {Zhang}, J., \& {He}, M. 2023, Transient
  Name Server AstroNote,
  \hypersetup{urlcolor=blue}\href{https://ui.adsabs.harvard.edu/abs/2023TNSAN.132....1Z}{132,
  1}

\end{thebibliography}



\end{document}